\documentclass[11pt,a4paper]{article}
\pdfoutput=1
\usepackage{mystylefile}
\usepackage{latexsym,amsfonts,amsmath,amssymb,textgreek,upgreek}
\usepackage{amsthm}
\usepackage{mathtools}
\usepackage{bbm}
\usepackage{graphicx}
\usepackage{color}
\usepackage{amsfonts}
\usepackage{caption}
\usepackage{subcaption}
\usepackage[normalem]{ulem}
\usepackage{comment}
\usepackage{url}
\usepackage{slashed}
\usepackage{bm}
\usepackage{hhline}
\usepackage{physics}
\usepackage{booktabs}
\usepackage{siunitx}
\usepackage{tikz}
\usepackage{tabu}

\usepackage{footmisc}

\usepackage{tikz}
\usetikzlibrary{arrows}

\newcommand{\CJ}{\mathcal{J}}

\newcommand{\CH}{\mathcal{H}}

\renewcommand{\Im}{{\rm Im}}
\renewcommand{\Re}{{\rm Re}}
\renewcommand{\Tr}{\mbox{Tr}}

\newcommand{\sgn}{\mbox{sgn}}

\newcommand{\IR}{\mathbb{R}}
\newcommand{\IC}{\mathbb{C}}
\newcommand{\IZ}{\mathbb{Z}}

\newcommand{\IN}{\mathbb{N}}
\newcommand{\IH}{\mathbb{H}}

\def\J{{\mathcal J}}
\def\R{\mathbb{R}}
\def\Z{\mathbb{Z}}
\def\M{{\mathcal M}}
\def\L{{\mathcal L}}
\def\C{{\mathbb{C}}}

\newcommand\be{\begin{equation}}
\newcommand\ee{\end{equation}}
\newcommand\bea{\begin{eqnarray}}
\newcommand\eea{\end{eqnarray}}

\renewcommand{\=}{\;= \;}

\renewcommand{\d}{\mathrm{d}}

\renewcommand{\a}{\alpha}

\newcommand{\s}{\sigma}
\renewcommand{\t}{\tau}
\renewcommand{\k}{\kappa}

\newcommand{\wt}{\widetilde}

\renewcommand{\Re}{\text{Re}}
\renewcommand{\Im}{\text{Im}}

\newcommand{\p}{\partial}

\newcommand{\ndt}{\noindent}

\renewcommand{\i}{i}

\newcommand{\vth}{\vartheta}

\newcommand{\defeq}{\; \coloneqq \;} 

\def\vth{\vartheta}
\renewcommand{\Im}{\mbox{Im}}
\renewcommand{\Re}{\mbox{Re}}

\newcommand{\zbar}{\overline{z}}
\newcommand{\qbar}{\overline{q}}
\newcommand{\pbar}{\overline{\partial}}
\newcommand{\taubar}{\overline{\tau}}

\newcommand{\sutwogen}[1]{t^{#1}}
\newcommand{\sltwogen}[1]{r^{#1}}

\newcommand{\JA}{\CJ}
\newcommand{\JAtl}{\wt \CJ}

\newcommand{\wst}{v^0}
\newcommand{\wsx}{v^1}
\newcommand{\wsy}{v^2}
\newcommand{\coeq}{\sim}

\newcommand{\lmp}{\rho}
\newcommand{\lmpx}{\rho_1}

\newcommand{\lmpt}{\rho_0}

\newcommand{\AL}{A^L}
\newcommand{\AR}{A^R}

\newcommand{\alphaL}{\alpha_L}
\newcommand{\alphaR}{\alpha_R}

\newcommand{\alphabarL}{\overline{\alpha}_L}
\newcommand{\alphabarR}{\overline{\alpha}_R}

\def\H{\CH}
\def\barL{{\overline{L}}}
\def\bar{\overline}
\def\tilde{\widetilde}
\def\h{\widehat}
\def\CPT{{\sf CPT}} 
\def\k{{\rm k}}
\def\la{\langle}
\def\ra{\rangle}

\title{Localization of strings on group manifolds}

\author{Sameer Murthy$^1$ and Edward Witten$^2$}

\affiliation{${}^1$ Department of Mathematics, King's College London, The Strand, London WC2R 2LS, UK}
   
\affiliation{${}^2$ School of Natural Sciences, Institute for Advanced Study, Princeton, NJ, USA}

\emailAdd{sameer.murthy@kcl.ac.uk, witten@ias.edu}

\abstract{We compute the partition function of the WZW model with target a compact Lie group $G$ by 
adapting a method used by Choi and Takhtajan to compute the heat kernel of the group manifold.   
The basic idea is to compute the partition function of a supersymmetric version of the WZW model 
using a form of supersymmetric localization and then use the fact that, since the fermions of the 
supersymmetric WZW model are actually decoupled from the bosons, this also determines the partition
function of the purely bosonic WZW model.   
The result is a formula for the partition function  as a sum over contributions from abelian classical
solutions. We verify for $G=SU(2)$ that this formula agrees with the result for the same partition function 
that comes from the Weyl-Kac character formula. We extend the method of supersymmetric localization 
to certain related models such as the $SL(2,\R)$ WZW model and a Wick-rotated version of this model in 
which the target space is hyperbolic three-space $H_3^+$.
}

\begin{document}

\maketitle 

\section{Introduction and Review}

In a generic supersymmetric theory, the fermions interact with the bosons and neither bosons nor fermions 
are free fields. However, in the minimal supersymmetric quantum mechanics with target a group manifold, 
the fermions are actually free fields, and in particular are completely decoupled from the bosons.
Choi and Takhtajan~\cite{Choi:2021yuz} defined a novel supersymmetric
localization argument that takes advantage of this, and used it to deduce an exact formula for the heat kernel of
a group manifold as a sum over classical orbits.   
What is remarkable about this result is that a purely bosonic model, namely the quantum mechanics with target 
a group manifold, was analyzed by using the fact that this model, plus a collection of completely decoupled 
free fermions, has a supersymmetric structure.

The basic idea of the present article is to exploit the existence of another supersymmetric model in which 
the fermions are again free fields, completely decoupled from the bosons.
This is the supersymmetric WZW model in two dimensions~\cite{DiVecchia:1984nyg}, with $\mathcal N=1$ 
supersymmetry. We will exploit this unusual property of the supersymmetric WZW model, together with the 
Choi-Takhtajan argument, to obtain exact formulas for certain natural traces in the purely bosonic WZW 
model that generalize the heat kernel of a group manifold  (for details, see eqn.~(\ref{eq:defZbasic}) below).

Consider a free fermion field $\psi_i(t)$ in one dimension or $\psi_i(t,x)$ in two dimensions. If $\psi_i$ 
satisfies periodic boundary conditions,
as is appropriate in computing an index, then we can define the average $\chi_i=\int \d t \, \psi_i(t)$ 
or $\chi_i=\int\d t\,\d x\,\psi_i(t,x)$.
If $\psi_i$ is a free field, then $\chi_i$ is a zero-mode in a very strong sense: it does not appear 
in the action at all, regardless of the values of the bosonic fields.  

Hence in a theory with bosonic and fermionic fields $X$ and $\psi$ and an action $S[X,\psi]$ in which the 
fermions are decoupled and free, we have
\be \label{eq:WitInd0}
  \int  \d \chi_i \, \exp \bigl(-S[X,\psi] \bigr) \= 0 \,, \qquad  i\=1,\dots , n \,.
\ee 
In particular, the equation~\eqref{eq:WitInd0} implies that the supersymmetric index, which is the 
integral over all the variables of~$e^{-S}$, vanishes. 
However, we can potentially get a nonzero result by inserting the $\chi_i$ in the functional integral, and defining
\be
Z \= \int  [DX \, D \psi] \; \chi_1 \dots \chi_n \, \exp(-S) \,.
\ee
In such a model, $Z$ is the product of a path integral in a purely bosonic theory
and a fermionic path integral for which one can give an explicit formula because the fermions are free.
Choi and Takhtajan described conditions under which $Z$ can be computed by supersymmetric localization, 
even though it is not an index in any standard sense.  
For a general introduction to localization methods in path integrals see~\cite{Pestun_2017}.

The procedure is as follows.
Letting $\delta$ denote the action of a linear combination of the supersymmetries, so that $\delta S=0$, 
we deform the action by a $\delta$-exact operator~$\delta V[X,\psi]$ (with fermionic $V$).  
We assume that $\delta$ is chosen so that the supersymmetry algebra generated
by $\delta$ closes off-shell, and that $V$ is such that
\be
\delta^2 V \= 0.
\ee 
In practice, this will be achieved as follows:   $\delta^2$ will be a translation generator, and $V$ is translation-invariant.
We furthermore impose two conditions on the nature of the fermion zero-modes and the deformation. 
Firstly, we assume that the variation of the product of all fermion zero-modes does not contain
a term proportional to that product, or more briefly that
\be \label{eq:delpsinozm}
 \int \d \chi_1 \, \dots \d \chi_n \; \delta (\chi_1 \, \dots \chi_n)  \= 0 \,.  
\ee
In fact, in the models that we study, the following stronger statement is true:
\be \label{eq:delpsinozm1}
 \int \d \chi_i \, \delta \chi_j \;+\;  \int \d \chi_j \, \delta \chi_i \= 0 \,, \qquad  i,j\=1,\dots , n \,.  
\ee
Secondly, we assume that~$V$ and~$\delta V$ do not depend on the fermion zero-modes, i.e.,
\be \label{eq:delVnozm}
 \int \d \chi_i \, V \= 0 \,, \qquad \int \d \chi_i \, \delta V \= 0 \,, \qquad  i\=1,\dots , n \,.   
\ee

Now we consider the deformed path integral 
\be\label{useful}
Z_V(\lambda) \= \int  [DX \, D \psi] \; \chi_1 \dots \chi_n \, \exp \bigl(-S - \lambda \, \delta V \bigr) \,,
\ee
for~$\lambda \ge 0$, and show that it is independent of~$\lambda$. Indeed, 
\be
\begin{split}
\frac{\d}{\d \lambda} Z_V(\lambda) 
& \= - \int  [DX \, D \psi] \; \chi_1 \dots \chi_n \, \delta V \, \exp \bigl(-S - \lambda \, \delta V \bigr) \,, \cr
& \= (-1)^{n+1} \int  [DX \, D \psi] \; \delta \Bigl( \chi_1 \dots \chi_n \,  V \, \exp \bigl(-S - \lambda \, \delta V \bigr) \Bigr) \cr 
& \qquad + (-1)^{n} \int  [DX \, D \psi] \; \delta (\chi_1 \dots \chi_n) \, V \, \exp \bigl(-S - \lambda \, \delta V \bigr) \cr 
& \qquad - \int  [DX \, D \psi] \; \chi_1 \dots \chi_n \, V \, \delta \bigl( \exp \bigl(-S - \lambda \, \delta V \bigr) \bigr) \cr
& \= 0 \,.
\end{split}
\ee
Here, as in the usual proof of localization, the first line on the right-hand side of the second equality
vanishes because it is a total derivative and the third line 
vanishes because of the invariance of the original action as well as the deformation. 
The second line is new compared to the usual proof of localization, 
and this also vanishes because of the conditions~\eqref{eq:delVnozm},
\eqref{eq:delpsinozm} above.
In more detail, the condition  \eqref{eq:delVnozm}  implies that $ V \, \delta \bigl( \exp \bigl(-S - \lambda \, \delta V \bigr) \bigr)$
is independent of the $\chi_i$, and given this, the condition \eqref{eq:delpsinozm} implies the vanishing of 
$\int \d\chi_1\cdots \d\chi_n\, \delta ( \chi_1 \dots \chi_n) \,  V  \exp \bigl(-S - \lambda \, \delta V \bigr) $.

The fact that the integral does not depend  on $\lambda$  implies that $Z = Z(0) = Z(\infty)$
(where  $Z(\infty)=\lim_{\lambda \to \infty} Z(\lambda)$).  If $V$ has been suitably chosen,
$Z(\infty)$  can be evaluated exactly as a sum over the critical points of~$\delta V$. 
Thus we arrive at the formula, 
\be \label{eq:Zlocgen}
Z \= \sum_{\alpha}  \, e^{-S[X_\alpha]} \, Z_\text{1-loop} ,
\ee
where $\alpha$ labels the critical points of $\delta V$, and 
 the expression~$Z_\text{1-loop}$ stands for the integral over  
quadratic fluctuations of~$S+ \lambda \,\delta V$ around each critical point, 
after performing the bosonic as well as the fermionic zero mode integrals in eqn.~\eqref{useful}.

Choi and Takhtajan used this procedure, with suitable  boundary conditions, to compute 
the partition function $\Tr\,e^{-\beta H}$  and more generally a twisted partition function
$\Tr\, g \, e^{-\beta H}$, $g\in G$ (here $g$ acts on $G$ on, say, the left),  
for a purely bosonic theory of a particle moving on a compact group manifold $G$~\cite{Choi:2021yuz}.
Equivalently, they computed the heat kernel of the group manifold.
The partition function and the heat kernel of any quantum mechanical model can be expressed in 
terms of a sum over eigenstates of the Hamiltonian.   For the particular case of a sigma-model with 
target a group manifold, the localization procedure leads to an alternative formula in terms of a sum 
over geodesics. Relationships of this type between a sum over quantum states 
and a sum over classical orbits are important in chaos theory, as studied by Gutzwiller and others, but it is quite
exceptional to have an {\it exact} formula of this nature, such as the one that in the case of a group manifold
comes from supersymmetric localization.

The expression for the heat kernel of a group manifold as a sum over geodesics actually was previously 
known~\cite{Eskin}. However, the derivation of this formula via supersymmetric localization is extremely 
illuminating from a physical point of view.

For our purposes in the present article, the importance of the method of Choi and Takhtajan is that 
it can be applied to any supersymmetric theory in which
the fermions are decoupled and free.  In particular, as we noted at the outset,
 we will apply the same procedure to the supersymmetric WZW model
with target a compact Lie group $G$. 

In that application, we will aim to compute, in the purely bosonic WZW model of $G$, traces of the 
form
\be \label{eq:defZbasic}
\Tr_{\CH}  \; g_L \, g_R^{-1} \, e^{-2 \pi \tau_2 H + 2 \pi i \tau_1 P} \,, 
\ee 
with~$\tau = \tau_1 + i \tau_2 \in \IH$ (the complex upper half plane), and 
\be  
g_L = e^{\alphaL^{}t}  ,\;  g_R = e^{\alphaR^{}\wt t} \, \in \, G  \,, 
\qquad t, \wt t \, \in \, \mathfrak{g} \,, \qquad \alphaL^{}, \alphaR^{} \, \in \, \IC \,.
\ee   
Here $g_L$ and $g_R$ act on $G$ on, respectively, the left and the right; $H$ and $P$ are the Hamiltonian 
and the momentum. 
The analogous observable for the particle moving on the group manifold~$G$
has been calculated by Choi and Takhtajan in~\cite{Choi:2025iqp}. 

The trace (\ref{eq:defZbasic}) can of course be expressed in terms of a sum over a basis of joint eigenvectors 
of the operators~$H$ and~$P$.   Alternatively, we can represent this trace by
the Euclidean functional integral of the WZW model on  a torus.  
Adding free fermions to make the model supersymmetric and applying supersymmetric localization leads 
to an  alternate formula for the trace~\eqref{eq:defZbasic},
in terms of a sum over classical solutions.  These classical solutions are classified by the momentum 
and winding of the string.   As far as we know, the formula
for these traces as a sum of classical solutions is new, although it can certainly be proved directly starting 
with the Weyl-Kac character formula.   In fact, for $G=SU(2)$, we will 
show this 
by comparing the formula in terms of a sum over classical solutions to a formula that comes from a sum over states.

There have been various other applications of supersymmetric observables that are defined by  
absorbing fermion zero-modes in a situation in which the supersymmetric index vanishes.  
Examples include the ``new" supersymmetric index~\cite{Cecotti:1992qh}, the modified elliptic genus of the 
torus~\cite{Maldacena:1999bp},  and helicity supertraces~\cite{Kiritsis:1997gu}. These calculations have 
generally been  done in a  Hamiltonian framework. It is interesting to ask if some of those analyses 
could be made using a variant of the Choi-Takhtajan procedure.    There have also been previous
localization calculations based on decoupling of fermions, for example, in two-dimensional Yang-Mills 
theory \cite{Witten:1992xu}, but not in sigma-models.

The organization of this article is as follows.
In section~\ref{two}, we will practice with the abelian case $G=U(1)=S^1$.  
In this example, the bosons are free as well as the fermions, so there is no surprise
to be able to get an exact formula in terms of a sum over classical orbits.   
But  obtaining this formula via supersymmetric localization helps in understanding the localization procedure.
In section~\ref{sec:su2}, we analyze in detail the
case $G=SU(2)$.  The general case of a compact nonabelian Lie group is similar, while requiring 
more complicated notation.  It is very briefly discussed in section~\ref{sec:othergroups}.  
In section \ref{sec:hthreeplus}, we discuss in a similar way  the $SL(2,\R)$ WZW model and various close cousins of this
such as models with target space $H_3^+$ (hyperbolic three-space) or $H_3^+/\Z$.

Some important technical details are the following.  First of all, the overall sign of the measure in a fermion path integral
involves an arbitrary choice, because reversing the order of two fermions reverses the sign of the measure.   We will make
an arbitrary choice of this overall sign.   The sign choice cancels when we divide the path integral of the supersymmetric WZW model
by the analogous path integral of the decoupled free fermions, so it does not affect our computation of the path integral of the bosonic WZW model.
We will have to be careful with a relative sign between the contributions of different fixed points in the localization calculation, but we treat the overall
sign of the fermion measure as an arbitrary convention.
Second, a real Clifford algebra of rank $k$ with generators $\chi_1,\cdots, \chi_k$  satisfying $\{\chi_i,\chi_j\}=2\delta_{ij}$ has a well known mod 8 periodicity: if and 
only if $k$ is a multiple of 8, the Clifford algebra has an irreducible $\Z_2$-graded representation by real matrices, in which the $\Z_2$ grading
is generated by the real matrix $(-1)^F=\chi_1\chi_2\cdots \chi_k$.   When $k$ is not a multiple of 8, the Clifford algebra
does not have that property, and has what physically would be called an anomaly.  This will be important for us, because the $U(1)$ and $SU(2)$
models that we study have do have an anomaly, corresponding to $k=2$ mod 4.  (The value of $k$ mod~8 will not be important.)
To explain that $k=2$ mod 4, note that the $U(1)$ and $SU(2)$ models have a pair of zero-modes $\psi_0, \widetilde\psi_0$ that have zero momentum on the torus,
and, in the case of the $SU(2)$ model, zero charge under the Cartan subalgebra that is used in the localization procedure.   All other modes come in an even
number of pairs, because a mode of~$\psi$ that has nonzero momentum or nonzero charge is paired in the action with a mode of~$\psi$ of opposite momentum
and charge, and complex conjugation maps this pair to a pair of modes of $\widetilde\psi$.   
So all fermion modes in the path integral comprise an even number of pairs except for the single zero-mode pair $\psi_0,\widetilde\psi_0$.  Our way of dealing with the anomaly is to include a factor of $i$ for every pair of fermion modes
in the fermion measure and also,
in a Hamiltonian approach, in the definition of the operator $(-1)^F$ (so that it always has square 1 and generates the $\Z_2$-grading of fermions;
see a further discussion at the end of Appendix~\ref{orbifoldtheory}).   For zero-modes, this means that the measure is~$i d\psi_0\,d\widetilde\psi_0$ and the operator $(-1)^F$
in a two-dimensional Hilbert space in which $\psi_0$ and $\widetilde\psi_0$ act irreducibly 
is~$i  \psi_0\widetilde\psi_0$.   For non-zero modes, we get an even number of factors
of $i$ in the measure and in the definition of $(-1)^F$,  giving  signs $\pm 1$, which for our purposes we absorb in  arbitrary signs of the fermion measure and of the
operator $(-1)^F$ that generates the $\Z_2$ grading.

\section{Strings on~$S^1$}\label{two}

In this section we consider the two-dimensional superconformal field theory on a torus with 
bosonic target space a circle, corresponding to~$G=U(1)$. 
This example is relatively trivial because the group manifold $U(1)$ is flat and
the fields are all free fields.  But analyzing this example by localization
is useful to set up the various calculations that we need  later 
to study  non-abelian groups.

\medskip

We parametrize  the  worldsheet torus~$T^2$ with coordinates~$(\wsx,\wsy)$ and give it  a flat 
metric~$\d s^2 = (\d v^1)^2+(\d v^2)^2$.
Equivalently, we use complex coordinates~$z=\wsx+ i \wsy$, $\zbar = \wsx-i \wsy$. 
We define~$\p_z \equiv \p = \frac12 (\p_{\wsx} - i \p_{\wsy})$, 
$\p_{\zbar} \equiv \pbar=\frac12 (\p_{\wsx} + i \p_{\wsy})$,
so that~$\p z = \pbar \zbar =1$. 
We define the torus by identifications~$z \coeq z+2\pi \coeq z+ 2 \pi\tau$. 
In terms of real coordinates $v^1,v^2$, the equivalences are
\be\label{realeq} 
(\wsx,\wsy) \; \coeq \; (\wsx+2\pi ,\wsy)\,, \qquad (\wsx,\wsy) \; \coeq \; (\wsx+2\pi \tau_1,\wsy+2\pi \tau_2) \,. 
\ee
The volume form is $\d^2 z \equiv 2 \, \d\wsx \wedge \d\wsy$ 
so that $\frac{1}{4\pi}\int_{T^2} \d^2 z = 2 \pi \t_2$.

The $U(1)$ group manifold is a circle, which we take to have circumference $2\pi R$.  
We can parametrize this manifold by a real bosonic field~$X$ that takes values 
in~$S^1 = \IR/2 \pi \,\IZ$; the metric of the circle is $R^2\,\d X^2$.
The superpartner of $X$ is a  Majorana fermion field with chiral components $(\psi, \wt \psi)$.
Upon going around either cycle of the torus, the boson can come back to itself up to shifts of 
integer multiples of~$2 \pi$. 
For the fermions, we choose a spin structure such that $\psi$ and $\wt\psi$ are periodic around each cycle.

\medskip

The worldsheet action is~\cite{Polchinski:1998rr}
\be \label{eq:wsaction}
S[X,\psi,\wt \psi] \= \frac{R^2}{4\pi} \int_{T^2} \d^2z \, \bigl( \p X \, \pbar X + \psi \, \pbar \, \psi + \wt \psi \, \p \, \wt \psi \, \bigr) \,.
\ee
It is convenient, however, to absorb a factor of~$R$ in all the fields so that~$X$ has period~$2 \pi R$, 
i.e.,  for~$m,w, \in \IZ$, it obeys
\be \label{eq:bosperiodic}
X(z+2\pi,\zbar+2\pi) \= X(z,\zbar) \,+\, 2 \pi  R w \,, \qquad 
X(z+2\pi\tau,\zbar+2\pi \taubar) \= X(z,\zbar) \,+\, 2 \pi R  m \,.
\ee
The action for the rescaled fields is
\be \label{eq:wsaction}
S[X,\psi,\wt \psi] \= \frac{1}{4\pi} \int_{T^2} \d^2z \, \bigl( \p X \, \pbar X + \psi \, \pbar \, \psi + \wt \psi \, \p \, \wt \psi \, \bigr) \,.
\ee

The classical equations of motions are solved when~$\p X, \psi$ are holomorphic, and~$\pbar X, \wt \psi$ 
are anti-holomorphic.\footnote{We also use the terminology left- and right-moving for 
holomorphic and anti-holomorphic, respectively.}    On a flat torus, this implies that $\p X$, etc., are all constant.
The solutions of the equations of motion for $X$
that respect the periodicity conditions~\eqref{eq:bosperiodic} are given (up to a possible additive constant) 
by\footnote{In describing classical solutions, 
we set the fermions to zero.   Zero-modes of the fermions
will be taken into account when we expand around a purely bosonic classical solution.}
\be \label{eq:classsolns}
X_{m,w}(z,\zbar) \= \frac{R}{2 i \t_2} \bigl( m(z - \zbar) + w( \tau \zbar - \taubar z) \bigr) \,, 
\ee
or~$X_{m,w} = R \bigl( m\wsy/\t_2 + w(\wsx-\t_1\wsy/\t_2) \bigr)$ in  Cartesian variables.  
The action of these configurations is 
\be \label{eq:classact}
S[X_{m,w}] \= \frac{1}{4 \pi} \int_{T^2} \d^2z \, \frac{R^2}{4\, \t_2^2} \, |m-w \t|^2 
\= \frac{\pi R^2}{2\, \t_2} \, |m-w \t|^2 \,.
\ee

\bigskip

The action~\eqref{eq:wsaction} is invariant under separate left- and right-moving supercharges 
with the following action,
\be \label{eq:susyvars}
\delta_L X \= - \psi \,, \quad \delta_L \psi \= \p X \,, \qquad \qquad 
\delta_R X \= - \wt \psi \,, \quad \delta_R \wt \psi \= \pbar X \,. 
\ee
The supersymmetric index~$\int  [DX \, D \psi] \, \exp(-S)$  vanishes because of the two fermionic 
zero-modes\footnote{This normalization of the zero-modes differs
by a factor of $\tau_2$ from the normalization stated in the introduction.  That normalization will cancel 
out of the eventual formula for traces in the purely bosonic model
with target $U(1)$.}
\be
\psi_0 \= \oint \d z \, \psi \,, \qquad \wt \psi_0 \= \oint \d \zbar \, \wt \psi \,.
\ee
The modified or reduced partition function obtained by absorbing the fermion zero-modes, 
\be \label{eq:ZT2def}
Z^{\rm{susy}}_{{\mathfrak u}(1)} (R;\tau) \= 
\int  [DX \, D \psi] \; i \, \psi_0 \, \wt \psi_0 \, \exp(-S) 
\ee
is the most basic observable of interest.
We have included a factor of $i$, so that in a Hamiltonian description  the insertion~$\i  \, \psi_0\wt\psi_0$ corresponds
to a hermitian operator (which in fact is $(-1)^F$ in a two-dimensional Hilbert space in which $\psi_0,\wt\psi_0$ act irreducibly;
see the discussion at the end of the introduction).

\subsection{Localized path integral \label{sec:locPIU1}}

Our goal is to calculate the path  integral~\eqref{eq:ZT2def} by localization. 
In the localization argument, we need the supersymmetry algebra to hold off-shell.
The algebra 
\be
\delta_L^2 \= - \p \,, \qquad \delta_R^2 \= - \pbar 
\ee
does hold off-shell.\footnote{
The additional relation $\{ \delta_L \,, \delta_R \} = 0$ only holds on-shell. 
We could also use~$\delta_L + \delta_R$ for localization after introducing additional auxiliary fields 
to ensure that $\{\delta_L,\delta_R\}=0$ off-shell.}  
We use the supercharge~$\delta = \delta_R$, and choose the localizing deformation to be
\be \label{eq:defVU1}
V 
\= -  \int_{T^2} \d^2z \;  \p \wt \psi \; \p \pbar X\,,
\ee
so that
\be  \label{eq:defdelVU1}
\delta V \=  \int_{T^2} \d^2z \, \bigl(\, (\p \pbar X )^2 +   \p \wt \psi \; \p \pbar \wt \psi\, \bigr) \,.
\ee
This deformation obeys the conditions laid out in the previous section. 
Note that, although the WZW model is conformally invariant, in the localization computation 
we explicitly break conformal invariance. The localizing term $\delta V$ has dimensions of $(\rm{length})^{-2}$.    
We could make this explicit by including a factor $\ell^{-2}$ in eqns.~\eqref{eq:defVU1}, \eqref{eq:defdelVU1}, 
and elsewhere, where $\ell$ is a parameter with dimensions of length.  
However, we will just work in units with~$\ell=1$. 

The zeroes of the bosonic part of $\delta V$ are precisely the solutions of the original classical equation 
of motion ~$\p \pbar X = 0$.
The solutions that respect the periodicity conditions were already described in~\eqref{eq:classsolns},
and their action was given in~\eqref{eq:classact}.

\bigskip

In order to calculate the localized path integral~\eqref{eq:Zlocgen} we need to 
calculate the one-loop integral around the above solutions.   Taking $\lambda\to\infty$
and rescaling the non-constant modes of $X$ and $\wt \psi$ to compensate, we arrive at 
\be \label{eq:Z1loopU1}
Z^{(m,w)}_\text{1-loop}
\= \int [DX] \exp \Bigl(- \int_{T^2} \d^2z \, (\p \pbar X )^2 \Bigr)
\int [D\psi] \,  [D\wt \psi] \,
\exp \Bigl(- \int_{T^2} \d^2z \, \bigl( \psi \; \pbar \, \psi +\p \wt \psi \; \p \pbar  \, \wt \psi \, \bigr) \Bigr)\,.
\ee
There is no rescaling of the zero-modes of $X$ and $\wt \psi$, which remain as zero-modes for any $\lambda$,
and likewise there is no rescaling of $\psi$, because again its action does not depend on $\lambda$.
The fact that one of the fermion fields requires no rescaling is not typical of calculations based on 
supersymmetric localization, and it is possible because the fermions are free.  Because
the rescaling applies to the non-constant modes of one bosonic field and one fermionic field, 
it produces no change in the measure.  
This statement holds on a mode-by-mode basis.  Supersymmetry
pairs the non-constant modes of $X$ and $\wt\psi$. For each pair $u|\zeta$ consisting of a bosonic variable $u$
and a fermionic variable $\zeta$, the measure is a multiple of $[\d u|\d \zeta]$ and is invariant under the joint scaling
$u|\zeta\to\lambda u|\lambda\zeta$.   We discuss
the UV regulator in more detail  in defining the one-loop determinant in Section~\ref{renormalization}.

The integral~\eqref{eq:Z1loopU1} runs over the field fluctuations 
around any classical solution labeled by~$(m,w)$ as  in~\eqref{eq:bosperiodic}.  
Because the kinetic operators in eqn.~(\ref{eq:Z1loopU1}) do not depend on $(m,w)$, the path integral
governing the small fluctuations does not depend on $(m,w)$ and we will drop the superscript
from $Z_{\text{1-loop}}$.

Let us first look at the integral over the bosonic field~$X$ in~\eqref{eq:Z1loopU1}. 
This field can be separated into a zero-mode $x$ and non-zero-modes that are 
orthogonal to it: $X(z,\zbar) = x+ X'(z,\zbar)$, with the 
functional measure~$[DX] = \d x \, [DX']$.
The integral over the bosonic zero-mode gives the volume of the circle~$2 \pi R$. 
The integral over the non-zero-modes gives
\be
\frac{1}{ \text{det}^{\,'}  \, \p \, \pbar } \int  [D\delta X'] \, \exp \Bigl(-\frac{1}{4\pi} \int_{T^2} \d^2z \, (\delta X')^2 \Bigr) \,,
\ee
where the first factor is regarded as an infinite product over the non-zero eigenvalues of the 
differential operator~$\p \, \pbar $.
The second factor is fixed by the argument of ultralocality, i.e., by normalizing 
the Gaussian integral of the fluctuations of the full field around the vacuum to one:
\be
\int_{-\infty}^\infty \d x \, \exp \Bigl(-\frac{1}{4\pi} \int_{T^2} \d^2z \, x^2\Bigr) \; 
\int  [D\delta X'] \, \exp \Bigl(-\frac{1}{4\pi} \int_{T^2} \d^2z \, (\delta X')^2 \Bigr) \= 1 \,.
\ee
Recalling that~$\frac{1}{4 \pi} \int_{T^2} \d^2z = 2 \pi \t_2$, the first integral is calculated to be
\be
\int_{-\infty}^\infty \d x \, \exp \Bigl(-\frac{1}{4\pi} \int_{T^2} \d^2z \, x^2\Bigr) 
 \= \frac{1}{\sqrt{2\t_2}} \,.
\ee
Thus we obtain, including the zero-mode integral,  
\be
\int [DX] \exp \Bigl(- \int_{T^2} \d^2z \, (\p \pbar X )^2 \Bigr)
\= \frac{2 \pi R \, \sqrt{2\t_2}}{ \text{det}^{\,'}  \, \p \, \pbar } \,.
\ee

\bigskip

The fermionic integral in~\eqref{eq:Z1loopU1} is also calculated by a similar set of steps. 
The fermionic field~$\psi(z,\zbar)$ is separated into the zero-modes $\psi_0$ and non-zero-modes that are 
orthogonal to it as $\psi (z,\zbar) = \psi_0+ \psi'(z,\zbar)$ (and similarly for~$\wt \psi \,$), with the  
functional measure~$[D\psi] [D\wt \psi] = i \, \d \psi_0 \,  \d \wt \psi_0 \, [D\psi'] [D\wt \psi']$.
The integral over the non-zero-modes gives
\be
 \Bigl( \text{Pf}^{\,'} \, \bigl( i\pbar \, \bigr) \; \text{Pf}^{\,'} \bigl(-i\p^2 \pbar \, \bigr) \, \Bigr)
\int  [D\psi'] \, [D\wt \psi'] \,\exp \Bigl(\frac{i}{4\pi} \int_{T^2} \d^2z \, \bigl( \psi' \, \psi' +  \wt \psi' \, \wt \psi' \,  \bigr) \Bigr) \,.
\ee
The first factor equals, up to an overall sign, $\bigl(\text{det}^{\,'}   \p^2 \; \pbar{}^2 \bigr)^\frac12 = \text{det}^{\,'} \p \; \pbar$. 
The second factor is fixed by the argument of ultralocality, i.e., by normalizing the fermionic fluctuations as: 
\be
\int_{-\infty}^\infty i \, \d \psi_0 \,  \d \wt \psi_0 \, \exp \Bigl(\frac{i}{4\pi} \int_{T^2} \d^2z \, \psi_0 \, \wt \psi_0\Bigr) \; 
\int  [D\psi'] \, [D\wt \psi'] \,\exp \Bigl(\frac{i}{4\pi} \int_{T^2} \d^2z \, \bigl( \psi' \, \psi' +  \wt \psi' \, \wt \psi' \,  \bigr) \Bigr) 
  \= 1 \,.
\ee
Including the zero mode integral, we obtain
\be
\begin{split}
& \int [D\psi] \,  [D\wt \psi] \, i \psi_0 \, \wt \psi_0 \, 
\exp \Bigl(- \int_{T^2} \d^2z \, \bigl( \psi \; \pbar \, \psi +\p \wt \psi \; \p \pbar  \, \wt \psi \, \bigr) \Bigr) \\
& \qquad \= \bigl(\text{det}^{\,'}  \p \; \pbar \bigr) \, 
  \biggl( \int i \, \d\psi_0 \, \d \wt \psi_0  \, i \psi_0 \, \wt \psi_0 \biggr)
    \biggl( \int i \, \d\psi_0 \, \d \wt \psi_0  \, \exp \Bigl(\frac{i}{4\pi} \int_{T^2} \d^2 z \, \psi_0 \, \wt \psi_0 \Bigr) \biggr)^{-1}\\
&  \qquad\=  \frac{ \text{det}^{\,'}  \p \; \pbar }{2\pi \t_2} \,. 
\end{split}
\ee

\bigskip

Now we put together the bosonic and fermionic pieces. There is a mode-by-mode cancellation 
of the fermionic Pfaffian and the bosonic determinant that give the path integral over non-zero modes,
leading  to  
\be \label{eq:ans1loop}
Z_\text{1-loop} 
\= \frac{\sqrt{2} R}{\sqrt{\t_2}} \,.
\ee
So  the localization formula~\eqref{eq:Zlocgen} leads to 
\be\label{u(1)answer}
Z^\text{susy}_{{\mathfrak u}(1)}(R;\t) \= 
\sqrt{\frac{2}{\t_2}}  \, R
\sum_{m,w \in \IZ} \exp \Bigl(-\frac{\pi R^2}{2\t_2} \, |m-w \t|^2\Bigr) \,.
\ee

\bigskip

Note that we could calculate the determinants for bosons and fermions separately 
(and observe the cancellation) by using a basis of eigenfunctions of the 
above differential operators and regulating the resulting infinite products. 
Indeed, this calculation is needed if our goal is to obtain the purely bosonic functional integral,
which can be obtained from eqn.~(\ref{u(1)answer}) by dividing by the fermionic partition function
with the insertion of~$(-1)^F$.
As we explain below, in the Hamiltonian formalism, the latter result is $2|\eta(\tau)|^2$, with 
the factor of~$|\eta(\tau)|^2$ coming from the fermionic oscillator modes, and the factor~2 from the fermionic 
zero modes.
So, after dividing by the fermion path integral, we learn that the partition function of the purely bosonic
model is
\be\label{u(1)bosanswer}
Z_{{\mathfrak u}(1)}(R;\t) \= 
\frac{R}{\sqrt{2 \t_2}} \frac{1}{|\eta(\t)|^2}
\sum_{m,w \in \IZ} \exp \Bigl(-\frac{\pi R^2}{2\t_2} \, |m-w \t|^2\Bigr) \,.
\ee
In a conventional approach, without ever introducing the fermions, the factor of~$1/4 \pi \tau_2 |\eta(\tau)|^2$ would come 
from integrating over the nonzero (oscillator) modes of $X$. 
The answer~\eqref{u(1)bosanswer} agrees, for example, with eqn.~(8.2.11) of \cite{Polchinski:1998rq}.

\subsection{Twisting \label{sec:twisting}}

Next, we analyze a twisted version of the same functional integral.
We impose the condition that~$X$ is shifted by~$2a, 2b \in \IR$ in going around the 
space and time circles.\footnote{We include a factor of 2 that looks rather arbitrary
for the $U(1)$ theory discussed here, but which will lead to a smoother match with
 the~$SU(2)$ partition function derived in section \ref{sec:su2}. }
Thus~$X$ now obeys 
\be \label{eq:bostwisted}
X(z+2\pi,\zbar+2\pi) \= X(z,\zbar) \,+\, 2 \pi R (w-2a) \,, \quad 
X(z+2\pi\tau,\zbar+2\pi \taubar)  \= X(z,\zbar) \,+\, 2 \pi R (m+2b) \,.
\ee with $(m,w)\in\IZ$.
We define $\alpha=a\tau+b$ and denote as~$Z^\text{susy}_{{\mathfrak u}(1)} (R;\t,\a)$  
the path integral as in~\eqref{eq:ZT2def} but with the twisted periodicity condition on $X$.
This twisting could be achieved by coupling the model to a background $U(1)$ gauge field $A$ that
gauges the symmetry of shifting $X$ by a constant.   If $A$ is flat but has appropriate holonomies 
around 1-cycles in~$T^2$, coupling to $A$ is equivalent to the twisting defined in \eqref{eq:bostwisted}.   
We will adopt that point of view in discussing sigma-models
with non-abelian target space, but for the free theory with~$U(1)$ target, this is not necessary.

The solutions of the classical equations of motion with these new periodicity conditions are 
\be \label{eq:twistedsolns}
X^{\a}_{m,w}(z,\zbar) \= \frac{R}{2i \t_2} \bigl( (m+2b)(z - \zbar) + (w-2a)( \tau \zbar - \taubar z) \bigr) \,.
\ee
The fluctuations are, as before, periodic on the torus. 
The effect of this twist is equivalent to the 
replacements~$\p X \mapsto \p X +  R\, \overline{\alpha}/i\t_2$, $\pbar X \mapsto \pbar X -  R\, \a/i\t_2$, 
with $\a = a \t+b$, in the original action~\eqref{eq:wsaction}. 
The action of the twisted theory, therefore, is invariant under the transformation given by 
implementing the same replacements\footnote{This modification of the supersymmetries to account 
for the twists looks more natural if the twists are interpreted in terms of a coupling to a background 
gauge field, as mentioned in the last paragraph.}  in the original supersymmetry transformations~\eqref{eq:susyvars}.

The action of these twisted solutions is
\be
S[X^\a_{m,w}] \= \frac{\pi R^2}{2\t_2} \, (m-w \t + 2\alpha)(m-w \taubar +2 \overline{\alpha}) \,.
\ee
The fluctuations around such a classical solution are periodic, so the one-loop contribution to the localized path
integral  is independent of~$\a$,
and equals the result at~$\a=0$ given in~\eqref{eq:ans1loop}. 
Thus we reach the final answer for the path integral, 
\be \label{eq:ZT2PI}
Z^\text{susy}_{{\mathfrak u}(1)} (R;\t,\a) \= 
\sqrt{\frac{2}{\t_2}} \,  R
 \sum_{m,w \in \IZ} \exp \Bigl(-\frac{\pi R^2}{2\t_2} \,|m-w \t + 2 \alpha|^2 \Bigr) \,.
\ee

Once again, the partition function of the purely bosonic twisted model is obtained by multiplying this 
by $1/2|\eta(\tau)|^2$, to cancel the one-loop contribution from the fermions.
At special values of the radius of the circle, there is an enhancement of symmetry, 
e.g.~the bosonic theory at~$R=\sqrt{2}$ is equivalent to the bosonic theory~$su(2)_{k=1}$. 
We show explicitly in Appendix~\ref{app:u1su2eq} that the twisted partition functions of the two models 
are equal.

\subsection{Equivalence to Hamiltonian trace formula and the sum over characters \label{sec:U1Hamtrace}}

In the Hamiltonian formalism, the untwisted boson carries left- and right-momentum
\be \label{eq:kLRppbarX}
k_L \= \frac{R}{2\pi} \oint \d z \, \p X \,, \qquad k_R \= -\frac{R}{2\pi}  \oint \d \zbar \, \pbar X \,,
\ee
which are related to the momentum and winding as
\be \label{eq:klkrdefs}
k_L \= \frac{n}{R} + \frac{wR}{2} \,, \qquad 
k_R \= \frac{n}{R} - \frac{wR}{2} \,.  
\ee

\medskip

The twists around the space and time directions are treated in different fashions in the Hamiltonian
formalism. The twist in the spatial direction (the $\text{Re}(z)$ direction) is implemented by 
changing the Hilbert space to a twisted version~$\CH_a$ obtained by quantizing fields 
that satisfy $X(z+2\pi)=X(z)-4\pi a $ (mod $2\pi$).
This shifts the winding  in~\eqref{eq:klkrdefs}  by $w \mapsto w -2a$ (still with integer $w$)
without affecting  the oscillator modes of the bosons or fermions. 
The twist in the time direction is incorporated by including an explicit 
factor of~$e^{- 4 \pi  ib P}$, which is the operator that generates a shift in~$X$ by~$-4\pi b$.
Thus  with~$q=e^{2 \pi i \tau}$, $\a=a\t+b$ as before, the partition function is a trace, 
\be \label{eq:Ztrace}
\begin{split}
Z^\text{susy}_{{\mathfrak u}(1)}(R;\t,\a) & \=  
\text{Tr}_{\CH_a} (-1)^{F} \, i\, \psi_0 \, \wt \psi_0 \, q^{L_0} \, \qbar^{\overline{L}_0} \, e^{-4 \pi i \, b \, n}  \\
& \= 2 \, \text{Tr}'_{\CH_a} (-1)^{F} \, q^{L_0} \, \qbar^{\overline{L}_0} \, e^{-4 \pi i \, b \, n} \,.
\end{split}
\ee 
Here the prime means that the trace runs only over non-zero fermionic modes; we 
separated the fermionic zero-modes in the first line of~\eqref{eq:Ztrace} and evaluated 
$\text{Tr}_{\psi_0, \wt \psi_0} \, i \, \psi_0 \, \wt \psi_0 \, (-1)^{F} = 2$.  
That equality holds because quantization of the fermionic modes $\psi_0, ~\wt \phi_0$ 
gives a two-dimensional Clifford algebra and hence a  two-dimensional Hilbert space.
In this space, the operator 
$i \, \psi_0\,\wt\psi_0$, since it anticommutes with both $\psi_0$ 
and $\wt\psi_0$, acts  as $(-1)^F$.   
So $i\, \psi_0 \, \wt \psi_0 \, (-1)^{F} =\left((-1)^F\right)^2=1$,
and the trace of this operator is just the dimension of the Hilbert space, which is~2.  (See the end of the introduction for some relevant details.)

The trace~\eqref{eq:Ztrace} is fairly easy to evaluate; the oscillator modes of the boson give the 
factor\footnote{We have $\eta(\tau)=q^{1/24}\prod_{n=1}^\infty (1-q^n)$ with $q=e^{2\pi i \tau}$.   
The partition function of a chiral boson is $1/\eta(\tau)$, where  $\prod_{n=1}^\infty (1-q^n)^{-1}$ 
is the partition function for the nonzero right-moving modes of the bosons, and the factor $q^{-1/24}$ 
comes from the Casimir energy, that is the ground state energy of a massless scalar field.    
For  a chiral fermion, the partition function is $\eta(\tau)^{+1}$, 
where now a factor  $\prod_{n=1}^\infty (1-q^n)$ comes from the sum over oscillator modes, 
and a factor $q^{+1/24}$ comes from the fermion Casimir energy.} $1/\eta(\tau) \eta(\taubar)$, 
while the oscillator modes of the fermions with 
the factor~$(-1)^F$ present give precisely the inverse of this factor. 
(A pedagogical treatment of such calculations can be found in the textbook~\cite{DiFrancesco:1997nk}.)
Thus we are left with the winding and momentum modes of the boson, which give the following contribution,
\be\label{eq:Hamcont}
L_0 \= \frac12 \, \biggl(\frac{n}{R} + \frac{(w-2a)R}{2} \biggr)^2 \,, \qquad 
\overline{L}_0 \=  \frac12 \, \biggl(\frac{n}{R} - \frac{(w-2a)R}{2} \biggr)^2\,.  
\ee
The above trace then evaluates to  
\be \label{eq:HamansS1}
Z^\text{susy}_{{\mathfrak u}(1)}(R;\t,\a) \= 
2 \sum_{n,w \in \IZ}   \exp\biggl( - \pi \t_2 \Bigl(2\frac{n^2}{R^2} +
 \frac{1}{2}(w-2a)^2 R^2\Bigr) + 2\pi i \t_1 n(w-2a) -4 \pi i n b \biggr) \,.
\ee

We use the Poisson summation formula as given in~\cite{Polchinski:1998rq},
\be \label{eq:Poisson}
\sum_{n \in \IZ} \exp( - \pi A n^2 + 2 \pi i B n) \=
\frac{1}{\sqrt{A}} \, \sum_{m \in \IZ} \exp(- \pi (m-B)^2/A) \,.
\ee
It is easy to check that, by a Poisson summation on~$n$, the expression~\eqref{eq:HamansS1} 
can be expressed as the sum
\be
Z^\text{susy}_{{\mathfrak u}(1)}(R;\t,\a) \=
\sqrt{\frac{2}{\t_2}} \, R
 \sum_{m,w \in \IZ} \exp \Bigl(-\frac{\pi R^2}{2\t_2} \, |m-w \t + 2\alpha|^2 \Bigr) \,,
\ee
which equals the path integral answer~\eqref{eq:ZT2PI}.

\bigskip

Finally, when~$R^2 =2k$, $k \in \IN$ , 
we can write this answer in a third way in terms of chiral~${\mathfrak u}(1)$ characters,\footnote{Something similar 
and slightly more complicated happens at other rational values of $R^2$.   We consider
$R^2=2k$ because it leads to a diagonal modular invariant, similar to what happens for simple and simply-connected 
nonabelian Lie groups such as $SU(2)$.}  by which we mean
characters of 
 the chiral algebra of all holomorphic fields of the (bosonic) model.  For generic $R^2$, the only holomorphic fields
are $\partial X$ and its derivatives, but the condition $R^2=2k$ ensures  that holomorphic fields  
exist with \hbox{$(m,w)\not=(0,0)$}, in fact
with $m=kw$ and any $w\in\IZ$.
The characters of the extended ${\mathfrak u}(1)$ chiral algebra are
\be \label{eq:fthetarel}
\chi_{{\mathfrak u}(1),\ell} \bigl(\sqrt{2k};\tau, \a \bigr)  \= \frac{\vartheta_{k,\ell}(\tau,\a)}{\eta(\tau)} 
\ee 
in terms of the index~$k$ theta functions 
\be \label{eq:defthml}
 \vartheta_{k,\ell}(\tau,\a) \= \sum_{{r\in\Z} \atop {r\,\equiv\,\ell\,(\text{mod}\;2k)}} q^{r^2/4k} \, e^{2 \pi i \, \a \, r}
\= \sum_{n \in \IZ} \,q^{(\ell+2kn)^2/4k} \,e^{2 \pi i \a(\ell+2kn)} \,.
\ee 

The trace in the full theory~\eqref{eq:Ztrace} is obtained by taking the product of left and right chiral bosonic 
characters, summing over all characters, and multiplying this by the trace over the fermionic variables.
The fermionic trace cancels the~$\eta(\tau)$ in the denominator of~\eqref{eq:fthetarel}, 
so that the full trace only gets contributions from the momentum and winding modes of the boson.  Modulo one
detail explained momentarily, the result is 
\be  \label{eq:Zu1kChars}
Z^\text{susy}_{{\mathfrak u}(1)}(\sqrt{2k} \,;\t,\a)  \= 
2\, e^{-4 k \pi  \text{Im}(\a)^{2}/\t_{2}} \, \sum_{\ell \,( \text{mod} \; 2k)} 
\vth_{k,\ell} (\t,\a) \, \overline{\vth_{k,\ell} (\t,-\a)}  \,.
\ee
This is easily checked to be equal to the expression~\eqref{eq:HamansS1} when~$R^2=2k$.
We encounter similar sums of theta functions in the $SU(2)$ models studied in section \ref{sec:su2}.

Actually, to arrive at  eqn.~(\ref{eq:Zu1kChars}), we had to include  a 
prefactor~$e^{- 4\pi k (\text{Im}\,\a)^{2} /\tau_2}=e^{-2\pi R^2 a^2\tau_2}$.
This factor corresponds to a term $2\pi R^2 a^2$ in the Hamiltonian; 
this makes sense as a contribution to the Hamiltonian, because it only depends on the twist parameter $a$
in the spatial direction.   It is a ``constant'' contribution to the energy; every state is shifted in energy by this amount.
   Bearing in mind that the Hamiltonian as usually normalized is $H=2\pi (L_0+\barL_0)$, this contribution to the energy 
is visible in eqn.~\eqref{eq:Hamcont}.   A factor involving this contribution to the energy is not 
usually included in the definition of the chiral characters as this would spoil their holomorphy in $\alpha$.  
From the point of view of the expansion of the partition function in terms of chiral characters, 
the factor of $e^{-2\pi R^2 a^2\tau_2}$ is usually viewed as part of the hermitian inner product on 
the space of chiral characters, rather than part of the characters themselves.   We will say more about such factors
in section \ref{sec:HamTracesu2}.

\bigskip

\ndt {\bf Modular and elliptic symmetries} 
The unmodified Euclidean path integral over the fermions and bosons should give a Jacobi invariant 
function on the torus, that is to say, a function invariant under modular transformations 
\be(\t,\a) \mapsto \biggl(\frac{a\t+b}{c\t+d}, \frac{\a}{c\t+d} \biggr) \,,  \qquad 
\biggl( \,\begin{matrix} a & b \\ c & d \end{matrix} \, \biggr) \in SL(2,\IZ)\,,
\ee  
as well as elliptic transformations 
\be
\a \mapsto \a + \mu \t+\nu \,, \qquad  \mu, \nu \in \IZ \,.
\ee  
Note, however, that our observable~\eqref{eq:Ztrace} has an insertion 
of left- and right-moving fermion zero-modes, which makes it into a modular form of weight~$(\frac12,\frac12)$
(and still an elliptic invariant). It is easy to check that these properties are obeyed by~$Z^\alpha_{{\mathfrak u}(1)}(\t)$.

\bigskip

\ndt {\bf Point-particle limit}  In the point-particle limit, we can expect to recover the localization
computation of Choi and Takhtajan for the quantum mechanical model with target $S^1$.
In point-particle limit, the torus corresponding to the two-dimensional space-time of the string worldsheet 
degenerates to a circle corresponding to the one-dimensional time of the point-particle worldline. 
The Euclidean path integral is thus given by a sum over maps from the Euclidean time circle of 
length~$\tau_2$ to the target space circle, and twists around this circle are parameterized by~$\a \in \IR$. 
Since there is no spatial circle, we only obtain configurations with no winding number.
The localized path integral is therefore given by restricting to the~$w=0$ sector in the string 
path integral~\eqref{eq:ZT2PI}.
This gives the point-particle answer to be 
\be \label{eq:ZT2PIptpar}
\sqrt{\frac{2}{\t_2}} \, R \,  
\sum_{m \in \IZ} \exp \Bigl(-\frac{\pi R^2}{2\t_2} \, (m+ 2 \, \alpha)^2 \Bigr) \,.
\ee
Note that the answer, as expected, does not depend on~$\tau_1$. 
Similarly, in the Hamiltonian formalism, one restricts the sum in~\eqref{eq:HamansS1}
to the~$w=0$ sector and to~$\a \in \IR$. The resulting sum over~$n$ is equal to the 
path integral answer~\eqref{eq:ZT2PIptpar} by the Poisson summation formula.

\section{WZW model on the $SU(2)$ group manifold \label{sec:su2}}

Now we move to the WZW model for the Lie group~$G$. 
In this section we consider~$G=SU(2)$, and in the following section we more briefly  discuss other groups. 

The Euclidean functional integral of the WZW model on a torus $T^2$ is given by
\be \label{eq:PITorus}
\int [Dg] \, \exp(-S_\text{WZW}(g)) \,,
\ee 
where the integral runs over all maps
\be
g:T^2 \to G \,.
\ee
The measure~$[Dg]$ is an invariant measure on the group manifold.
The WZW action~\cite{Witten:1983ar} in Euclidean signature is
\be \label{eq:WZWaction}
S_\text{WZW}(g) \= S_0(g) + i k \, \Gamma(g) \,,
\ee
where the first term is the usual sigma-model action\footnote{By $\tr$ we mean the trace 
in the two-dimensional representation of $SU(2)$
(or more generally the $N$-dimensional representation of $SU(N)$).} 
\be \label{eq:PCact}
S_0(g) \= \frac{k}{4 \pi} \int_{T^2} \d^2z \, \tr \, \p g^{-1} \pbar g 
\= - \frac{k}{4 \pi} \int_{T^2} \d^2z \, \tr \, g^{-1} \p g \, g^{-1} \pbar g \,,
\ee
and the second term is the Wess-Zumino term
\be
\Gamma (g) \= -\frac{1}{12 \pi} \int_{B_3} \, \d^3 \sigma \, \varepsilon^{ijk} \, \tr \, g^{-1}  \p_i g \, 
g^{-1}  \p_j g \,  g^{-1} \p_k g.
\ee
Here~$B_3$ is a 3-manifold, with coordinates $\sigma$, whose boundary is~$T^2$, and~$\Gamma$ 
is independent  mod $2\pi {\Bbb Z}$ of the choice of $B_3$ and the extension
of $g$ over $B_3$.
$k$ should therefore be an integer to ensure that $\exp(-S_\text{WZW}(g))$ is single-valued.  
We require $k>0$ so that the real part of the action is positive.

\bigskip

The group~$G$ has a natural left and right action on itself, which we denote 
by \hbox{$g \mapsto U_L \, g \, U_R^{-1}$}.  
We denote this symmetry group as $G_L\times G_R$.
The holomorphic current~$j$ and the anti-holomorphic current~$\wt j$ of the WZW model are 
given by, respectively, 
\be \label{eq:WZWcurrents}
j \=   g^{-1}  \p g  \,, \qquad 
\wt j  \=  \pbar g \, g^{-1}  \,.
\ee
Here $j$ generates $G_R$ and $\wt j$ generates $G_L$.
Accordingly, $j$ transforms in the adjoint representation of $G_R$ 
and  $\wt j$ transforms in the adjoint representation of  $G_L$:
\be \label{eq:jLRaction}
j \mapsto U_R \, j \, U_R^{-1} \,, \qquad \wt j \mapsto U_L \, \wt j \, U_L^{-1} \,.
\ee

\bigskip

The following matrices, given in terms of Pauli matrices, are a basis over~$\IR$ 
for the Lie algebra $su(2)$,  
\be
\sutwogen{1} \= i\s^{1} \= \biggl( \, \begin{matrix} 0 & i \\ i & 0 \end{matrix} \, \biggr) \,, \qquad 
\sutwogen{2} \= i\s^{2} \= \biggl( \,\begin{matrix} 0 & 1 \\ -1 & 0 \end{matrix}  \, \biggr)  \,, \qquad  
\sutwogen{3} \= i\s^{3} \= \biggl( \, \begin{matrix} i & 0 \\ 0 & -i \end{matrix} \, \biggr) \,. 
\ee
An $SU(2)$ group element can be written (not uniquely) as~$\exp(\theta_a \sutwogen{a})$, $\theta_a \in \IR$, $a=1,2,3$.
The currents can be decomposed into components as $j = j^a \s^a$, $\wt j = \wt j^a \s^a$.

\subsection{The WZW model coupled to external gauge fields}\label{extgf}

We will introduce  external gauge fields $\AL$ and $\AR$
taking values in
the Lie algebra $\mathfrak g_L\oplus \mathfrak g_R$  of $G_L\times G_R$, 
because this gives a convenient way to define  a twisted version of  the WZW model.

  There is no completely gauge-invariant way to couple the WZW model to $G_L\times G_R$ gauge fields,
since the $G_L\times G_R$ global symmetry is anomalous (like the $U(1)\times U(1)$ global symmetry of 
section \ref{two}).   However, there is a choice that is the best possible in the sense that
(1) the violation of gauge invariance depends only on $\AL$ and $\AR$, and not on $g$; (2) related to
this, the action becomes completely gauge-invariant if one restricts $\AL$ and $\AR$ to any anomaly-free
subgroup of $G_L\times G_R$.   Here property  (1) is important for our application, because it is needed to interpret
the twisting  that we will study in terms of  coupling to background gauge fields.

In this sense, the action of the  WZW model with gauging of 
$G_L \times G_R$ is \cite{Gawedzki:1988hq,Gawedzki:1988nj,Witten:1991mm}
\be \label{eq:LRgauging}
\begin{split}
& S^\text{gauged}_\text{WZW} \bigl(g,\AL,\AR \bigr) \\
& \quad \= S_\text{WZW}(g) \\ 
& \qquad \quad - \, \frac{k}{2 \pi} \int_{T^2} \d^2z \; \text{tr} \,
\Bigl( \AL_{z} \, (\pbar g) g^{-1} -   g^{-1} \p g \, \AR_{\zbar}  
- \AL_{z} \, g  \, \AR_{\zbar}  g^{-1} + 
\frac12 \bigl( \AL_{z} \,  \AL_{\zbar}  + \AR_{z} \,  \AR_{\zbar} \, \bigr)  \Bigr) \,.
\end{split}
\ee
The gauge symmetry acts as~$g \mapsto U_L \, g \, U_R^{-1}$. 
The infinitesimal gauge transformation with gauge 
parameter~$(\lambda^L, \lambda^R) \in \mathfrak{g}_L \oplus \mathfrak{g}_R$ acts as follows, 
\be
\delta_\lambda g \= \lambda^L \, g - g \, \lambda^R \,, \qquad \delta_\lambda A^{L,R}_i \= -D_i \lambda^{L,R}  \,, \quad i \= z, \zbar \,,
\ee
where the covariant derivative is defined as
\be
D_i \, g \= \p_i \, g  + \AL_i \, g - g \AR_i \,, \qquad 
D_i \, \lambda^{L/R}\=\partial_i \, \lambda^{L/R}+[A_i^{L/R},\lambda^{L/R}] \,. 
\ee

The left- and right-moving currents in the gauged theory are the gauge-covariant 
generalizations of the currents~\eqref{eq:WZWcurrents}, and have the form  
\be \label{eq:ggeWZWcurrents}   
\begin{split}
\JA_z & \defeq g^{-1} \, D_z \, g  \= g^{-1} \, \p \, g + g^{-1} \, \AL_{z} \, g- \AR_{z}   \,, \\
\JAtl_{\zbar} & \defeq D_{\zbar} \, g \, g^{-1} \= \pbar \, g \, g^{-1} + \AL_{\zbar} - g \,\AR_{\zbar} \, g^{-1}  \,. 
\end{split}
\ee
The left- and right-moving currents transform under the adjoint action of the right and left gauge group, 
respectively, as in~\eqref{eq:jLRaction}.    The classical equations of motion for the field $g$ 
are\footnote{The curvature of a gauge field $A$ is as usual $F=d A+A\wedge A$.   
A gauge field is said to be flat if and only if $F=0$.}
\be\label{classeqns} D_{\bar z} \, \JA_z+F_{\bar z z}^R\= D_z\,\JAtl_{\bar z} -F_{z\bar z}^L\=0 \,. \ee

The $G_L\times G_R$ symmetry of the WZW model is anomalous, and accordingly, as remarked earlier,
 the general theory of \eqref{eq:LRgauging} is not completely gauge-invariant.
 However, this action is gauge-invariant if one restricts from $G_L\times G_R$ to any anomaly-free subgroup.
 A simple  example of an anomaly-free subgroup is a diagonally embedded copy of $G\subset G_L\times G_R$.
 We can restrict to such a subgroup by specializing to 
 $\AL=\AR=B$.  Making this substitution  in~\eqref{eq:LRgauging},  we obtain the following action 
\be\label{diagform}
S^\text{G/G}(g,B) \= S_\text{WZW}(g) \, - \, \frac{k}{2 \pi} \int_{T^2} \d^2z \; \text{tr} \,
\bigl( B_{z}  (\pbar g) g^{-1} - B_{\zbar} \, g^{-1} \p g + B_{z} \, B_{\zbar} -   B_{z}  g B_{\zbar} \, g^{-1}\bigr),
\ee which is completely gauge-invariant.
As an aside, that means that $B$ can be treated as a dynamical gauge field.  
If we do so, then (\ref{diagform}) becomes  the action of the so-called $G/G$ coset model 
 (see e.g.~eqn.~4.1 in~\cite{Witten:1991mm}). 
The  gauge symmetry in eqn.~(\ref{diagform})  acts as~$g \mapsto U \, g \, U^{-1}$ and the 
covariant derivative acts as
\be
D_i  \, g \= \p_i  \,g  + [B_i, g].
\ee

For our purposes, however,  $\AL$ and $\AR$ are just devices to describe a twisted version of the WZW model, 
and we treat them 
as background gauge fields (which are held fixed in the path integral), rather than dynamical ones
(which are variables to be integrated over).\footnote{The partition function of
gauged WZW models with dynamical gauge fields have been studied by other methods, and have led to 
a variety of applications; see~\cite{Dei:2024uyx} and references therein.}
To describe the WZW model twisted by elements of $G_L\times G_R$, we need to consider 
the case that $\AL$ and $\AR$ are flat, with specified holonomies.   
As  the fundamental group of the torus is abelian, we can assume
$\AL$ and $\AR$ to be valued in the Lie algebra $\mathfrak{h}_L \oplus \mathfrak{h}_R$ of a maximal torus
$H_L\times H_R\subset G_L\times G_R$.   Moreover, up to a gauge transformation, we can represent 
them by constant one-forms.  Thus for $G=SU(2)$, a sufficiently general form is
\be \label{eq:extAAbar}
\AL \=  \frac{\sutwogen{3}}{2i\t_2} \, \bigl( - \alphabarL^{}\, \d z + \alphaL^{}\, \d \zbar\, \bigr) \,  \,, \qquad 
\AR \=  \frac{\sutwogen{3}}{2i\t_2} \, \bigl( - \alphabarR^{}\, \d z + \alphaR^{}\, \d \zbar \, \bigr)  \,. 
\ee
For any $G$, we would assume a similar form, with $\AL$ and $\AR$ valued in the 
Lie algebra $\mathfrak h$ of a maximal torus $H$.

It is also useful to express $\AL$ and $\AR$ in Cartesian coordinates.  
With $z=\wsx+i \wsy$, $d z=d \wsx+i d \wsy$, we have
\be\label{acart} 
\AR\=\sutwogen{3}\left( \frac{{\rm Im}\,\alphaR}{\tau_2}  d \wsx-\frac{{\rm Re}\,\alphaR}{\tau_2}d \wsy\right).
\ee
From this it is clear that under $\alphaR\to \alphaR^{}+1$, the change in $\AR$ is the change produced 
by a single-valued gauge transformation $e^{ \wsy \sutwogen{3}/\tau_2}$.  Similarly under $\alphaR\to \alphaR^{}+\tau$, 
$\AR$ changes by a single-valued gauge transformation $e^{-(\wsx-(\tau_1/\tau_2)\wsy)\sutwogen{3}}$.   
Combining these statements, we find the transformation of $\alphaR$ and similarly of $\alphaL$ under
abelian gauge transformations with nontrivial ``winding'' around the $\wsx$ and $\wsy$ directions:
\be\label{wintrans}
\alphaL\to \alphaL^{}+\lambda\tau +\mu \,, \qquad   
\alphaR\to \alphaR^{}+\lambda' \tau+\mu' \,, \qquad 
\lambda,\lambda',\mu,\mu'\in\Z \,.
\ee

\bigskip 

We aim to calculate  the partition function of this model, 
\be \label{eq:PIgauged}
Z_\text{WZW} (\tau,\alphaL,\alphaR^{}) \= 
\int [Dg] \, \exp(-S^\text{gauged}_\text{WZW} \bigl(g,\AL,\AR \bigr)) \,. 
\ee 
Under localization, it will turn out that all localizing solutions have the property that the field $g$ 
is valued in the either the $U(1)$ subgroup of $SU(2)$ that is generated
by $\sutwogen{3}$, or in the conjugate of this by a Weyl transformation.   It is therefore illuminating to write 
the form that the $SU(2)$ gauged WZW model takes
when  specialized to a $U(1)$ subgroup.  We take
\be\label{abansatz}g=e^{\sutwogen{3} X(z,\zbar)},\ee for a circle-valued field $X$, and we make a similar 
abelian ansatz for the gauge fields~$\AL$,~$\AR$:
\be\label{abansatz2}
\AL\=a^L \, \sutwogen{3}\,, \qquad \AR\=a^R \, \sutwogen{3} \,. 
\ee
Here $a^L$, $a^R$ are $U(1)$ gauge fields.  With this ansatz, the gauged $SU(2)$ WZW model 
will reduce to a gauged $U(1)$ WZW model that we describe momentarily.
 Of course, the background gauge fields that we will actually study are obtained by the further 
 specialization~\eqref{eq:extAAbar}.
With the abelian ansatz of eqns.~(\ref{abansatz}), (\ref{abansatz2}), the gauge invariance reduces to 
\be\label{abgauge} 
{X} \to X+ \varepsilon^L-\varepsilon^R\,,\qquad 
a^L_i\to a^L_i-\partial_i \varepsilon^L\,,  \qquad 
a^R_i\to a^R_i-\partial_i \varepsilon^R\,.
\ee
We note that after this abelianization, the coupling to both $\AL$ and $\AR$
is redundant in the sense that $X$ transforms the same way under $\varepsilon^L$ or $-\varepsilon^R$.
The reason that this has happened is that $U(1)$ is abelian, 
so the left and right actions of $U(1)$ on itself are equivalent.   Because $SU(2)$ is nonabelian,
the left and right actions of the group $SU(2)$ on itself are not equivalent, 
and it is important to introduce both $\AL$ and $\AR$ in order to analyze the $SU(2)$ WZW model.
With the abelian ansatz of eqns.~(\ref{abansatz}), (\ref{abansatz2}), the action~\eqref{eq:LRgauging} 
reduces to\footnote{This action does not receive
a contribution from the Wess-Zumino term, which vanishes upon restriction to $U(1)$. 
(For~$\text{rank}(G)>1$, the Wess-Zumino term remains nontrivial after restriction to a maximal torus. 
We thank Yongchao L\"u for explaining this.)
However, it does receive a contribution from couplings of $A^{L/R}$ that
arise in gauging the Wess-Zumino term.} 
\be\label{abreduction} 
S_{\rm WZW}^{U(1)}\=\frac{k}{2\pi}\int  \d^2z \left(\partial_z X\partial_{\bar z} X+2a^L_z\partial_{\bar z}
X-2 a^R_{\bar z}\partial_zX +a^L_z a^L_{\bar z}+a^R_z a^R_{\bar z} -2 a^L_z a^R_{\bar z}
 \right) .
 \ee 
This abelian theory is actually a gauged version of the theory studied in section~\ref{two}; 
the field~$X$ is the same as the one used there and the radius $R$ in section~\ref{two} is given by~$R^2=2k$. 
The gauged action $S_{\rm WZW}^{U(1)}$ is not quite gauge-invariant;  under a gauge transformation, it transforms as 
\begin{align}\label{anomaly} \notag
\delta S_{\rm WZW}^{U(1)} &\=
\frac{k}{2\pi}\int_{T^2} \d^2z\left( \varepsilon^L(\partial_z a^L_{\bar z}-\partial_{\bar z} a^L_z)
+\varepsilon^R(-\partial_z a_{\bar z}^R+\partial_{\bar z} a_z^R)
\right) \\ 
& \=-\frac{ik}{2\pi}\int_{T^2} \left( d\epsilon^L\wedge a^L-d\epsilon^R\wedge a^R\right).
\end{align} 
The first formula is the one that comes directly from this derivation.  It is valid for topologically trivial 
gauge transformations (that is, it is valid if $\epsilon^{L/R}$ are single-valued functions)
regardless of whether the gauge fields are topologically trivial (that is, regardless of the vanishing 
or not of the first Chern classes $\frac{1}{2\pi}\int f^{L/R}$, with $f^{L/R}=d a^{L/R})$. 
The second formula is valid if the first Chern classes vanish (so that $a^{L/R}$ can be globally 
defined as one-forms).   When both conditions are satisfied, the two formulas are
equal by integration by parts. (With more care, it is possible to describe the anomaly functional 
when neither condition is satisfied, but we will not need this.)
The background gauge fields (\ref{eq:extAAbar}) that we consider in this paper are flat, so the 
first formula shows that we will always have gauge invariance under topologically trivial
gauge transformations.   However, the second formula shows that the partition function will {\it not} 
be invariant under topologically non-trivial gauge transformations of $a^{L/R}$,
that is under the shifts (\ref{wintrans}), and instead shows how the partition function will transform 
under such gauge transformations.

This anomaly of eqn.~(\ref{wintrans}) cancels if we restrict to either $a^L=a^R$ or $a^L=-a^R$, 
with a corresponding restriction on the gauge parameters.
The case $a^L=a^R$ corresponds to the anomaly-free subgroup $G\subset G_L\times G_R$ 
that can be defined for a WZW model based on any Lie group $G$, as in eqn.~(\ref{diagform}).   
But the gauge-invariant theory with $a^L=-a^R$ only exists after restricting from $SU(2)$ to $U(1)$. 
The generalization of this for any compact 
 Lie group~$G$ is to restrict from $G$ to a maximal torus $T\subset G$,
 and then take $a^L$ to differ from~$a^R$ by a Weyl transformation.

We will see that localization of the $SU(2)$ WZW model leads to the abelian theory of 
eqn.~(\ref{abreduction}), or more precisely to the disjoint sum of two copies of it,
 embedded  in $SU(2)$ in ways  that are related by a Weyl transformation.

\subsection{The supersymmetric theory}

The action~\eqref{eq:WZWaction} can be supersymmetrized by adding left-moving and right-moving 
chiral fermions $\psi$, $\wt \psi$ valued, respectively, in  $\mathfrak g_L$ and $\mathfrak g_R$.
The  fermions can be  decoupled by a chiral rotation,  leading to a theory that is described by adjoint-valued 
free fermions~$\psi$, $\wt \psi$ togther with a purely bosonic WZW model~\cite{DiVecchia:1984nyg}.
In this description, which we will employ, since twisting has been incorporated by including the 
flat gauge fields $\AL$ and $\AR$, the fields $g$, $\psi$, and $\wt\psi$ are all periodic 
in both directions around $T^2$.  

We can easily promote the supersymmetric WZW model to a supersymmetric version of the gauged theory
by coupling the fermions minimally to the gauge fields. 
The action of the supersymmetric gauged theory is 
\be \label{eq:gaugedsusyact}
S^\text{susy}(g, \AL, \AR, \psi) \= S^\text{gauged}_\text{WZW} (g, \AL, \AR)+ 
i \frac{k}{4\pi} \int_{T^2} \d^2z \, \text{tr} \, \bigl( \psi \, D_{\zbar} \, \psi + \wt \psi \, D_z \, \wt \psi \, \bigr) \,.
\ee
Covariant derivatives are
\be \label{eq:fercharges}
D_i \, \psi \= \p_i \psi + \bigl[\AR_i,\psi \bigr] \,, \qquad 
D_i \, \wt \psi \= \p_i \wt \psi + \bigl[\AL_i, \wt \psi  \,\bigr].
\ee

Assuming that  $\AL$ and $\AR$ are flat and so can locally be gauged away,
the action~\eqref{eq:gaugedsusyact} is invariant under left- and right-moving supersymmetry transformations
obtained by replacing derivatives in the ungauged theory by covariant derivatives.
This has been discussed for  anomaly-free embeddings in~\cite{Witten:1991mk,Henningson:1993nr}.    
The coupling to gauge fields provides a convenient way to study the twisting of the model by elements 
of the symmetry group $G_L\times G_R$.

For localization, we will  use the following right-moving supersymmetry,
\be
\delta g \= i  \wt \psi \, g \,, \qquad 
\delta \wt \psi  \= \JAtl_{\zbar} + i \wt \psi \, \wt \psi \,, \qquad \delta \AL \= \delta \AR \= 0 \,,
\ee
which satisfies $\delta^2=i D_{\bar z}$ off-shell.
Note that the supersymmetry variation of the covariant right-moving current is 
\be
\begin{split}
\delta \JAtl_{i} 
& \= i  \p_{i} \, \wt \psi + i \bigl[\wt \psi \, , \, \p_{i} \, g \, g^{-1} \bigr] 
- i  \bigl[\wt \psi \, , \, g \,\AR_{i} \, g^{-1}\bigr] \cr
& \= i  D_{i} \, \wt \psi + i \bigl[\wt \psi \, , \, D_{i} \, g \, g^{-1} \bigr] \cr
& \= i  D_{i} \, \wt \psi + i \bigl[\wt \psi \, , \,  \JAtl_{i}  \bigr] \,,
\end{split}
\ee
so that, indeed,  
\be
\delta^2 \wt \psi \= i  D_{\zbar} \, \wt \psi \,, 
\ee
with the covariant derivative defined in~\eqref{eq:fercharges}. 

\smallskip

This gauged supersymmetric WZW theory obeys the key condition  described in the introduction:
 the fermions are  completely decoupled from the bosonic fields. 

\subsection{Renormalization of the level?}\label{renormalization}

Before attempting any calculations, we will ask the following deceptively simple question:
what is the effective value of the level in the purely bosonic  WZW model of a compact simple Lie group $G$ at ``level $k$''?

One  version of the question is the following.   Consider the WZW model with the overall coefficient of the action 
being a positive integer $k$, as in~\eqref{eq:WZWaction}.  The theory is conformally invariant and has holomorphic 
and antiholomorphic currents generating the $G_L\times G_R$
symmetry.    The current algebra (or affine Lie algebra) generated by the currents has a central term with 
an integer coefficient $k'$.   Taking quantum corrections into account, what  is the relation between $k$ and $k'$?

Another version of the question is this: do loop corrections renormalize the effective value of $k$?

One simple observation is that a possible renormalization of $k$ can only arise at one-loop order.   
A $g$-loop contribution to the effective action is proportional to $k^{1-g}$,
and since the coefficient of the Wess-Zumino term in the effective action must be an integer for consistency, 
only a one-loop diagram can possibly contribute a correction
to this coefficient.

Another simple observation is the following.   Any attempt to define precisely the one-loop correction 
to the effective action will involve some sort of regularization procedure.
Such a procedure often depends on arbitrary choices that can be varied continuously.   
Because $k'$ must be an integer, it will be invariant under any continuously
variable choices in the regularization procedure.   However,  two different regularization 
procedures that are {\it not} continuously connected
to each other can conceivably lead to different values of $k'$.

We claim the following.  If one defines the one-loop correction to the WZW effective action by a 
standard method such as Pauli-Villars  or zeta function regularization,  
there is no one-loop correction to the level and $k'=k$.    However, the localization procedure developed 
in the present paper involves a different method of defining
the one-loop determinant.  For a compact simple Lie group $G$ with dual Coxeter number $h$, the 
localization procedure leads to $k'=k-h$.  For $G=SU(2)$, this is $k'=k-2$.

Let us first explain why standard ways of defining the theory lead to $k'=k$.
The classical equation of motion of the gauged WZW model is 
$D_{\bar z}(g^{-1}D_z  g)+F_{\bar z z}^R=0$, or equivalently $D_z(D_{\bar z} g \,g^{-1})-F_{z\bar z}^L=0$. 
Perturbing around this equation by $\delta g = g \alpha$, where $\alpha$ is an adjoint-valued 
field that describes the small fluctuations, the linearized equation obeyed by $\alpha$ is 
\be\label{linearized} 
D_{\bar z} \left(D_z\alpha+[g^{-1}D_z g,\alpha] \right)\=0 \,.
\ee
(Here $\alpha$ is $G_L$-invariant but transforms in the adjoint representation of $G_R$, 
so $D\alpha=d\alpha+[\AR,\alpha]$.)
The one-loop correction involves the determinant of the operator $M$
that appears in eqn.~(\ref{linearized}):
\be\label{defop}
M\alpha \=- D_{\bar z}\left(D_z\alpha+[g^{-1}D_z g,\alpha]\right) \,. 
\ee

With standard methods of regularization, there is no anomaly in $\det M$.  The basic reason for this is that $M$ 
maps a space (namely the space of adjoint-valued functions on the worldsheet) to itself, while anomalies are 
part of index theory and occur for operators that map one space to another.  In more detail, 
$M$ is an elliptic operator, so on a compact two-manifold such as the torus considered in the present article, 
it has a complete set of eigenfunctions with a discrete spectrum.   The eigenvalues of $M$ are not real as $M$ 
is not self-adjoint, but they are asymptotically positive, since the term in $M$ of highest degree is just the Laplacian
$-\partial\bar\partial$, which is positive.  The eigenvalues of $M$ tend to $+\infty$ at the same rate as the eigenvalues 
of the Laplacian.   
One can define $\det M$ by zeta-function regularization,\footnote{This is slightly subtle as the eigenvalues of $M$ are
complex numbers.   Let $S$ be any finite set of eigenvalues of $M$ that includes all eigenvalues of, say, nonpositive real part.     
One can define a 
zeta function of $M$ summing only over eigenvalues $\lambda_i$ that are not in $S$: 
$\zeta_S(s)=\sum_{i\notin S} \lambda_i^{-s}$.   This is well-defined as the $\lambda_i$ that appear in the sum have
positive real part; the sum converges and defines an analytic function of $s$ if ${\rm Re}\,s$ is sufficiently large.    
After continuing to $s=0$ (and adding to $\zeta(s)$ a local counterterm
to cancel the pole  at $s=0$, whose residue is the divergent part of the quantum effective action), the determinant of $M$
can be defined as  $\det\,M=\left(\prod_{i\in S}\lambda_i \right) e^{-\zeta_S'(0)}$.    
This is easily seen to be independent of $S$, so it behaves well when the fields $g,\AL,\AR$ that
appear in the definition of $M$ are varied, even though the number of eigenvalues with non-positive real part may change.   
This gives a satisfactory definition of $\det\,M$.}
 using only the eigenvalues of $M$  as input.
Because the eigenvalues are gauge-invariant, $\det M$ defined in this way  is also gauge-invariant, 
with no anomaly under $G_L\times G_R$ gauge transformations.

Alternatively, we can consider Pauli-Villars regularization, which is based on adding to $M$ a higher order 
term with a small coefficient $\epsilon$.      In the present case, exploiting the fact that $M$ maps a space to itself, 
one can add a higher order term that is parity-invariant and positive-definite:
\be\label{zillow} M_\epsilon\=\epsilon(D_z D_{\bar z}+D_{\bar z} D_z)^2+M \,. \ee
Anomalies in the determinant of an elliptic operator such as $M_\epsilon$  always involve only the leading term 
of the operator.   In the present case, the leading term is parity-even and as the anomaly is parity-odd, 
there is no anomaly in the effective action derived from $M_\epsilon$.   This then automatically remains true as $\epsilon\to 0$.

It is instructive to consider how these arguments fail in the case of an operator that really does have an anomaly, 
for example the operator $D_z$ on a Riemann surface that
maps functions in some representation of the gauge group (in our application this will be the adjoint representation) 
to $(1,0)$ forms valued in the same representation.
The operator $D_z$ maps a vector space to a different vector space, so it does not have eigenvalues and 
zeta-function regularization is not relevant.   What about Pauli-Villars
regularization?    To try to regularize $D_z$, we want to add to it a higher order operator of some sort.    
We cannot add to $D_z$ a parity-symmetric operator such as the Laplace operator
$-(D_z D_{\bar z}
+D_{\bar z}D_z)$, as this maps functions to themselves, while $D_z$ maps functions to $(1,0)$-forms.   
An example of a perturbation of $D_z$ by a higher order term that does make
sense is
\be\label{highpert} D_z\to \epsilon D_z D_{\bar z} D_z+D_z. \ee
The higher order term that we have added is not parity-symmetric, so there is no obvious reason that 
the regularized operator has no anomaly.   On the contrary, the assertion
that the anomaly in $\det\,D_z$ is unavoidable and physically meaningful means, in particular, that the 
anomaly of $\det(\epsilon D_z D_{\bar z} D_z+D_z)$ is independent of $\epsilon$.
We will actually essentially encounter the operator $\epsilon D_z D_{\bar z} D_z+D_z$ in the localization 
calculation in section \ref{sec:locPIcalc}.   The role of $\epsilon$
will be played by the localization parameter $\lambda$.   Since multiplying an operator by a constant also 
does not affect the chiral anomaly, the operator
$\lambda D_z D_{\bar z} D_z+D_z$ has the same anomaly as $D_z D_{\bar z} D_z+\frac{1}{\lambda}D_z$.   
Since this anomaly is moreover independent of $\lambda$,
we can take the limit $\lambda\to\infty$ and conclude that the operators $D_z D_{\bar z} D_z$ and $D_z$ 
have the same anomaly.   This fact will be relevant in understanding the
localization calculations.

Now that we have understood that with some natural approaches to understanding the  path integral, 
there is no quantum renormalization of the WZW level, let us discuss
the procedure that we will actually follow in the present paper, which does lead to such a renormalization. 
Here we use the fact that $M$ is a product, $M=M_1M_2$, where $M_2$ is the operator on functions 
in the adjoint representation of $G_R$  defined by
\be\label{timbo}M_2(\alpha)\=D_z\alpha+[g^{-1} D_z g,\alpha] \ee
and $M_1$ is the operator on  $(1,0)$-forms in the adjoint representation of $G_R$ defined by
\be\label{limbo} M_1(\beta)= D_{\bar z} \beta=\partial_{\bar z}\beta+[A_{z}^R,\beta] \,. \ee 
So we can interpret $\det M$ as a product $\det M_1 \det M_2$.    With that procedure, the one-loop correction to the effective
action of the WZW model comes from
\be\label{tellmex}\frac{1}{\sqrt{ \det M}}\=\frac{1}{\sqrt{ \det M_1}}\frac{1}{\sqrt {\det M_2}} \,. \ee
This is the approach that is natural in the localization calculation that we will perform starting in the next section.
With that approach, any renormalization of the level $k$ in the bosonic WZW model will be a sum of effects 
coming from $M_1$ and from $M_2$.

Since $M_1$  does not depend on $\AL$ at all, it has no anomaly for $\AL$. But it has an anomaly 
under $\AR$ gauge transformations. Indeed, $M_1$ is essentially a chiral Dirac operator, acting on 
fermions in the adjoint representation of $G_R$.
Accordingly, $\det M_1$ has an anomaly coefficient $2h$ (where $h$ is the dual Coxeter number of $G$, 
and the value $2h$ comes from the trace of the second Casimir operator in the adjoint representation).   
Hence the anomaly coefficient in $1/\sqrt {\det M_1}$ is $-h$.
This suggests that if $\det M$ is defined by the procedure~(\ref{tellmex}), 
then the effective value of the level is actually $k'=k-h$.  For $SU(2)$, one has $h=2$ and $k'=k-2$.

To confirm this, we look at $M_2$.
$M_2$ has a similar anomaly under $\AL$, though this is slightly less obvious.  
Indeed, $M_2\alpha=[g^{-1}(\partial_z+[\AL,\cdot])g,\alpha]$.  It is convenient to define
$\alpha'=g\alpha g^{-1}$, which transforms in the adjoint representation of $G_L$ and trivially under $G_R$, 
and similarly to set $M_2'=g M_2 g^{-1}$.   
So $M_2'\alpha'=\partial_z\alpha' +[\AL,\alpha']$.  Thus $M_2$ up to conjugation by $g$
 is essentially an opposite chirality Dirac operator coupled to $\AL$, so $1/\sqrt{\det M_2}$ has an anomaly for $\AL$ that
is equal and opposite to the anomaly of $1/\sqrt{\det M_1}$ for $\AR$.   This is again consistent with the claim 
that with this approach to defining the path integral,
the one-loop correction shifts the effective level from $k$ to $k-h$, since the classical WZW model has equal 
and opposite anomalies for $G_L$ and $G_R$.

In general, one expects that different regularizations are equivalent up to a renormalization of parameters.  
The only parameter of the bosonic WZW model is the level $k$,
so we expect that standard regularizations of the one-loop determinant and a regularization based on the 
factorization of $M$ (which we use in this article) differ only by the shift in $k$.

\subsection{Calculation of localized path integral \label{sec:locPIcalc}}

Since we will choose  the external gauge field to be in 
the~$\sutwogen{3}$ direction in~\eqref{eq:extAAbar}, 
the fermions~$\psi^{1,2}$, $\wt \psi^{1,2}$ are charged and, for a generic twisting  
(more precisely, when~$2 \alphaL, 2 \alphaR \notin \mathbb{Z} \tau + \mathbb{Z}$),
do not have any zero-modes. 
On the other hand, the fermions~$\psi^{3}$, $\wt \psi^3$ are neutral and have one zero-mode each.
We will study the partition function with  these zero-modes absorbed
\be \label{eq:ZWZW}
Z_\text{SWZW}(\tau,\alphaL, \alphaR^{}) \= \int  [Dg \, D \psi] \; i \, \psi^3_0 \, \wt \psi^3_0 \, 
\exp \bigl(-S^\text{susy}(g,\AL,\AR,\psi)\bigr) \,,
\ee and
with the external gauge fields given in~\eqref{eq:extAAbar}.

\bigskip

\ndt {\bf Localizing term and critical points} \\
A  non-abelian, gauge-invariant generalization of the localizing term~\eqref{eq:defVU1} is given by
\be \label{eq:defVSU2}
V \= \int_{T^2}  \d^2z \;  \text{tr} \, D_z \wt \psi  \; (D_z \, \wt \JA_{\zbar})^\dagger 
\=  - \int_{T^2}  \d^2z \;  \text{tr} \, D_z \wt \psi  \; D_{\zbar} \, \wt \JA_{z}  \,.
\ee
The corresponding contribution to the action is
\be 
 \delta V \= \int_{T^2}  \d^2z \;  \text{tr} \, D_z \bigl(\JAtl_{\zbar} + i \wt \psi \, \wt \psi \ \bigr) \; 
 (D_z \, \wt \JA_{\zbar})^\dagger - i \int_{T^2}  \d^2z \;  \text{tr} \, D_z \wt \psi  \; D_{\zbar} \,
  \bigl( D_{z} \, \wt \psi + \bigl[\wt \psi \, , \,  \JAtl_{z}  \bigr] \bigr)
\ee
The bosonic part of ~$\delta V$ is non-negative and vanishes if and only if
$D_z \, \wt \JA_{\zbar} =0$ or, in more detail, 
\be \label{eq:covconstcurrent}
 \p_z \wt \CJ_{\zbar} + \bigl[ \AL_z, \wt \CJ_{\zbar} \bigr] \= 0 \,,
\ee
where~$\wt \CJ_{\zbar} \in \mathfrak{g}$ is well-defined on the torus. 
The localizing term was chosen so that this  equation is precisely the classical equation of motion of the theory, 
in the presence of the given background gauge fields.
So what we will learn from the localization procedure is that the WZW partition function in the presence of those 
fields can be written as a sum of contributions from classical solutions.

For a general Lie group $G$, with $\AL$, $\AR$  assumed to be constants valued in the Lie 
algebra $\mathfrak h$ of a maximal torus $H$, 
the classical equation (\ref{eq:covconstcurrent}) actually implies  that $\wt \CJ_{\zbar} $ is valued in $ \mathfrak{h}$ 
and is a constant (see Appendix~\ref{sec:locsol}).   Concretely, then, 
\be  \label{eq:constcurrent}  
\JAtl_{\zbar} \=
D_{\zbar} \, g \, g^{-1} \= \p_{\zbar} \, g \, g^{-1} + \AL_{\zbar} - g \,\AR_{\zbar} \, g^{-1}  
\ee  must be $\mathfrak h$-valued and constant.
The general solution to this condition consistent with  periodicity on the torus is 
analyzed in  Appendix~\ref{sec:locsol}, with the result that, for generic twisting, up to a Weyl group 
element $\omega$, $g$ is valued in $H$
and takes a very simple form.   For $G=SU(2)$, with $\AL$, $\AR$ parametrized as in 
eqn.~\eqref{eq:extAAbar}, the general localizing solution has the form
\be \label{eq:Weyltwist}
g_{m,w}^{\, \omega}(z,\zbar) \= 
  \omega \, h_0 \exp \Bigl(\frac{\sutwogen{3}}{2 i \t_2} \bigl( m (z - \zbar) +w( \tau \zbar - \taubar z) \bigr) \Bigr) \,, 
  \qquad m, w \in \IZ \,. 
\ee 
Here $h_0$ is a constant element of $H$, and~$\omega$ is an element of~$\text{Weyl}(SU(2))$.  
There are two possible choices of $\omega$, 
namely the identity and a non-trivial element that can be represented by an $SU(2)$ element that 
anticommutes with $\sutwogen{3}$, for example 
 $ \biggl( \, \begin{matrix} 0 & -1 \\ 1 & 0 \end{matrix} \, \biggr) \equiv p$.  
The analogous formula for a general compact simple Lie group can be found in eqn.~\eqref{eq:WeyltwistsuN}.

The localizing solutions and the appearance here of the Weyl group have a  particularly simple interpretation in the case of $SU(2)$.
First of all, the $SU(2)$ manifold is a three-sphere, which we can parametrize by real variables $x_1,\cdots, x_4$ satisfying
\be\label{eq:sphere} x_1^2+x_2^2+x_3^2+x_4^2\=1. \ee
These coordinates can further be chosen so the identity element of the group is the point $(x_1,x_2,x_3,x_4)=(1,0,0,0)$
and  $\sutwogen{3}$, 
acting on $SU(2)$ on the left or right, is represented respectively by
\begin{align}\label{eq:coords} 
\sutwogen{3}_L& \= \left(x_1\frac{\partial}{\partial {x_2}}-x_2\frac{\partial}{\partial {x_1}}\right)- 
 \left(x_3\frac{\partial}{\partial{x_4}}-x_4\frac{\partial}{\partial {x_3}}\right) \\
\sutwogen{3}_R&\= \left(x_1\frac{\partial}{\partial {x_2}}-x_2\frac{\partial}{\partial {x_1}}\right)+ 
 \left(x_3\frac{\partial}{\partial{x_4}}-x_4\frac{\partial}{\partial {x_3}}\right) \, .
\end{align}
In particular, both $\sutwogen{3}_L$ and $\sutwogen{3}_R$  generate  $U(1)$ symmetries of the 
circle $S_1$  defined by $x_3=x_4=0$ and also of the circle $S_p$ defined by $x_1=x_2=0$.   
So the background gauge fields that we have chosen  couple to  symmetries of these circles.
The names that we have given to these two circles are motivated by the fact that $S_1$  
contains the identity element of the group, and the non-trivial Weyl group element $p$  of $SU(2)$
exchanges $S_1$ with $S_p$.  If we restrict the WZW field $g\in SU(2)$ to be valued in 
either $S_1$ or $S_p$, the $SU(2)$ WZW model, coupled to the specific
background gauge fields that we have chosen,
reduces to the twisted  $U(1)$ theory studied in section \ref{two} (whose formulation in terms of 
gauge fields was presented in eqn.~(\ref{abreduction})), and the localizing term (\ref{eq:defVSU2}) 
of the $SU(2)$ model reduces to the corresponding localizing term of the $U(1)$
model.   Thus the localizing solutions of the $SU(2)$ model are simply the localizing solutions of 
the $U(1)$ model, embedded in $SU(2)$ via either $S_1$ or $S_p$.
It also follows that the supersymmetric $SU(2)$ and $U(1)$ models have very similar reduced 
partition functions (with zero-modes removed), except that the result for $SU(2)$ involves
a sum over the Weyl group.

\medskip

\ndt {\bf One-loop determinant} \\
Now we discuss the one-loop determinant for fluctuations around these classical solutions.   
We start with the basic solution with~$m=w=0$.
We can parameterize the fluctuations as~$g= \omega \, e^{y} $ with $y\in \mathfrak{h}$; moreover, 
$y$ is perturbatively small except for a constant mode valued
in $\mathfrak h$ (corresponding to the arbitrary constant $h_0\in H$ in eqn.~(\ref{eq:Weyltwist})). 
The functional measure equals~$[Dg]=[ \d y]$ on this space.

For any element $x\in \mathfrak{h}$, we define $x^\omega =\omega(x)$.
For $G=SU(2)$, this just means that 
\be
x^\omega \= 
\begin{cases} 
+ x \,, & \omega \= 1 \,, \\
- x \,, & \omega \= p \,, 
\end{cases} \,.
\ee
In particular, we use the above equation for the gauge fields $A^{L/R}$
as well as the fluctuations~$y$ and the 
parameters~$\alpha_{L/R}$.  
With this notation, and thinking of~$\omega$ as an element of~$SU(2)_R$, 
we have, for $g=  \omega \, e^{y}$, 
\be \label{eq:Jbar}
\wt \CJ_{\zbar} \= D_{\zbar} \, g \, g^{-1} \= 
\p_{\zbar} \, y^{\omega} + \AL_{\zbar} - (\AR_{\zbar})^{\omega} +  [(\AR_{\zbar})^{\omega},y^{\omega}] + \dots 
\ee
and
\be \label{eq:DJbar}
\begin{split}
D_{z} \, \wt \CJ_{\zbar} &\= 
\p_z \wt \CJ_{\zbar} + \bigl[ \AL_z, \wt \CJ_{\zbar} \bigr]  \\
   &\=  \p_z \, \p_{\zbar} \, y^{\omega}  +  [ (\AR_{\zbar})^{\omega} ,\, \p_z y^{\omega} ] +  [ \AL_z , \, \p_{\zbar} \, y^{\omega}  ] 
+ \bigl[ \AL_z \,, [(\AR_{\zbar})^{\omega}\,, y^{\omega} ]\bigr] + \dots \,,
\end{split}
\ee
where higher order terms in $y$ are omitted and we have used that our choice of the 
left and right external gauge fields commute with each other.

Upon expanding the localizing action to quadratic order, we obtain 
\be  \label{eq:delVsu2bos}
\delta V \big{|}_\text{bos} \= \int_{T^2}  \d^2z \;  \text{tr} \, 
\big| \bigl(\, \p_z + \AL_{z} \bigr) \bigl(\, \p_{\zbar} + (\AR_{\zbar})^{\omega} \bigr) y^{\omega} \big|^2 \,,
\ee
and 
\be \label{eq:delVsu2fer}
\delta V \big{|}_\text{fer} \= -i \int_{T^2}  \d^2z \;  \text{tr} \,  \bigl(\, \p_z +  \AL_{z} \bigr)  \, \wt \psi \;
 \bigl(\, \p_{\zbar}  +  \AL_{\zbar} \bigr)\, \bigl(\, \p_z  +   (\AR_z)^{\omega} \bigr) \,\wt \psi    \,.
\ee
This deformation obeys the conditions described in the introduction.
For localization, we can take the quadratic action for fluctuations to be the sum of $\delta V \big{|}_\text{bos}$ and $\delta V \big{|}_\text{fer} $, plus 
the original action  in~\eqref{eq:gaugedsusyact} for the free left-moving fermion $\psi$.

The one-loop determinant is calculated as in section~\ref{sec:locPIU1}, and is given 
by\footnote{Here, when we write a gauge field it represents its action on the field that follows 
according to the charge of the field, e.g.~$\AL_{\zbar} \, \wt \psi \equiv [\AL_{\zbar}, \wt \psi \,] $.} 
\be  \label{eq:Z1loopsu2}
\begin{split}
&Z_\text{1-loop}  \= \\
 & \quad\frac{\text{Pf}^{\,'}  \bigl(\,\p_{\zbar}  +   \AR_{\zbar} \bigr) \; \; \text{Pf}^{\,'}  \bigl(\, \p_z  +  \AL_z \bigr) 
 \bigl(\, \p_{\zbar} +  \AL_{\zbar} \bigr) \, \bigl(\, \p_z  +  (\AR_{z})^{\omega} \bigr)}{\bigl(\text{det}^{\,'} 
	 \bigl(\,\p_{\zbar}  +   (\AR_{\zbar})^{\omega} \bigr) \, \bigl(\, \p_{\zbar}  +  \AL_{\zbar} \bigr) 
 \bigl(\, \p_z  +  \AL_{z} \bigr) \, \bigl(\, \p_z  +  (\AR_{z})^{\omega} \bigr) \bigr)^\frac12} \, \times \,
\frac{1}{(2 \pi^2\t_2)^{1/2}}  \, \times \, \text{Vol}(H) \,.  
\end{split}
\ee
The pfaffians in the numerator and the determinants in the denominator come from the integral over the non-zero-modes. 
The remaining factor ${\mathrm {Vol}}(H)/{(2 \pi^2\t_2)^{1/2}}$ comes from the integral over the $\mathfrak h$-valued 
zero-mode of $y$ and of the fermions, with the measure determined 
by ultralocality, as in section \ref{two}.  
The bosonic and fermionic zero-modes are the same as those of the $U(1)$ model (embedded in either
$S_1$ or $S_p$), so the calculation of this factor reduces to that of section \ref{two}.  
In particular, since $H=U(1)$, we can borrow the result~$\text{Vol}(H) = 2 \pi R = 2 \pi \sqrt{2k}$ 
from the~$U(1)$ theory.   
One has ${\rm{Pf}}'(\partial_{\zbar}+A_{\zbar}^R)={\rm{Pf}}'(\partial_{\zbar}+(A_{\zbar}^R)^p)$, 
since $A_{\zbar}^R$ and $(A_{\zbar}^R)^{\omega}$ differ by the constant gauge transformation $\omega$.  
Taking this into account, and since the pfaffian is the same as the square root of the determinant up to sign,
the infinite products over the non-zero-modes of bosons and fermions manifestly 
cancel up to a factor~$\pm 1$.  This sign is discussed momentarily.

\bigskip

Now we will calculate the fluctuation determinant around the localizing solution with general $(m,w)$.
We can expand the WZW field $g$ around this solution 
as~$g_{m,w}^\omega \, e^{y}  $, where~$g_{m,w}^\omega $ is defined in~\eqref{eq:Weyltwist}.    
As in the derivation of  eqn.~(\ref{wintrans}), a single-valued
$G_R$ gauge transformation by 
\be\label{gtrans}e^{\sutwogen{3}(w(\wsx-(\tau_1/\tau_2)\wsy)+m\wsy/\tau_2)}\ee
adds $(m,w)$ to the winding numbers of the localization solution and simultaneously shifts~$\AR$ as
\be \label{eq:gaugefieldshift}
\AR \mapsto \AR + \frac{\sutwogen{3}}{2 i \t_2} \bigl( (m-w\tau) d\overline{z} - (m-w\taubar) \d z \bigr) \,,
\ee  
with no change in $\AL$.  The shift in $\AR$ modifies the expression~\eqref{eq:Jbar} for $\wt \CJ_{\zbar} $ 
and therefore modifies the actions
\eqref{eq:delVsu2bos}, \eqref{eq:delVsu2fer} that govern fluctuations in $g$ and $\widetilde\psi$.    There
is no such shift in the action for $\psi$.     
Thus, in the ratio of determinants~\eqref{eq:Z1loopsu2}, 
there is no  shift in $(\AR_{\zbar})^{\omega}$ in the first factor $\text{Pf}^{\,'}  \bigl(\,\p_{\zbar}  +  (\AR_{\zbar} )^p\bigr)$ 
 in the numerator (which comes from the $\psi$ path integral)
though there are such shifts where $(\AR_{\zbar})^{\omega}$ appears elsewhere in the numerator or denominator.
This does not affect the cancellation between the numerator and denominator, because the relevant factor 
$\text{Pf}^{\,'}  \bigl(\,\p_{\zbar}  +   (\AR_{\zbar})^p \bigr)$ is anyway invariant under the shift in question, 
which shifts the lattice of charged modes by a lattice element.

For a slightly different explanation, since  a solution with any $(m,w)$ 
in the bosonic model can be reached from a solution with $(m,w)=(0,0)$ by the  
gauge transformation~(\ref{gtrans}), we can supersymmetrize the model  using the new gauge 
field from the very beginning.   Repeating the previous calculation with a different
gauge field in the starting point, we clearly get the same cancellation as before.

\bigskip

\ndt {\bf The~$\omega$-dependence of the sign} \\
The denominator of~\eqref{eq:Z1loopsu2}, coming from the bosonic determinant, is positive, while the phase of
the numerator needs to be determined more carefully.  
We can determine the dependence on~$\omega$ of the overall phase of~$Z_\text{1-loop}$ as follows.
First recall that for~$SU(2)$, the Weyl group has two elements, which, for our choice of Cartan subgroup, 
we can take to be the following elements of $SU(2)$:  $\omega=1$ and 
and~$\omega=p\equiv\biggl(\, \begin{matrix} 0 & 1 \\ -1 & 0 \end{matrix} \, \biggr)$. 
We assign~$\text{sgn}(1) = +1$ and~$\text{sgn}(p) = -1$.

From eqn.~(\ref{eq:Weyltwist}), we see that we can view $\omega$ as an element of $G_L$ (or $G_R$).   
As such it is a global symmetry of the fermion measure.  Indeed, if we view $\omega$ as a linear transformation
of~$\frak g_L$, then its determinant is $+1$, and therefore, for each value of the momentum, 
the fermion measure is invariant.  However, we have to take into account
the zero-mode insertion $i\, \psi_0^3\, \wt\psi_0^3$.   Of course, for $\omega=1$, this insertion is invariant.  
But  $\omega=p$ reverses the sign
of $i\, \psi_0^3\, \wt\psi_0^3$, since, for $p\in G_L$, $\psi_0^3$ is odd under $p$ and $\wt\psi_0^3$ is even.   
So the fermion path integral with the zero-mode insertions transforms under $\omega$ as $\text{sgn}(\omega)$.

This result has a straightforward generalization to any simple Lie group~$G$, say of rank~$r$, with maximal torus~$H$.   
The zero-mode insertion in the twisted model is then
$i^r \, \psi_0^1\psi_0^2\cdots \psi_0^r \, \wt\psi_0^1\wt\psi_0^2\cdots \wt\psi_0^r$, where the superscript runs over a basis of 
the Cartan subalgebra~$\mathfrak h$.
Under a general Weyl group element $\omega$, which can be regarded as an element of $G_L$, the~$\wt\psi_0^i$ are invariant.   
The change in the measure comes from the action of $\omega$ on $\psi_0^1\psi_0^2\cdots \psi_0^r$, or equivalently, 
from the determinant of $\omega$ regarded as a linear transformation of $\mathfrak h$.
The Weyl group of $G$ is generated by $r$ elementary reflections, each of which acts on $\mathfrak h$ with 
determinant~$-1$.   If we factor $\omega$ as $q_1\cdots q_s$, where the $q_i$ are elementary reflections,
then the determinant of $\omega$ is what we may call $\text{sgn}(\omega)=(-1)^s$.   
So in the representation of the path integral as a sum over classical solutions, we must weight the solution
of given~$\omega$ by a factor~$\text{sgn}(\omega)$.

\bigskip
\ndt {\bf Classical action} \\
The action~\eqref{eq:gaugedsusyact} of the solutions~\eqref{eq:Weyltwist} with~$\omega=1$
and the gauge fields given in~\eqref{eq:extAAbar} is  
\be  \label{eq:actvalue}
S^{(1)}_{m,w}(\tau,\alphaL,\alphaR^{})  \= \frac{\pi\,k}{\t_2} 
\Bigl(\bigl(m - w\,\tau + 2\alphaR^{}\bigr) \bigl(m-w \,\taubar - 2\alphabarL\bigr) 
+  2 \alphaR^{}\, \alphabarL^{}+  \bigl( \alphaL^{}\, \alphabarL^{}+  \alphaR^{}\, \alphabarR^{}\bigr)
\Bigr)\,.
\ee
The only effect on the action of a non-trivial Weyl element~$p$  is ~$\AL \mapsto (\AL)^{p}$,
and therefore~$\alphaL^{}\mapsto \alphaL^{p}=-\alphaL$. 
For general $\omega$, the action~\eqref{eq:gaugedsusyact} of the solutions~\eqref{eq:Weyltwist} is therefore
\be \label{eq:actvaluep}
S^{(\omega)}_{m,w}(\tau,\alphaL,\alphaR^{})  \= \frac{\pi\,k}{\t_2} 
\Bigl( \bigl(m - w\,\tau + 2\alphaR^{}\bigr) \bigl(m-w \,\taubar - 2\alphabarL^\omega \bigr) 
+ 2 \alphaR^{}\, \alphabarL^\omega +  \bigl( \alphaL^\omega \, \alphabarL^\omega +  \alphaR^{}\, \alphabarR^{}\bigr) 
\Bigr)\,.
\ee
For $G=SU(2)$, the only nontrivial Weyl element $p$ satisfies $(A^L)^p=-A^L$
so the classical actions~$S^{(\omega)}$ for~$\omega=1$ and~$\omega=p$
given in~\eqref{eq:actvalue} and~\eqref{eq:actvaluep}, respectively, are related as
\be \label{eq:S1Sprel}
S^{(1)}_{m,w}(\tau,\alphaL,\alphaR^{}) \= S^{(p)}_{m,w}(\tau,-\alphaL,\alphaR^{}).
\ee

\ndt {\bf Full result} \\
Upon putting everything together, we obtain 
\be \label{eq:PIResultsu2}
Z_\text{SWZW}(\tau,\alphaL, \alphaR^{}) \= 
\frac{2\pi\sqrt{2k}}{(2 \pi^2\t_2)^{1/2}}  
 \sum_{\omega \in W_G}  \, \sgn(\omega) \, 
\sum_{m,w \in \IZ}  \exp \bigl(-S^{(\omega)}_{m,w}(\tau,\taubar,\alphaL,\alphaR^{})  \bigr) \,,
\ee
with the action~$S^{(\omega)}_{m,w}$ given in~\eqref{eq:actvalue}, \eqref{eq:actvaluep}. 

\bigskip

\ndt {\bf Modular invariance}: 
The effect of~$(\t,\alphaL,\alphaR^{}) \mapsto (\t+1,\alphaL,\alphaR^{})$ is the shift~$(m,w)\mapsto (m-w,w)$ 
in the above sum and the effect of~$(\t,\alphaL,\alphaR^{}) \mapsto \bigl(-\frac{1}{\t},\frac{\alphaL}{\t},\frac{\alphaR}{\t} \bigr)$ 
is the shift~$(m,w)\mapsto (w,-m)$. 
This shows that the expression~\eqref{eq:PIResultsu2} is modular invariant.

\bigskip

\ndt {\bf Gauge invariance}: 
As in eqn.~(\ref{wintrans}), abelian gauge transformations can shift $\alphaL$ by $\alphaL\to \alphaL^{}+\lambda\t+\mu$, 
with $\mu,\lambda\in\IZ$, and similarly for $\alphaR$. However, we do not expect full gauge invariance 
because the gauged WZW model is anomalous.   As explained in the discussion of eqn.~(\ref{anomaly}),
full gauge-invariance is expected if we restrict to a ``diagonal'' theory with $\alphaL=\alphaR$ or to an 
``anti-diagonal'' theory with $\alphaL=-\alphaR$.
If~$\alphaL=\alphaR=\alpha$, the action~\eqref{eq:actvalue} takes the form 
\be  \label{eq:actvaluepp}
S^{(1)}_{m,w} (\tau,\a,\a) \= \frac{\pi\,k}{\t_2} \Bigl( \bigl|m - w\,\tau\bigr|^2  - 2 \bigl(m - w\,\tau\bigr) \overline{\a}
+ 2 \a \bigl(m-w \,\taubar\bigr) \Bigr)\,,
\ee
In this case, the effect of~$(\t,\a) \mapsto (\t,\a + \lambda \t + \mu)$
is to shift the action by~$4 \pi i (\lambda m - \mu w)$, so that~$e^{-S_{m,w}}$ remains invariant.
For~$-\alphaL=\alphaR=\alpha$, the action is
\be  \label{eq:actvaluepm}
S^{(1)}_{m,w}(\tau,-\a,\a)  \= \frac{\pi\,k}{\t_2}   \bigl(m - w\,\tau +2 \a \bigr) \bigl(m-w \,\taubar + 2\overline{\a}\bigr) \,.
\ee
In this case, the effect of~$(\t,\a) \mapsto (\t,\a + \lambda \t + \mu)$ is~$(m,w)\mapsto (m+2\mu,w-2\lambda)$,
so that the corresponding lattice  sum in~\eqref{eq:PIResultsu2} is gauge-invariant.    

Actually, these formulas show that for $\alpha_L=\pm \alpha_R=\alpha$,  there is a 
symmetry $\a\to \a+\lambda\t+\mu$ for $\lambda,\mu\in\frac{1}{2}\Z$, not just $\lambda,\mu\in\Z$.   
The reason is that $SU(2)$ has a non-trivial center $\Z_2= \{\pm 1 \}$, and a central element of $SU(2)$ 
acts in the same way on the $SU(2)$ group manifold whether it acts on the left or the right.  
Accordingly, the faithfully acting global symmetry of the $SU(2)$ WZW model is actually not
$SU(2)_L\times SU(2)_R$ but $(SU(2)_L\times SU(2)_R)/\Z_2$, and  a maximal abelian subgroup 
of this  is really
$(U(1)_L\times U(1)_R)/\Z_2$, not $U(1)_L\times U(1)_R$.    Consequently, the partition function 
is unchanged if we multiply  the holonomies  of both left and right $U(1)$ gauge fields  by
the same central elements $\pm 1$ in going around one-cycles in $T^2$.   A transformation $\a\to \a+\lambda\t+\mu$
with $\lambda,\mu\in\frac{1}{2}\Z$ modifies the holonomies in precisely such a fashion, accounting for why this is a symmetry.   
In view of \eqref{eq:S1Sprel}, the same  is true for $p\not=1$.

\bigskip

Finally, in order to obtain the partition function of the bosonic theory,
we should divide the above expression for~$Z_\text{SWZW}(\tau,\alphaL, \alphaR^{})$ 
by the fermionic partition function~$Z_\text{fer}(\tau,\alphaL, \alphaR^{})$, 
\be \label{eq:ZWZWbos}
Z_\text{WZW}(\tau,\alphaL, \alphaR^{}) \= \frac{Z_\text{SWZW}(\tau,\alphaL, \alphaR^{})}{Z_\text{fer}(\tau,\alphaL, \alphaR^{})} \,.
\ee
For an explicit formula for $Z_{\rm fer}(\tau,\alpha_L,\alpha_R)$, see eqn.~(\ref{fermcontx}) below.

The equation~\eqref{eq:S1Sprel} and the analysis of the one-loop corrections imply the following simple 
relation between the diagonal and anti-diagonal theory in the supersymmetric case
\be
Z_\text{SWZW}(\tau,\a, \a) \= - Z_\text{SWZW}(\tau,-\a, \a).
\ee
This relation reflects the fact that a non-trivial Weyl transformation in $SU(2)_L$ 
 exchanges the diagonal and anti-diagonal theories while reversing the sign of $\alpha_L$.
The minus sign comes from the action of the Weyl transformation on the fermion measure.   
Dividing out the fermion path integral eliminates this minus sign, so
the diagonal and anti-diagonal versions of the purely bosonic theory have equal partition functions, 
\be
Z_\text{WZW}(\tau,\a, \a) \= Z_\text{WZW}(\tau,-\a, \a) \,. 
\ee

\bigskip

\subsection{Equivalence to Hamiltonian trace and sum over characters \label{sec:HamTracesu2}}

Roughly speaking, the partition function~\eqref{eq:ZWZW} that we have computed from path integrals
has a Hamiltonian interpretation as a  trace
\be \label{hilbtr}
Z_\text{SWZW}(\tau ; \alphaL, \alphaR^{}) \overset{?}{\=}
\text{Tr}_{\CH} (-1)^{F} \, i\, \psi^3_0 \, \wt \psi^3_0 \,
q^{L_0} \, \qbar^{\overline{L}_0} \, e^{2\pi i \alphaR^{} Q_{3,R}} \,  e^{2\pi i \alphabarL^{}Q_{3,L}}
\ee 
over the Hilbert space $\CH$ of the supersymmetric WZW model; here $Q_{3,L/R}$ are conserved charges
associated to the diagonal generator 
$\sigma_3=\biggl(\, \begin{matrix}  1&0\cr 0&-1\end{matrix} \, \biggr)$ of $SU(2)_{L/R}$.
However, this is oversimplified, as we will discuss.

 In developing a Hamiltonian formalism, we will treat $\wsx$ as the ``space'' coordinate on $T^2$ and $\wsy$
as a ``time'' coordinate.   Looking back to eqn.~(\ref{acart}), we see that the holonomy of the gauge 
field $a_{L/R}$ around the circle $S$ defined by $0\leq \wsx \leq 2\pi$ is $\varphi_{L/R}\, \sutwogen{3}$ with
\be\label{holcircle}
\varphi_{L/R}\=2\pi  \frac{{\rm Im}\,\alpha_{L/R}}{\tau_2} \,. 
\ee  
The SWZW fields on $S$ are twisted in a way that depends on $\varphi_{L/R}$.
We write $\H_{\varphi_{L/R}}$ for the Hilbert space of the  model, quantized in the presence of this twist.   
Then the path integral on the torus can be computed as a trace in $\H_{\varphi_{L,R}}$. 
In computing this trace, the state is propagated forwards in imaginary
time by an amount $2\pi \Im\,\tau$, rotated by an angle $2\pi \Re\,\tau$, and finally 
it undergoes an $SU(2)_L\times SU(2)_R$ transformation  that is determined
by the component of the background gauge field in the $\wsy$ direction.   
As usual, the rotation and imaginary time translation give a factor $q^{L_0}\, \overline q^{\barL_0}$ in the trace.
From eqn.~(\ref{acart}), we see that the component of the background gauge
field in the $\wsy$ direction is $-\sutwogen{3} \, {\rm Re}\,\alpha_{L/R}/\tau_2$.   Consequently, the $SU(2)_L\times SU(2)_R$ 
transformation that appears in the trace is $g_L \, g_R^{-1}$ with $g_{L/R}=\exp(2\pi i Q_{3,L/R} \Re\,\alpha_{L/R})$.  
Altogether, the Hamiltonian interpretation of the torus path integral is actually
\be \label{hilbtr2}
Z_\text{SWZW}(\tau ; \alphaL, \alphaR^{}) \=
\text{Tr}_{\CH_{\varphi_{L/R}}} (-1)^{F} \, i\, \psi^3_0 \, \wt \psi^3_0 \, 
q^{L_0} \, \qbar^{\overline{L}_0} \, e^{2\pi i {\rm Re}\,\alphaR^{} Q_{3,R}} \,  e^{2\pi i {\rm Re}\,{\a}_L Q_{3,L}} \,.
\ee 
If $\alphaR$ and $\alphaL$ are real, then the twist angles $\varphi_{L/R}$ vanish, and eqn.~(\ref{hilbtr2}) reduces 
to the trace (\ref{hilbtr}) in the original untwisted Hilbert space~$\H$.  
In general, though, as we will discuss, the two formulas differ by certain elementary but important factors 
that encode the shifts in the ground state energy and charge that results from twisting.

To begin with, we assume that $\alphaR^{}$ and $\alphaL$ are real, so that the two formulas coincide.
We will first analyze the partition function of the positive chirality fermions of the SWZW model.
This partition function, including the insertion~$(-1)^F$ and the integral over   
the charged fermion zero-modes (but suppressing for the moment the neutral zero-mode), is given by
\be \label{trace1}
  q^{1/8}  (\zeta-\zeta^{-1})   \prod_{m=1}^\infty (1-q^m) (1-q^m \zeta^2)(1-q^{m}\zeta^{-2}) \,,
\ee
with~$q=e^{2 \pi i \tau}$, $\zeta=e^{2\pi i \alphaR}$.   
The infinite product comes from a straightforward count of the states of the positive
chirality fermions, using  powers of $q$ and $\zeta$ to keep track of the values of $L_0$ and $Q_{3,R}$  
(the infinite product is a function of $\zeta^2$ since the charged fermions, being the off-diagonal components of the adjoint representation,  have $Q_{3,R}=\pm 2$).    
The remaining factors in eqn.~(\ref{trace1}) are more subtle.
The factor $q^{1/8}$ reflects the ground state energy of three (untwisted) Majorana fermions in the Ramond sector.    
Quantizing two fermion zero-modes of charges $\pm 2$ gives a pair of quantum 
states with $Q_{3,R}$ charges $\pm 1$, and with eigenvalues
$\pm 1$ of the operator $(-1)^F$, accounting for the factor  $(\zeta-\zeta^{-1})$.
We have not included the neutral fermion zero-mode in this calculation because an odd
number of Majorana zero-modes do not have a natural Hilbert space quantization. 
See the discussion at the end of the introduction and in Appendix~\ref{orbifoldtheory}. 
When we combine positive and negative chirality modes, there will be two neutral fermion zero-modes, and their 
quantization will give two states, contributing a factor of 2 in the trace~(\ref{hilbtr}).

By an identity due to Jacobi, the function in eqn.~(\ref{trace1}) can be more concisely written 
as $\vartheta_1(\tau,2\,\alphaR^{})$, where 
\be\label{eq:Jactheta1}
\vartheta_1(\tau,z)\=\sum_{n\,\in \, \Z+1/2}(-1)^n \, q^{n^2/2} \, e^{2\pi i n z}
\ee
is the odd Jacobi theta function.   Including also the  negative chirality fermions, whose trace is computed in the same way,
 and including a factor of 2 from the neutral fermion zero-modes, we find that, if $\alphaL$ and $\alphaR$ are real, 
the contribution of the fermions to either~(\ref{hilbtr}) or~(\ref{hilbtr2}) is a factor
\be\label{fermcont} 
2 \, \vartheta_1(\tau,2\,\alphaR^{})\, \overline{\vartheta_1(\tau,2\,\alphaL^{})} \,. 
\ee

In the case of~(\ref{hilbtr}), the generalization away from real $\alpha_{L/R}$ is straightforward.   
The trace in eqn.~(\ref{hilbtr}) is holomorphic in $\alphaR$ and antiholomorphic
in $\alphaL$; equivalently, it is holomorphic in $\alphaR^{}$ and ${\overline \alpha}_L$.   
The function in eqn.~(\ref{fermcont}) has the same property, so it represents the fermion contribution 
to (\ref{hilbtr}) regardless of whether  $\alpha_{L/R}$ are real.   
But the trace that we really want, namely the one in eqn.~(\ref{hilbtr2}), is slightly different.  
Similarly to the derivation of eqn.~(\ref{trace1}), the sum over oscillator modes in
eqn.~(\ref{hilbtr2}) can be straightforwardly computed by counting the oscillator modes weighted by 
the appropriate values of $L_0$, $\barL_0$, and $Q_{3,L/R}$.
This actually gives a result consistent with eqn.~(\ref{hilbtr}).   However, to evaluate the trace in eqn.~(\ref{hilbtr2}), 
we also need to the know the energy and charges of the ground state.   The twisting by $\varphi_{L/R}$ shifts 
the ground state quantum numbers in a way that is widely used in string theory (for example, in applications to orbifolds)
and is briefly reviewed in Appendix~\ref{orbifoldtheory}.   The upshot is that twisting by $\varphi_{L,R}$ shifts 
each of\footnote{The shifts in $L_0$ and $\barL_0$ are equal, since there is no shift in $L_0-\overline{ L}_0$, 
whose eigenvalues are integers.}  $L_0$ and $\barL_0$ by an amount $\frac{1}{4\pi^2}(\varphi_L^2+\varphi_R^2)$ 
that is not reflected in eqn.~(\ref{hilbtr}).  This shifts
$\exp(-2\pi \tau_2(L_0+\barL_0))$ by a factor 
$\exp(-\frac{1}{\pi} \tau_2(\varphi_L^2+\varphi_R^2))=\exp(-4\pi( ({\rm Im}\,\alphaL)^2+ ({\rm Im}\,\alphaR)^2)/\tau_2)$, 
where we used eqn.~(\ref{holcircle}).   
In addition, twisting by $\varphi_{L,R}$ shifts $Q_{3,R}$ by $\varphi_R/\pi$ and shifts $Q_{3,L}$ by $-\varphi_L/\pi$.  
Using eqn.~(\ref{holcircle}) again, the  result is to multiply the operator 
$e^{2\pi i {\rm Re}\,\alphaR^{} Q_{3,R}} \,  e^{2\pi i {\rm Re}\,{\a}_L Q_{3,L}}$ that appears in~(\ref{hilbtr2}) 
by a factor $\exp\left(\frac{4\pi i}{\tau_2}({\rm Re}\,\alphaR^{}\,{\rm Im}\,\alphaR^{}-{\rm Re}\,\alphaL^{}\,
{\rm Im}\,\alphaL^{})\right)$.   
The product of this with the factor that comes from the shift of $L_0$ and $\barL_0$ is
\be\label{zeldico} 
C_{\rm fer}\=\exp\left(\frac{4\pi i}{\tau_2}(\alphaR^{}\, {\rm Im}\,\alphaR^{}-\overline \alpha_L^{}\, {\rm Im}\,\alphaL^{})\right) \,.
\ee
Finally, the contribution of the fermions to the Hamiltonian formula (\ref{hilbtr2}) for the partition function is
\be\label{fermcontx} 
Z_{\rm fer}(\t, \alphaL, \alphaR) \=2 \, C_{\rm fer}  \, \vartheta_1(\tau,2\,\alphaR^{}) \, \overline{\vartheta_1(\tau,2\,\alphaL^{})} \,. 
\ee

The fermionic partition function~\eqref{fermcontx} is a modular form of weight~$(\frac12,\frac12)$ and 
has the following behavior under elliptic transformations, with~$\lambda, \mu, \lambda', \mu' \in \frac12 \IZ$, 
\be\label{eq:ferelliptic3} 
Z_{\rm fer} \bigl(\t, \alphaL+\lambda\t+\mu, \alphaR+\lambda'\t+\mu' \bigr) 
\= e^{\frac{4 \pi i}{\t_2} \text{Im} ( \alphaR(\lambda'\taubar+\mu')- \alphaL(\lambda\taubar+\mu))}
Z_{\rm fer}\bigl(\t, \alphaL, \alphaR \bigr)  \,.
\ee
This behavior under topologically non-trivial gauge transformations follows from the second 
formula\footnote{Since the computation of $C_{\rm fer}$ in Appendix \ref{orbifoldtheory} uses bosonization 
of fermions, the formula (\ref{wintrans}) for the anomaly is directly applicable.} in (\ref{wintrans}).
From \eqref{eq:ferelliptic3}, it is clear  that when~$\a_R=\a_L=\a$ or when~$\a_R=-\a_L=\a$,
the fermionic partition function is completely gauge-invariant.
Indeed, for these values, it takes the following familiar modular and elliptically invariant form, 
\be\label{eq:ZferAnomFree} 
Z_{\rm fer}(\t, \pm \alpha, \alpha) \= 
2 \, e^{-\frac{8\pi}{\t_2}\text{Im}(\a)^2} \, \vartheta_1(\tau,2\,\a) \, \overline{\vartheta_1(\tau,\pm 2\,\a)} \,.
\ee

The anomaly can also be interpreted as an obstruction to holomorphy.  Looking back at the underlying 
Lagrangian \eqref{eq:gaugedsusyact} of
the fermion system, since $\psi$ couples to $\AR_{\bar z}$ only and not to $\AR_z$, while 
conversely $\widetilde \psi$ couples to $\AL_z$ only but not
to $\AL_{\bar z}$, one would naively expect that the resulting partition function would be holomorphic in $\alphaR$ and $\overline\alpha_L$.
Given this, since eqn.~(\ref{hilbtr}) is valid for real $\alpha_{L/R}$, it would be valid for all $\alpha_{L/R}$.   
That is not the case, because holomorphy is obstructed by an anomaly.
 The  action of (\ref{abreduction}) (which for $k=1/2$ gives a bosonized description of the fermions, 
 as exploited in Appendix \ref{orbifoldtheory}) is not holomorphic in $a^L_z$ and $a^R_{\bar z}$. 
But  this is obstructed only by terms that are homogeneous and quadratic in $a^{L/R}$ and depend on $a^{L/R}$ only.   
Hence the holomorphically factorized function (\ref{hilbtr}) differs from the correct formula (\ref{hilbtr2}) 
by the exponential of a homogeneous quadratic function of $\alpha_{L/R}$, as found in the last paragraph.  

The analysis of the bosonic contribution to the trace (\ref{hilbtr2}) is similar, except that one has to use 
the characters of affine Lie algebras, rather than free fermion partition functions.
The characters of the affine extension $\mathfrak{su}(2)_{k-2}$ of $SU(2)$ at level $k-2$
are the functions $\chi(\tau,\alpha)=\Tr_R\, q^{L_0}e^{2\pi i \alpha Q_3}$, where $R$ is an irreducible 
unitary representation of the algebra at that level.   Up to equivalence,
there are $k-1$ such representations $R_\ell$, $\ell=1,2,\cdots,k-1$.
Their
characters are given by the Weyl-Kac character formula
\be \label{eq:su2char}
\chi^{k-2}_\ell (\tau,\alpha) \= \frac{\vth_{k,\ell} (\t,\a) -\vth_{k,\ell} (\t,-\a)}{\vth_1(\t,2\a)}
\ee
in terms of the theta functions~\eqref{eq:defthml} and the odd Jacobi theta 
function~\eqref{eq:Jactheta1}.  
The untwisted Hilbert space of the WZW model is $\H=\oplus_{l=1}^{k-1} \overline R_\ell \otimes R_\ell$
where $\overline R_\ell$ and $R_\ell$ are representations respectively of the antiholomorphic and 
holomorphic current algebras of the model.    So as long as $\alpha_{L/R}$ are real,
the torus partition function of the gauged WZW model is
\be\label{wzwpart1} 
Z(\tau;\alphaL,\alphaR^{})\=\sum_{\ell=1}^{k-1}\chi^{k-2}_\ell(\tau,\alphaR^{}) \, \overline{\chi^{k-2}_\ell(\tau,\alphaL^{})}\,.
\ee
This formula is analogous to eqn.~(\ref{hilbtr}) for the fermions.  For the same reasons as in that case, 
if $\alpha_{L/R}$ are not real, in a Hamiltonian formulation, the torus partition function  of the gauged WZW 
model is a trace not in the ``ordinary'' Hilbert space $\H$, but in a ``twisted'' Hilbert space $\H_{\varphi_{L/R}}$ 
that is defined in the presence of background gauge fields with holonomy around the spatial circle $S$.
Accordingly,  this torus path integral is not simply given by eqn.~(\ref{wzwpart1}) but has an additional factor that
results from a shift in the ground state quantum numbers.  This factor is actually determined by the same computation 
(described in Appendix~\ref{orbifoldtheory}) that leads to eqn.~(\ref{zeldico}).  
Indeed, the adjoint-valued fermions whose analysis leads to eqn.~(\ref{zeldico}) provide a representation 
of the~$\mathfrak{su}(2)$ current algebra at level 2.
Replacing level 2 by some other level $\kappa$  multiplies all anomalies by $\kappa/2$ and multiplies 
the exponent in eqn.~(\ref{zeldico}) by $\kappa/2$.   So the analog of eqn.~(\ref{zeldico})
at level $k-2$ is 
\be\label{bosan}
C_{\rm bos}\=\exp\left(\frac{2\pi i(k-2)}{\tau_2}(\alphaR^{}\, {\rm Im}\,\alphaR^{}-\overline{\alpha}_L^{}\, {\rm Im}\,\alphaL^{})\right) \,. 
\ee
From a Hamiltonian point of view, therefore, the expected torus partition function of the gauged WZW model is
\be \label{eq:bossu2}
Z_\text{WZW}(\tau ; \alphaL, \alphaR^{})  
 \= C_{\rm bos}\sum_{\ell=1}^{k-1} \chi^{k-2}_\ell (\alphaR^{}) \, \overline{\chi^{k-2}_\ell (\alphaL^{})} \,.
\ee

A Hamiltonian formula for  the  torus path integral of the gauged SWZW model is then
simply the product of the factors that we have computed for bosons and fermions in
eqns.~(\ref{fermcont}) and~\eqref{eq:bossu2}.   
So we combine the anomalous factors from bosons and fermions
\be\label{proddef}
C_k\=C_{\rm fer} \, C_{\rm bos} \= 
\exp\left(\frac{2\pi i k}{\tau_2}(\alphaR^{}\, {\rm Im}\,\alphaR^{}-\overline{\alpha}_L\, {\rm Im}\,\alphaL^{})\right)\,, 
\ee
and, after a nice cancellation of factors of $\vartheta_1$, we find that the Hamiltonian prediction 
for the torus path integral of the SWZW model is 
\be \label{eq:susysu2Ham}
\begin{split}
& Z_\text{SWZW}(\tau ; \alphaL, \alphaR^{})  \\  
&\;  \=  C_k(\alphaL,\alphaR^{}) \! \! 
\sum_{\ell \, (\text{mod} \; 2k)} \bigl(\vth_{k,\ell} \bigl(\t,\alphaR\bigr) -\vth_{k,\ell} \bigl(\t,-\alphaR^{}\bigr) \bigr)
\bigl(\overline{ \vth_{k,\ell}  \bigl(\t,\alphaL^{}\bigr)} - \overline{\vth_{k,\ell}  \bigl(\t,-\alphaL^{}\bigr)} \bigr)  \, .
\end{split}
\ee
Here we have expressed~$ Z_\text{SWZW}$ as a sum over~$\ell \, (\text{mod} \; 2k)$ 
(compared to a sum from~$1$ to~$k-1$ in~\eqref{eq:bossu2}) using the theta function 
identities~$\vth_{k,\ell}(\t,-z) = \vth_{k,-\ell}(\t,z) = \vth_{k,2k-\ell}(\t,z)$, and absorbing 
the additional factor of~2 in~\eqref{fermcont}.
Of the four products of theta functions in~\eqref{eq:bossu2}, 
the two terms where~$\alphaL$ and~$\alphaR$ appear in the arguments with the same signs 
are equal and the other two where they appear with opposite signs are also equal. 
Hence~\eqref{eq:susysu2Ham} reduces to two independent sums as follows, 
\be \label{eq:susysu2Hamfin}
\begin{split}
& Z_\text{SWZW}(\tau ; \alphaL, \alphaR^{})  \\  
&\quad  \=  2 \, C_k(\alphaL,\alphaR^{}) \! \! 
\sum_{\ell \, (\text{mod} \; 2k)}  \vth_{k,\ell} \bigl(\t,\alphaR\bigr) \, 
\overline{ \vth_{k,\ell}  \bigl(\t,\alphaL^{}\bigr)} \\
& \; \quad \qquad   - 2 \, C_k(\alphaL,\alphaR^{}) \! \! 
\sum_{\ell \, (\text{mod} \; 2k)}  \vth_{k,\ell} \bigl(\t,\alphaR^{}\bigr) \,  \overline{\vth_{k,\ell}  \bigl(\t,-\alphaL^{}\bigr)}   \, .
\end{split}
\ee

Now we compare the formula~\eqref{eq:susysu2Hamfin} to the formula~\eqref{eq:PIResultsu2} 
discussed in the previous subsection.\footnote{A detailed discussion of such lattices sums will be given in~\cite{MurZag}.}    
Both formulas contain two sums which are naturally associated, respectively, with the two 
elements~$\omega =1$ and~$\omega=p$ of the Weyl group. 
By using the Poisson summation formula, we now show that the first sums in the  two formulas are equal.  
This implies that the two formulas agree since, 
 modulo $\alpha_L\to -\alpha_L$,
the same argument applies to the  second pair of sums.

The first sum in~\eqref{eq:susysu2Hamfin} is  
\be \label{eq:defZ1alar}
Z_1(\t,\alphaL,\alphaR^{}) \= 2 \, C_k
\sum_{\ell \, ( \text{mod} \; 2k)} \vth_{k,\ell} (\t,\alphaR^{}) \; \overline{\vth_{k,\ell} (\t,\alphaL^{})} \, .
\ee
Writing out the theta functions as  sums over integers as in~\eqref{eq:defthml}, we obtain 
\be \label{eq:Zplmindef}
Z_1(\t,\alphaL,\alphaR^{}) \= 
2 \, C_k\sum_{n_1,n_2 \in \IZ \atop n_1 \,\equiv \, n_2 \; \text{mod}\, 2k}
\exp \Bigl( \, 2 \pi i \, \frac{n_1^2 \, \tau}{4k}  - 2 \pi i \, \frac{n_2^2\, \taubar}{4k} 
+ 2\pi i (n_1 \alphaR^{}- n_2 \, \alphabarL^{})  \Bigr) \,.
\ee
The congruence conditions can be solved by writing~$n_1 = r+ks$, $n_2=r-ks$, $r,s \in \IZ$,
and the summation variables can then be replaced by the two unconstrained integers~$r,s$.
This gives
\be \label{eq:defZLR}
Z_1(\t,\alphaL,\alphaR^{}) \= 2\, C_k
\sum_{r,s \in \IZ}
\exp \Bigl( \, - \frac{\pi \t_2}{k} \, (r^2+ k^2 s^2) + 2\pi i \t_1 sr 
+ 2\pi i r (\alphaR^{}-\alphabarL^{}) + 2\pi i ks (\alphaR^{}+\alphabarL^{}) \Bigr) \,.
\ee
Poisson resummation of the sum over~$r$ then leads to 
\be \label{eq:ZminPoisson}
\begin{split}
& Z_1(\t,\alphaL,\alphaR^{}) \\
& \quad \= 2\,C_k\sqrt{\frac{k}{\tau_2}} \sum_{s,m \in \IZ}
\exp \Bigl( \, - \pi \t_2 k \,s^2 + 2\pi i ks (\alphaR^{}+\alphabarL^{}) - \frac{\pi k}{\t_2} 
\bigl(m-s\t_1 - (\alphaR^{}-\alphabarL^{}) \bigr)^2 \Bigr) \\
& \quad \= 2\,C_k \sqrt{\frac{k}{\tau_2}}
\exp \Bigl( \, - \frac{\pi k}{\t_2} \, (\alphaR^{}-\alphabarL^{})^2 \Bigr) \\
& \qquad \qquad \sum_{s,m \in \IZ}
\exp \Bigl( \, - \frac{\pi k}{\t_2} \, \bigl(s^2 \t_2^2 + (m-s\t_1)^2 +  2(m-s\t_1)(\alphaR^{}-\alphabarL^{})
+ 2 i \t_2 s (\alphaR^{}+\alphabarL^{}) \bigr) \Bigr) \,.
\end{split}
\ee 
This agrees with the term~$\omega=1$ in~\eqref{eq:PIResultsu2} upon using~$C_k$ given in~\eqref{proddef}.
This completes the verification that the formula for the torus path integral that we obtained from localization 
agrees with a Hamiltonian description using the Weyl-Kac character formula.

\section{General Compact Lie Groups}\label{sec:othergroups}

The formula~\eqref{eq:PIResultsu2} obtained in the previous section for~$SU(2)$ largely goes 
through without essential change for any compact and simply-connected simple Lie group $G$,
The localizing solutions in eqn.~\eqref{eq:Weyltwist} are generalized by replacing $\sutwogen{3}$
by a basis of generators $h^i$, $ i=1,\cdots,\rm{rank}(G)$ of a maximal torus $H\subset G$, 
and replacing the pair of integers $(m,w)$
by a corresponding collection of integers  $(m_i, w_i)$, $i=1,2,\dots, \text{rank}(G)$.   
See the discussion in Appendix~\ref{sec:locsol}.
The quantity~$\omega$ is now valued in the Weyl group of $G$.
Assuming that the $h^i$ are normalized to $\langle h^i,h^j \rangle=-2\delta^{ij}$ 
(where $\langle~,~\rangle$ is the invariant inner product on the Lie algebra $\mathfrak{g}$, normalized 
in a conventional fashion\footnote{For $G=SU(N)$, $\langle~,~\rangle $ is defined to agree with
the trace in the fundamental representation; for  any simple $G$, 
it is defined so that its restriction to a minimal $SU(2)$ subgroup agrees with the 
trace in the fundamental representation.}), the generalization of \eqref{eq:Weyltwist} is
\be \label{eq:WeyltwistsuN}
 g^\omega_{M,W} (z,\zbar) \= 
   \omega \, h_0\, \exp \Bigl( \frac{1}{\t_2} \bigl( M (z - \zbar) + W ( \tau \zbar - \taubar z) \bigr) \Bigr) 
  \qquad h_0\in H\,,~~~~\omega \in \text{Weyl}(G)\,.
\ee 
where~$M=\sum_{i=1}^\text{rk(G)} m^i  h^i$, $W=\sum_{i=1}^\text{rk(G)} w^i  h^i$, 
with ~$m^i, w^i \in \IZ$.  As before, $h_0$ is an $H$-valued constant.

By reasoning similar to what we have explained for $SU(2)$, the  partition function is given by\footnote{The factor 
$(-1)^\zeta$  was omitted in the original version of this paper. (This factor is trivial for even $k$  since $\la h_i,h_j\ra\in\Bbb Z$ and also for $G=SU(2)$  
since in that case $\la h_i,h_j\ra$ is even, but in general it is nontrivial.)
Boan Zhao argued  that the formula without this factor is valid for even~$k$  (and for $G=SU(2)$) but  not true
in general~\cite{Zhao:2025hen}. 
Yongchao L\"u~\cite{Lu:2025} determined the missing factor, showed that it arises by evaluating the Wess-Zumino 
term for the localizing solutions, and gave a complete proof.
}
\be \label{eq:PIResultsuN}
\begin{split}
& Z_\text{SWZW}(\tau,\alphaL, \alphaR^{}) \= \\
& \quad \frac{\text{Vol}(H)}{(2 \pi^2\t_2)^{\text{rk}(G)/2}}  
 \sum_{\omega \in W_G}  \, \sgn(\omega)
\sum_{m^i,w^i \in \IZ}   \, (-1)^\zeta \, 
\exp \bigl(-S^{(\omega)}_{M,W}(\tau,\alphaL,\alphaR^{})  \bigr) \,,
\end{split}
\ee
with the action~$S^{(\omega)}_{m,w}$ given by
\be \label{eq:actvaluepsuN}
\begin{split}
&S^{(\omega)}_{M,W}(\tau,\alphaL,\alphaR^{})  \\
& \qquad \= \frac{k \, \pi}{\t_2} 
\text{tr} \Bigl(  \bigl(M - W\,\tau + 2\alphaR^{}\bigr) \bigl(M-W \,\taubar - 2\alphabarL^\omega \bigr) 
+  2\alphaR^{}\, \alphabarL^\omega +  \bigl( \alphaL^\omega \, \alphabarL^\omega 
+  \alphaR^{}\, \alphabarR^{}\bigr) 
\Bigr) \\
& \qquad \= \frac{k \, \pi}{\t_2} \sum_{i=1}^\text{rk(G)}
\Bigl(  \bigl(m^i - w^i\,\tau + 2\alphaR^i \bigr) \bigl(m^i-w^i \,\taubar - 2\alphabarL^{\omega\,i} \bigr) 
+  2 \alphaR^i \, \alphabarL^{\omega \, i} +  \bigl( \alphaL^{\omega\,i} \, \alphabarL^{\omega\,i} 
+  \alphaR^i \, \alphabarR^i \bigr) 
 \Bigr) ,
\end{split}
\ee
and with
\be\label{signfactor}(-1)^\zeta \= (-1)^{k \sum_{i,j} m^i w^j \langle h_i,h_j \rangle}.\ee

Though we will not prove this,
we  expect this formula to be equivalent  to a formula
for the same function computed by summing over states in a Hamiltonian 
approach.\footnote{See~\cite{Zhao:2025hen,Lu:2025} for this.}
In that approach, the partition function is expressed as a sum of products of holomorphic and
antiholomorphic characters, based on the diagonal modular invariant for these characters.
The characters are given by the Weyl-Kac character formula, and one has to include 
a non-chiral shift in the ground state quantum numbers.

If $G$ is compact but not simply-connected, then on the path integral side there are additional classical solutions,
corresponding to maps from $T^2$ to $G$ that are not homotopic to a constant map. The solutions are still given by
\eqref{eq:WeyltwistsuN}, but $m_i$ and $w_i$ no longer have to be integers. On the Hamiltonian side,
what happens is the following.   Let $\widehat G$ be the simply-connected cover of $G$.   
Then in a Hamiltonian approach, the partition function of the WZW model
with target space $G$ is constructed in terms of the $\widehat G$ affine characters, 
but now with a non-diagonal modular invariant.

\section{Models Obtained By Analytic Continuation}\label{sec:hthreeplus}

A number of interesting models can be described, at a classical level,  by analytic continuation starting 
with the $SU(2)$ WZW model. To see this, we can return to the picture of section \ref{sec:locPIcalc}, 
where $SU(2)\cong S^3$ was described by real functions $x_1,\cdots,x_4$ satisfying $\sum_{i=1}^4 x_i^2=1$.   
We can obtain new models by ``Wick-rotating'' some of the $x_i$ to imaginary values.  
The techniques discussed in this paper allow us to calculate the partition functions based 
purely on the corresponding symmetry structures, as we now proceed to discuss. 
It would be interesting to compare the path-integral results for partition functions that we obtain 
in this section with results that have been obtained by considering considering characters 
of the relevant chiral algebras, but we will not try to do so here.

\subsection{The $SL(2,\Bbb R)$ WZW Model}\label{sltwo}

Taking $x_3$ and $x_4$ to be imaginary,
say with $x_3=i y_3$, $x_4=i y_4$, we get  a model whose target space is parametrized by $x_1,x_2,y_3,y_4$ with
\be\label{eq:zeldo} x_1^2+x_2^2-y_3^2-y_4^2 \=1 \,. \ee
This analytic continuation of the $SU(2)$ WZW model  is a new model whose  target space,  described in eqn.~(\ref{eq:zeldo}), is 
in fact the $SL(2,{\Bbb R})$ group manifold.  To see this, 
we note that an element of $SL(2,\Bbb R)$ can be represented as a $2\times 2 $ real-valued matrix 
$\biggl( \,\begin{matrix} a&b\\ c&d\end{matrix} \, \biggr)$ with $ad-bc=1$.
The quadratic form $ad-bc$ has signature $++--$, so by a linear change of variables it can be put in the 
form $x_1^2+x_2^2-y_3^2-y_4^2$, leading to the
description of the $SL(2,\R)$ manifold in eqn.~\eqref{eq:zeldo}.
The model obtained this way is in fact simply the $SL(2,\R)$ WZW model.  In particular, Wick rotating an 
even number of the $x_i$ -- as we have done to go from $SU(2)$ to $SL(2,\R)$ -- preserves the fact that 
the Wess-Zumino term is real on a Lorentz signature worldsheet and imaginary on a Euclidean signature 
worldsheet, as expected in a WZW model based on any Lie group.

From the standpoint of localization, the $SL(2,\R)$ WZW model has the drawback that on a worldsheet 
of Euclidean signature, because of the indefiniteness of the target space metric,
the bosonic part of the action is not positive-definite and therefore
the Euclidean path integral is not well-defined.  In fact, the target space of the model has signature $-++$ if $k<0$ or $+--$ if $k>0$.
However, on a worldsheet of Lorentz signature, the model makes sense and (with $k<0$)  is considered to be an important
example of a sigma-model with Lorentz signature target space~\cite{Maldacena:2000hw}.
As an example of a Lorentz signature worldsheet, we can consider a two-torus with a flat metric of Lorentz signature.  
Such a torus depends on moduli that can be regarded as analytic continuations from the familiar moduli of a flat torus of
Euclidean signature;  see section \ref{formulas} for details.   The path integral of the
$SL(2,\R)$ WZW model is expected to make sense on such a torus, though as explained momentarily 
it has to be interpreted as a distribution on the parameter space of the torus
(not simply as a function of the torus moduli).    
The localization procedure that we have explained for $SU(2)$ can be applied to this $SL(2,{\Bbb R})$ model, 
with localizing term $\lambda\delta V$, with $V$ defined by the same formula as in eqn.~\eqref{eq:defVSU2} 
and $\lambda$ real.   Taking $\lambda\to \infty$, the Lorentz signature path integral becomes highly oscillatory 
and localizes as before  on the critical points of $\delta V$. 
For the analogous case of a quantum mechanical
model with $SL(2,{\Bbb R})$ target  (as opposed to the two-dimensional WZW model with the same target),  
this has already been done by Choi and Takhtajan~\cite{Choi:2023pjn}.\footnote{The path integral of the bosonic 
quantum mechanical model on~$SL(2,\mathbb{R})$ has been studied by 
another method in~\cite{Ashok:2022thd}.}

Since Lorentz signature partition functions in quantum field theory may be somewhat unfamiliar, 
we will make some general remarks.  In general, for any quantum mechanical system
or any quantum field theory defined on a compact spatial manifold, the real time partition 
function $\Tr\,e^{-i H t}$ should be understood as a distribution.   This means that
$\Tr\,e^{-i H t}$ is not defined for a  real value of $t$, but if $f(t)$ is a smooth function of compact support, 
then the integral $\int \d t \, f(t) \,\Tr\,e^{-i H t}$ makes sense,  provided that
one integrates over $t$ before taking the trace; in other words, the quantity that really makes sense is 
\be\label{intfirst} \Tr\,\int \d t \, f(t) \, e^{-i H t} \,.\ee
 For
example, for a  harmonic oscillator, $\Tr \,e^{-i Ht}$ has poles at certain real values of $t$ and is not defined 
as a complex number for all $t$.   For a single harmonic oscillator, there are only
discretely many bad values of $t$, but for a pair of harmonic oscillators
with incommensurate frequencies, the bad points are dense on the real axis.  For a quantum field theory, 
the structure on the real axis is  even more complicated.  So it is not satisfactory to view $\Tr\,e^{-i Ht}$ as
a ``function'' of real $t$.   However, if $H$ is bounded below, $\Tr\,e^{-i Ht}$ is holomorphic in the 
half-plane ${\rm Im}\,t\leq 0$, and (in quantum field theory on a compact spatial manifold
of any dimension)  it is bounded by a power of $1/|{\rm Im} \,t|$ as 
${\rm Im}\,t\to 0$. These properties imply that on the real $t$ axis, $\Tr\,e^{-i H t}$ 
makes sense  as a distribution.\footnote{For a proof, see proposition 4.2 in~\cite{Bostelmann_2009}.}   
This argument does not apply as stated to the $SL(2,\R)$ WZW model,
because the Hamiltonian of that model is not bounded below.  However, the localization formula that we 
will obtain shows (at least in the supersymmetric case) that the torus partition function of this model
does make sense as a distribution.   Hopefully, this can be proved directly.

Relative to $SU(2)$, the $SL(2,\Bbb R)$ WZW model on a torus has   fewer localizing solutions.
Suppose  that we twist the model by a compact
maximal torus of $SL(2,\Bbb R)\times SL(2,\Bbb R)$, which we can take to be a subgroup isomorphic 
to $U(1)\times U(1)$, acting by rotations of the pair $(x_1,x_2)$ and
also by rotations of the pair $(y_3,y_4).$     Equivalently, we twist the model by elliptic elements 
of $SL(2,\R)$ (an elliptic element by definition is an element contained in a compact maximal torus).
In our study of the $SU(2)$ WZW model, the localizing solutions were all supported at 
either $x_1=x_2=0$ or at $x_3=x_4=0$.   The same analysis applies for $SL(2,\R)$,
with the substitutions $x_3=i y_3$, $x_4=i y_4$.  So localizing solutions will be at $x_1=x_2=0$ or $y_3=y_4=0$.   
But in $SL(2,\R)$, the condition $x_1=x_2=0$ defines the empty set $-y_3^2-y_4^2=1$,  
So the localizing solutions of the $SL(2,\R)$ WZW model
are all at $y_3=y_4=0$; they are just the localizing solutions of the $U(1)$ model studied in section \ref{two},
with $U(1)$ embedded in $SL(2,\R)$ at $y_3=y_4=0$.   In particular, the localizing solutions
are characterized, as in \eqref{eq:Weyltwist}, by a pair of winding parameters $m,w$ as well as a zero-mode $h_0$ that
is valued in $U(1)$.
Hence,  the partition function of the $SL(2,\Bbb R)$
WZW model, twisted by  elliptic elements of $SL(2,\Bbb R)\times SL(2,\Bbb R)$, is similar to the partition function of the
$SU(2)$ model, but without the sum over the Weyl group.

This twisted  partition function  can be generalized to depend on an additional parameter. For this purpose, 
we replace $SL(2,\Bbb R)$ with a quotient
that is locally isomorphic to $SL(2,\Bbb R)$.    Observe first that if we set $x_1=r\cos \theta$, $x_2=r\sin\theta$,
then \eqref{eq:zeldo} becomes $r=\sqrt{1+y_3^2+y_4^2}$, so the $SL(2,\Bbb R)$ manifold is freely parametrized 
by $y_3,y_4$ and~$\theta$ and is topologically $\Bbb R^2\times S^1$,
with the first factor parametrized by $y_3,y_4$ and the 
second factor parametrized by $\theta$.   Here of course $\theta$ has period $2\pi$.  Alternatively, we can 
take the universal cover, taking $\theta$ to be real-valued.   This gives a group $\widetilde{SL}(2,\Bbb R)$ that is 
the universal cover of $SL(2,\Bbb R)$.  The partition function of the $\widetilde{SL}(2,\Bbb R)$ WZW model 
on a Lorentz signature torus, twisted by elliptic elements,
can be analyzed as we have done for $SL(2,\R)$.   Because $\theta$ is now real-valued instead of circle-valued, 
there are no winding parameters analogous to $m$ and $w$ and the localizing solution
depends only on the bosonic zero-mode $h_0$ that appears in eqn.~\eqref{eq:Weyltwist}.   But as $\theta$ is 
now real-valued, the constant $h_0$  is real-valued and the integral over $h_0$ 
diverges, implying that the torus
partition function of the $\widetilde{SL}(2,\Bbb R)$ model  is divergent, even after  twisting.   
However, there is a more general family of spaces $SL_\gamma(2,\Bbb R)$ that are locally isomorphic
to $SL(2,\Bbb R)$:  we can take $\theta$ 
to have an arbitrary period~$\gamma>0$.   
Then winding parameters do appear, because $\theta$ is circle-valued, and for the same reason the zero-mode 
integral gives a finite result.  So the twisted $SL_\gamma(2,\Bbb R)$ 
partition function is finite  and is a sum over a pair of winding parameters $m,w$.

If $\gamma=2\pi $, then $SL_\gamma(2,\R)$ is simply $SL(2,\R)$.   If $\gamma=2\pi n$ for $n$ a positive integer, then
$SL_\gamma(2,\R)$ is an $n$-fold cover of $SL(2,\R)$ and in particular is a Lie group.   For any Lie group,
the fermions of the supersymmetric WZW model are decoupled from the bosons, so the supersymmetric WZW partition
function is the product of a bosonic WZW partition function and the fermion partition function.  Hence, for $\gamma=2\pi n$,
the bosonic partition function of the $SL_\gamma(2,\R)$ model can be found by computing the partition function of the
supersymmetric version of the model and then dividing by the fermion partition function, just as we have done for $SU(2)$.
For other values of $\gamma$, it is still possible to use 
supersymmetric localization to determine the partition function of the purely bosonic model, but the procedure is 
more subtle.   In each twisted sector labeled by a pair $m,w$, the fermions are decoupled from
the bosons, but they are twisted in a way that depends on $m,w$, so the fermion partition function depends on $m,w$.  
Hence, instead of obtaining the bosonic partition function from the supersymmetric one by dividing by an overall factor 
coming from the fermion path integral, one must for each fixed $m,w$ divide out the fermion path integral
from the path integral of the supersymmetric model, and then sum over $m,w$.

Like $SL(2,\R)$, $SL_\gamma(2,\R)$ can be parametrized by $\theta,y_3,y_4$, the only difference being 
that now $\theta$ has period $\gamma$ rather than period $2\pi$.
Hence $SL_\gamma(2,\R)$ has first Betti number equal to 1, generated by the closed 
one-form $\Theta=\frac{1}{\gamma}d\theta$, whose periods are integers.   Accordingly,
the coupling of abelian gauge fields $a^L$, $a^R$ to a sigma-model with target $SL_\gamma(2,\R)$ is not unique.   
It can be modified by including in the Euclidean action a term
\be\label{modac}\Delta S = - i \int \Theta\wedge (u_L \, a^L+u_R \, a^R), \ee
with integers $u_L$, $u_R$.   
$SL_\gamma(2,\R)=\tilde{SL}(2,\R)/\Z$ can also be generalized to depend on one more parameter:   
one can take $\Z$ to act as a rotation of $y_3,y_4$ as well as a shift of $\theta$.

Instead of twisting  the torus path integral of the $SL(2,\Bbb R)$ (or $SL_\gamma(2,\Bbb R)$) WZW model 
by elliptic elements of $SL(2,\R)$, we can twist by hyperbolic elements of $SL(2,\R)$, which are elements of 
a non-compact maximal torus of $SL(2,\Bbb R)$, such as the group of diagonal matrices $\mathrm{diag}(e^s,e^{-s})$, 
$s\in \Bbb R$.
This group is isomorphic to $SO(1,1)\cong \Bbb R$, so the localization fixed points will be those of a sigma-model 
with target $\Bbb R$, not $U(1)$, and hence the twisted partition function will diverge.
Nonetheless, it is interesting to identify the localization fixed points.
We can choose coordinates so that one copy of $ SO(1,1)$  acts by Lorentz boosts of the
pair $(x_1,y_3)$ and the second by Lorentz boosts of $(x_2,y_4)$.   The localizing solutions will now come from 
localizing solutions of a sigma model with target $\Bbb R$,
rather than the $U(1)$ model studied in section \ref{two}, embedded via $x_1=y_3=0$ or  $x_2=y_4=0$.  
Since $\Bbb R$ is simply-connected, there are no integer-valued
winding parameters.    If $x_1=y_3=0$, the only additional discrete choice in the localizing solution is the sign of $x_2$,
and similarly for $x_2=y_4=0$, the only additional discrete choice is the sign of $x_1$.   
The zero-mode parameter $h_0$ of eqn.~\eqref{eq:Weyltwist} is real-valued, leading
to the divergence of the partition function.  

It is also possible to twist the $SL(2,\R)$ (or $SL_\gamma(2,\R)$) WZW model by  elliptic elements of the left action 
of $SL(2,\R)$ on itself and hyperbolic elements of the right action.
One can pick coordinates so that the elliptic elements act by simultaneous rotations of $x_1,x_2$ and $y_3,y_4$, 
while the hyperbolic elements act by simultaneous boosts of $x_1,y_3$ and $x_2,y_4$.    
We leave the reader to explore this case.

\subsection{The $H_3^+$ Model and the BTZ Black Hole}\label{hthree}

Alternatively, we can Wick rotate an odd number of the variables $x_1,\cdots , x_4$.   
This will give a model in which the Wess-Zumino term in the WZW action is imaginary if the worldsheet
has Lorentz signature or real if the worldsheet has Euclidean signature.   
In either case, this corresponds to a non-unitary quantum field theory.   Nevertheless, one of the models
that can be made this way is important and much-studied.

Wick rotating one of the $x_i$, say $x_4$,  gives a model with the target space
\be\label{minzo} x_1^2+x_2^2+x_3^2-y_4^2\=1 \,. \ee
This target space has Lorentz signature if the underlying $SU(2)$ model has $k>0$.  
This model has not been much studied.

Much more intensively studied, and with many applications, is the model obtained by Wick rotating three 
of the coordinates, say $x_2,x_3,x_4$ .   The target space is then
\be\label{linzo} x_1^2-y_2^2-y_3^2-y_4^2\=1 \,. \ee
This space has two components, differing by the sign of $x_1$.   
To get a model with connected target space, we restrict to $x_1>0$.   The resulting target space has a visible $SO(1,3)$
symmetry, and the subgroup of $SO(1,3)$ that leaves fixed a point, say the point with coordinates $(1,0,0,0)$, 
is a copy of $SO(3)$.   So the target space is $SO(1,3)/SO(3)$,
which is actually hyperbolic three-space, often denoted as $H_3^+$.  Replacing $SO(1,3)$ by its double 
cover $SL(2,\Bbb C)$ and $SO(3)$ by its double cover $SU(2)$,
the same target space can also be characterized as $SL(2,\Bbb C)/SU(2)$.  
We should stress that although this model is a sigma-model in which the target space is a homogeneous
space $SO(1,3)/SO(3)$ or $SL(2,\Bbb C)/SU(2)$, it is not an $SO(1,3)/SO(3) $ or $SL(2,\C)/SU(2)$ 
coset model in the sense of coset conformal field theory.\footnote{To make a $G/H$ coset conformal
field theory, one starts with a WZW model with target space $G$ and global symmetry $G_L\times G_R$, 
and then one gauges a diagonally embedded (and therefore anomaly-free) subgroup $H\subset G_L\times G_R$.
If we carry out this procedure with $G=SO(1,3)$, $H=SO(3)$, we get a model that does not 
have $SO(1,3)$ symmetry, as the $SO(1,3)_L\times SO(1,3)_R$ symmetry is
completely broken by gauging the diagonally embedded $SO(3)$ subgroup.   The model obtained by 
analytically continuing from the $SU(2)$ WZW model to a sigma-model
with target $H_3^+$ does, instead, have $SO(1,3)$ symmetry.}

For $k<0$, the target space metric of  the $H_3^+$   model has $+++$ signature.  The model is not a unitary 
two-dimensional field theory, because the analytic continuation from $SU(2)$ to $H_3^+$ multiplies
the Wess-Zumino term by a factor of $i$.  Nevertheless, the model is known to have many interesting applications.  
For one thing, $H_3^+$ can be viewed as the Euclidean version of ${\rm AdS}_3=SL(2,\R)$, and the $H_3^+$ model
has been studied as a sort of Wick-rotated version of the $SL(2,\R)$ WZW model \cite{Maldacena:2000kv}.   We will explore this relationship in section \ref{formulas}.
 
The action of the $H_3^+$ model on a worldsheet of Euclidean signature is real and non-negative, suggesting the possibility of a convergent 
path integral.  Actually, the noncompactness of the $H_3^+$ target space causes the partition function to diverge, 
even after twisting, as we will see shortly.   However, replacing $H_3^+$ with a quotient $H_3^+/\Z$ and twisting 
by the remaining symmetries does lead to a convergent partition function, as essentially 
computed in \cite{Gawedzki:1988nj,Gawedzki:1991yu,Maldacena:2000kv} and as we will analyze in 
section~\ref{formulas} using localization.

It is worth mentioning a few ways that the $H_3^+$ model differs from a  WZW model,  even though it can be 
viewed as an analytic continuation of the $SU(2)$ WZW model.   In general, the WZW model with target space a Lie group 
$G$ has $G_L\times G_R$ symmetry, where $G_L$ and $G_R$ are copies of $G$ acting  on $G$ respectively on the left
and right.   Moreover, with  suitable  conventions, in the supersymmetric version of the model,
the positive chirality fermions transform in the adjoint representation of $G_R$ and trivially under $G_L$, while the negative
chirality fermions transform trivially under $G_R$ and in the adjoint representation of $G_L$.     
By contrast, the $H_3^+$ model has only a single $SO(1,3)$ (or $SL(2,\Bbb C)$) symmetry.
There is no way to split the symmetry group of the $H_3^+$ model as a product of subgroups that act, 
in any sense, on $H_3^+$ on the left or on the right, or that act only on negative chirality fermions or positive chirality ones.

In somewhat more detail, this happens as follows.   The group $SO(1,3)$ is isomorphic to the Lorentz group 
in four dimensions.   As is well-known, this group is generated by rotation
generators $J_i$ and boost generators $K_i$, with the Lie algebra relations
\be\label{relns}
[J_i,J_j]\=i\epsilon_{ijk} J_k\,, \qquad [J_i,K_j]\=\i \epsilon_{ijk} K_k \,, \qquad [K_i, K_j]\=-i \epsilon_{ijk} J_k \,. 
\ee
By contrast, the symmetry group of the $SU(2)$ WZW model is $SO(4)=(SU(2)_L\times SU(2)_R)/\Bbb Z_2$. 
It is generated by operators $J_i$, $K'_i$ with the relations
\be\label{zelns}
[J_i,J_j]\=i\epsilon_{ijk}J_k\,, \qquad [J_i,K'_j]\=i\epsilon_{ijk}K'_k\,, \qquad [K'_i,K'_j]\=+i\epsilon_{ijk}J_k \,. 
\ee
From the $SO(4)$  relations, it follows that $J_i'{}^- =\frac{1}{2}(J_i-K_i)$ generates an $SU(2)$ subgroup, 
namely $SU(2)_L$, and $J_i'{}^+=\frac{1}{2}(J_i+K'_i)$ generates a second $SU(2)$
group that commutes with the first, namely $SU(2)_R$.   In particular,  $J_i'{}^-$ annihilates fermions of, say, 
positive chirality and $J_i'{}^+$ annihilates fermions of negative chirality.
The analytic continuation from the symmetry $SO(4)$ of the $SU(2)$ WZW model to the symmetry $SO(1,3)$ 
of the $H_3^+$ model can be made, at the level of group theory,
by setting $K'_i=i K_i$, mapping the commutation relations \eqref{zelns} to those of eqn.~\eqref{relns}.   
This maps $J_i'{}^+$ to $J_i^+=\frac{1}{2}(J_i+i K_i)$ and 
$J_i'{}^-$ to $J_i^-=\frac{1}{2}(J_i-i K_i)$.   Positive chirality fermions of the $H_3^+$ model are annihilated 
by $J_i^-$ and negative chirality fermions are annihilated  by $J_i^+$.
But because the $J_i^-$ and $J_i^+$ are not {\it real} linear combinations of the real generators $J_i,K_j$ 
of the $SL(2,\C)$ Lie algebra, it is not true that they generate proper subgroups
of $SL(2,\C)$.   Accordingly, it is not true that positive or negative chirality fermions of the $H_3^+$ model 
provide representations of proper subgroups of $SL(2,\C)$.   What is true instead
is the following.   View $SL(2,\C)$ as a complex manifold with a complex structure in which the matrix 
elements $a,b,c,d$ of an $SL(2,\C)$ element  
$\biggl(\,\begin{matrix}a & b\cr c&d\end{matrix} \, \biggr)$ 
are holomorphic functions on the $SL(2,\C)$ manifold.  This also determines a complex structure 
on $SO(1,3)=SL(2,\C)/\Z_2$.   A representation of  $SO(1,3)$ or $SL(2,\C)$  is called holomorphic or
antiholomorphic if matrix elements of group elements in that representation are holomorphic or antiholomorphic 
functions on the group manifold.   An irreducible finite-dimensional representation of $SL(2,\C)$ is
labeled by a pair $(p,q)$ of non-negative half-integers (such a representation can be viewed as a representation 
of $SO(1,3)$ if and only if $p+q$ is an integer).  The representations $(0,q)$ are holomorphic, the 
representations $(p,0)$ are antiholomorphic, and other representations are neither
holomorphic nor antiholomorphic.  A holomorphic representation  is annihilated by $J_i^-$ and an anti-holomorphic 
representation  is annihilated by $J_i^+$.
The positive  chirality fermions of the $H_3^+$ model transform in the holomorphic $(0,1)$ representation, 
while the negative chirality fermions transform in the antiholomorphic $(1,0)$ representation.

Analytic continuation from $SO(4)$ to $SO(1,3)$  does not spoil supersymmetry, so it maps the 
supersymmetric $SU(2)$ WZW model to a supersymmetric $H_3^+$ model.
Moreover, this analytic continuation does not affect any of the facts that make possible supersymmetric localization, 
so we can apply supersymmetric localization to the supersymmetric $H_3^+$ model just as we have done 
for the $SU(2)$ WZW model. Furthermore, analytic continuation does not affect the fact that the fermions are 
decoupled from the bosons.   So after using localization to compute the twisted partition function of the
supersymmetric $H_3^+$ model, we can divide by an appropriate fermion partition function and get the 
partition function of the twisted  bosonic $H_3^+$ model, exactly as for the $SU(2)$ WZW model.  

In contrast to  $SL(2,\R)$, which has both compact and noncompact maximal tori,  the symmetry group $SL(2,\C)$ 
of the $H_3^+$ model has only one type of maximal torus.
A convenient choice is the group of diagonal elements ${\rm diag}(e^\phi,e^{-\phi})$, where $\phi$ is complex-valued.  
Writing $e^\phi= e^t e^{i\psi}$ where $t\in \R$ and $\psi$ is an angular variable, we see that this maximal torus is equivalent 
to a product $\R\times U(1)$ or equivalently $SO(1,1)\times SO(2)$.  Coordinates can be chosen so that in the 
description (\ref{minzo}) of the $H_3^+$ manifold, the $SO(1,1)$ factor in the maximal torus acts by a Lorentz boost 
of the pair $x_1,y_2$, and the $SO(2)$ factor acts by rotations of $y_3,y_4$.
The localization equations are exactly as they were for the WZW model of $SU(2)$ except for the substitution 
$x_2=i y_2$, $x_3=i y_3$, $x_4=i y_4$.  So we can find the localizing solutions by borrowing the previous results.   
For $SU(2)$, we found that a localizing solution is supported at $x_1=x_2=0$ or at $x_3=x_4=0$.   In $H_3^+$, 
these conditions become $x_1=y_2=0$ and $y_3=y_4=0$.    However, setting $x_1=y_2=0$ leads to the empty 
set $-y_3^2-y_4^2=1$.   Hence a localizing solution will lie at $y_3=y_4=0$.
The locus $y_3=y_4=0$ in the manifold defined by $x_1^2-y_2^2-y_3^2-y_4^2=1$ is the hyperbola $x_1^2-y_2^2=1$, 
which has two branches.   But since $H_3^+$ was defined with a condition $x_1>0$, we only want the $x_1>0$ 
branch of the hyperbola. 
This branch is a copy of $\R\cong SO(1,1)$. We are actually in the same
situation that we encountered in section \ref{sltwo} in studying the WZW model of the group $\widetilde{SL}(2,\Z)$.  
The localizing solutions are just those of a sigma-model with target~$\R$,
embedded in $H_3^+$ at $y_3=y_4=0$, $x_1>0$.   Because $\R$ is simply-connected, the localizing solutions 
do not depend on momentum and winding parameters analogous to $m$ and $w$.
Rather, there is just one family of localizing solutions, labeled by a bosonic zero-mode, the constant $h_0$ 
in \eqref{eq:Weyltwist}.  Because $h_0$ is now valued in the non-compact group $SO(1,1)\cong \R$, the 
integral over $h_0$ diverges and therefore the twisted partition function of the $H_3^+$ model also diverges.

In section \ref{sltwo}, we observed that in the case of the WZW model with target $\widetilde{SL}(2,\R)$,
there is a simple cure for a similar divergence, namely to replace the group $\widetilde{SL}(2,\R)$ with its 
quotient $SL_\gamma(2,\R)$, depending on a free parameter $\gamma$.  
We can do much the same thing here.   We replace $H_3^+$ with a quotient $H_3^+/\Z$, where $\Z$ is a 
discrete subgroup of $SO(1,1)$, generated by an element that acts on $x_1,y_2$ by
\be\label{moraru} (x_1\pm y_2)\to e^{\pm \beta}(x_1\pm y_2) \ee
(with trivial action on $y_3,y_4$).  The quotient $H_3^+/\Z$ is none other than the BTZ black hole, 
with inverse temperature $\beta$ or $1/\beta$ (depending on how coordinates are chosen) and with
zero angular momentum (angular momentum can be included by accompanying the boost of \eqref{moraru} 
with a rotation of $y_3,y_4$).    
The localizing solutions are still valued in the set $y_3=y_4=0$, $x_1>0$, but now
this set is a circle $S^1_\beta$ of circumference $\beta$ rather than a copy of $\R$.   
So the localizing solutions are characterized by  winding parameters $m,w$ as well as a bosonic zero-mode
that now parametrizes the compact set $S^1_\beta$.   
That compactness implies that the twisted path integral of the $H_3^+/\Z$ model is finite.   
Moreover, if we set $\gamma=\beta$, then  there is an obvious 1-1 correspondence between
the localizing solutions of the $SL_\gamma(2,\R)$ model and those of the $H_3^+/\Z$ model.  
The upshot is that the partition function of the $SL_\gamma(2,\R)$ model
can be viewed as an analytic continuation of the partition function of the $H_3^+$ model.   
Just as for $SL_\gamma(2,\R)$, to get the partition function of the purely bosonic $H_3^+/\Z$ model
from the partition function of the supersymmetric model, one must divide out 
the fermion partition function before summing over $m,w$.

\subsection{Formulas}\label{formulas}

Now we present the formulas  that follow from localization for the partition functions of the bosonic and supersymmetric models
that were just described.

The following matrices, given in terms of Pauli matrices, are a basis over~$\IR$ 
for the Lie algebra $sl(2,\IR)$ and a basis over~$\IC$ for $sl(2,\IC)$,  
\be
\sltwogen{1} \= \s^{1} \= \biggl( \, \begin{matrix} 0 & 1 \\ 1 & 0 \end{matrix} \, \biggr) \,, \qquad 
\sltwogen{2} \= i\s^{2} \= \biggl( \,\begin{matrix} 0 & 1 \\ -1 & 0 \end{matrix}  \, \biggr)  \,, \qquad  
\sltwogen{3} \= \s^{3} \= \biggl( \, \begin{matrix} 1 & 0 \\ 0 & -1 \end{matrix} \, \biggr) \,. 
\ee
A group element is written as~$\exp(\theta_i \, \sltwogen{i})$,  $i=1,2,3$, with~$\theta_i \in \IR$ for~$SL(2,\IR)$
and~$\theta_i \in \IC$ for~$SL(2,\IC)$.

\bigskip

\ndt {$\bullet$ $SL(2)$}

\smallskip

We will compute path integrals on a two-torus~$T^2$ with a flat metric of Lorentz signature.   In this
paper, we have parametrized a flat torus of Euclidean signature by coordinates $\wsx,\wsy$ with metric 
$ds^2 =(\d \wsx)^2+(\d \wsy)^2$, a complex structure
defined by $z=\wsx+ i \wsy$, $\zbar = \wsx-i \wsy$, and equivalence relations $(z,\bar z) \coeq (z,\bar z)+2\pi(1,1) \coeq
(z,\bar z) +2\pi(\tau,\taubar)$.   
A Wick rotation $\wst= -i \wsy$ leads to the flat Lorentz signature metric~$ds^2 = -(\d v ^0)^2+(\d \wsx)^2$. 
The light-cone coordinates~$(v^-,v^+)=(\wsx - \wst, \wsx+\wst)$  are continuations of $(z, \bar z)$. 
In those coordinates,  $ds^2=\d v^+ \d v^-$ and~$\p_{v^\pm} = \frac12 (\p_{\wsx} \pm \p_{\wst})$.
We define the torus by identifications 
$(\wsx,\wst) \coeq (\wsx,\wst) + 2 \pi (1,0) \coeq (\wsx,\wst) + 2 \pi (\rho_1,\rho_0)$ or, equivalently, 
$(v^+, v^-) \coeq (v^+ + 2\pi,v^- + 2\pi) \coeq (v^+ + 2 \pi \lmp^+ , v^- +2 \pi \lmp^-)$, 
where $\lmp^\pm = \lmpx \pm \lmpt$. 
The volume form is $\text{vol}  = \d v^+ \wedge \d v^- = 2 \d \wst \wedge \d \wsx$  
so that $\frac{1}{4\pi}\int_{T^2} \text{vol} = 2 \pi \rho_0$.   
In continuation to Lorentz signatures, the complex conjugate modular parameters
$(\tau=\tau_1+i\tau_ 2, \, \bar\tau=\tau_1-i\tau_2)$  of the Euclidean torus have 
become the independent real parameters $(\rho^-,\rho^+)$.

\medskip

As explained in section~\ref{sltwo}, the localizing solutions of the $SL(2,\R)$ model are parameterized by the winding of the two cycles
of the worldsheet torus around the circle~$x_1^2+x_2^2=1$, and a constant element~$h_0 \in H$.
The~$SU(2)$ model has twist parameters~$\alphaL$ and~$\alphaR$ which couple to the Cartan generator of 
the left and right action of the gauge group.  
Recall from the discussion of the symmetries of the~$SU(2)$ model 
around~\eqref{eq:coords} that~$\alphaL^{}-\alphaR$ is the rotation angle of the circle~$x_1^2+x_2^2=1$
and~$\alphaL^{}+\alphaR$ is the rotation angle of the circle~$x_3^2+x_4^2=1$. 
In the~$SL(2)$ model the twist parameters correspondingly couple to the two rotations with~$(x_3,x_4)$
replaced by~$(y_3,y_4)$ as explained in section~\ref{sltwo}. 

A natural ansatz for the background gauge fields is a Lorentz signature analog of the ansatz \eqref{eq:extAAbar} that
we assumed in analyzing the $SU(2)$ WZW model:
\be \label{eq:sl2ALAR}
\AL \=  -\frac{\sltwogen{2}}{2 \lmpt} \, \bigl( - \alphaL^+ \, dv^- + \alphaL^- \, d v^+ \bigr) \,  \,, \qquad 
\AR \=  -\frac{\sltwogen{2}}{2 \lmpt} \, \bigl( - \alphaR^+ \, dv^- + \alphaR^- \, d v^+ \bigr).
\ee
Similarly to what happened for the modular parameters $\tau$ and $\bar\tau$, the complex parameter
$\alpha_L$ of the $SU(2)$ model and its complex conjugate $\bar\alpha_L$ are replaced by independent real
parameters $\alpha_L^-$ and $\alpha_L^+$; similarly $\alpha_R, \,\bar\alpha_R$ are replaced by 
independent real parameters~$\alpha_R^-, \alpha_R^+$

Adapting the results from $SU(2)$ to a Lorentz signature worldsheet, the localizing solutions for the $SL(2,\R)$ WZW model are
\be \label{eq:gLormw}
g(v^+,v^-) \= 
  h_0 \exp \Bigl(-\frac{\sltwogen{2}}{2 \lmpt} \bigl( m (v^- - v^+) +w( \lmp^- v^+ - \lmp^+ v^-) \bigr) \Bigr) \,, 
  ~~~ m, w \in \IZ \, ,~~~h_0\in H \,.
\ee 
The corresponding action is
\be \label{eq:sl2act}
S_{m,w}(\lmp^+,\lmp^-,\a^\pm_L,\a^\pm_R)  \= \frac{\pi\,k}{\lmpt} 
\Bigl( \bigl(m - w\,\lmp^- + 2\alphaR^- \bigr) \bigl(m-w \,\lmp^+ - 2{\a}^+_L \bigr) 
+  2 \alphaR^- \, {\a}_L^+ +  \bigl( \alphaL^- \, {\a}^+_L +  \alphaR^- \, {\a}_R^+ \bigr) 
\Bigr)\,.
\ee
To obtain this result for the action, we have simply taken the gauged $SU(2)$ WZW action as presented in section \ref{extgf},
Wick-rotated the worldsheet to  Lorentz signature, with the ansatz \eqref{eq:sl2ALAR} for the background gauge fields,
 and  analytically continued by $x_3=i y_3$, $x_4=i y_4$.   (The analytic continuation
has no effect in computing the action, since the relevant localization fixed points are anyway at $x_3=x_4=0$.)  
The $SU(2)$ WZW model only makes sense for positive integer $k$.   As explained in section \ref{sltwo}, 
the $SL(2,\R)$ WZW model is usually studied for $k<0$.    The  following analysis of localization,
however, makes sense for either sign of $k$.   We do not really know if the theory makes sense for $k>0$, 
but the one-loop partition function computed via localization does make sense.

Following the same steps as for $SU(2)$, the partition function of the gauged supersymmetric $SL(2,\R)$ 
WZW model turns out to be\footnote{The circle parametrized by $h_0$ in eqn.~(\ref{eq:gLormw}) has positive volume, 
proportional to $|k|$, not $k$.}
\be  \label{eq:Ssl2PF}
Z^{SL(2)}_\text{SWZW}(\lmp^+,\lmp^-,\a^\pm_L, \a^\pm_R) \= 
2\sqrt{\frac{|k|}{\lmpt}}  
\sum_{m,w \in \IZ}  \exp \bigl( \, i \, S_{m,w}(\lmp^+,\lmp^-,\a^\pm_L,\a^\pm_R) \,  \bigr).
\ee
The sum in~\eqref{eq:Ssl2PF}, as well as the other sums that 
appear presently in the discussion of  Lorentzian theories, have to be interpreted carefully.
They are not convergent sums and the partition function 
$Z^{SL(2)}_\text{SWZW}(\lmp^+,\lmp^-,\a^\pm_L, \a^\pm_R)$ is not a well-defined
complex number for real values of the arguments.    Rather, as anticipated in section \ref{sltwo}, it makes
sense as a distribution.   This will be demonstrated explicitly at the end of the present analysis.

To get the partition function of a purely bosonic gauged $SL(2,\R)$ WZW model, we have to divide by the
fermion partition function.
By analytically continuing the result in section~\ref{sec:HamTracesu2}, or by repeating all the steps, we find that 
the fermionic partition function, including the insertion~$(-1)^F$, the integral over   
the fermionic zero-modes, and a factor coming from shifts in ground state quantum numbers, is given by  
\be \label{eq:sl2ferdet}
Z^{SL(2)}_\text{fer}(\rho^+,\rho^-; \alpha^\pm_L, \alpha^\pm_R) \= 2 \, C_\text{2}(\alpha^\pm_L,\alpha^\pm_R)
\,\vth_1(\lmp^-,2\alphaR^-) \, \vth_1(-\lmp^+,-2\alphaL^+) \,.
\ee
with 
\be \label{eq:Cksl2}
C_k(\alpha^\pm_L,\alpha^\pm_R) 
\=  \exp\Bigl(-i\frac{\pi k}{\lmpt}  \,  \bigl( \alphaR^- (\alphaR^-  - \alphaR^+)  
+ \alphaL^+ (\alphaL^+ - \alphaL^-)  \bigr) \Bigr) \,. 
\ee  
Upon dividing by the fermionic partition function, we obtain the bosonic partition function, 
\be  \label{eq:sl2bos}
Z^{SL(2)}_\text{WZW}(\lmp^+,\lmp^-,\a^\pm_L, \a^\pm_R) 
\= \frac{Z^{SL(2)}_\text{SWZW}(\lmp^+,\lmp^-,\a^\pm_L, \a^\pm_R) }{Z^{SL(2)}_\text{fer}(\lmp^+,\lmp^-,\a^\pm_L, \a^\pm_R) } \,.
\ee
This diverges if $\alpha_R^-=0$ or $\alpha_L^+=0$.   
That happens because if $\alpha_R^-$ or $\alpha_L^+$ vanishes, the localization conditions
break down in the case $m=w=0$.   Indeed, if $g$ is a constant, then the localization equations 
reduce to $[\alpha_L^+,g \, \alpha_R^- \, g^{-1}]=0$.   If $\alpha_R^-$ or $\alpha_L^+$ vanishes,
this equation is satisfied for all $g$.   Integration over this constant $g\in SL(2,\R)$ causes the 
partition function to diverge.   One may wonder if this has an analog in the $SU(2)$ WZW model.
In that model, if  one of the complex parameters $\alpha_R$ and  $\alpha_L$ vanishes,  
the localization equations are satisfied for an arbitrary constant $g\in SU(2)$,
but this does not lead to a divergence because the $SU(2)$ group manifold is compact.  
In terms of the derivation by localization, if $\alpha_R^-$ or $\alpha_L^+$ vanishes,
the supersymmetric $SL(2,\R)$ partition function remains nonzero, but the fermionic partition function 
vanishes, so their ratio, which is the partition function of the bosonic $SL(2,\R)$ model,
diverges.   For $SU(2)$, if $\alpha_L$ or $\alpha_R$ vanishes, then the fermion determinant vanishes, 
but the partition function of the supersymmetric WZW model also vanishes,
because of a cancellation in the sum over the Weyl group.   
The ratio, which is the partition function of the bosonic $SU(2)$ model, remains finite.

\bigskip

Similarly to what we found for $SU(2)$, the  supersymmetric partition function can also be written as
\be 
Z_\text{SWZW}(\rho^+,\rho^- ; \alpha^\pm_L, \alpha^\pm_R)  
 \=  C_k(\alpha^\pm_L,\alpha^\pm_R) \! \! 
\sum_{\ell \, (\text{mod} \; 2k)}  \vth_{k,\ell} \bigl(\lmp^-,\alphaR^- \bigr)  \, \vth_{k,\ell}  \bigl(-\lmp^+,- \alphaL^+ \bigr) \,.
\ee
This function is a sum of finitely many terms, each of which is the product of a theta function 
with argument $\rho^-$ and a theta function
with argument $-\rho^+$.  It makes sense as a distribution but not as a function because
that is the nature of a theta function with real arguments.
To explain this point, we  consider for illustration the basic theta function
\be\label{thetab}\vartheta(\tau,z)\=\sum_{n\in\Z} e^{i\pi \tau n^2} e^{2\pi i n z}. \ee
For ${\Im}\,\tau>0$ if $k>0$, or for ${\Im}\,\tau<0$ if $k<0$, the sum is absolutely convergent. 
On the real $\tau $ axis, the sum does not converge, 
even if $z$ is also real, as we will assume.
However, for real $z$, the sum  is bounded by $1/\sqrt{\Im\,\tau}$ as $\Im\,\tau\to 0$, so by proposition 4.2 in
\cite{Bostelmann_2009},  the restriction of $\vartheta(\tau,z)$ to the real $\tau$ axis makes sense as a distribution 
in the variable~$\tau$, for any fixed real value of $z$.
Rather than appealing to the theorem, we can  show this explicitly.   Taking $\tau=\tau_1$ to be real, let $f(\tau_1)$ 
be a smooth function of compact support.   To show that $\vartheta(\tau_1,z)$ makes
sense as a distribution, we must show the convergence of\footnote{Here we integrate over $\tau_1$ before 
summing over $n$, just as in
eqn.~(\ref{intfirst}), one integrates over $t$ before taking the trace.} 
\be\label{uttu} \sum_{n\in\Z} \, \int_{-\infty}^\infty \d\tau_1 \,f(\tau_1) \, e^{i\pi \tau_1 n^2} e^{2\pi i n z}. \ee
Since $f(\tau_1)$ is a smooth function of compact support, its Fourier transform vanishes faster than any 
power of the momentum.   
In eqn.~(\ref{uttu}), the integral is the Fourier transform of $f(\tau_1)$ at momentum $\pi n^2$, so it vanishes 
for large $n$ faster than any power of $n$ and therefore the sum over $n$ converges, for real $z$.

\bigskip

\ndt {$\bullet$ $\widetilde{SL}(2,\R)$}

\smallskip

In this case the localization solutions do not contain any winding modes, and 
are labelled by only the bosonic zero-mode~$h_0$. 
The supersymmetric partition function is given by the restriction of the result for~$SL(2)$
to the~$m=w=0$ sector, and allowing for the divergent volume~$V$ of the 
zero-mode~$h_0$.
The result is
\be  \label{eq:wtsl2}
\begin{split}
Z^{\wt{SL}(2)}_\text{SWZW}(\lmp^+,\lmp^-,\a^\pm_L, \a^\pm_R) & \= 
\frac{V}{(2 \pi^2 \lmpt)^{1/2}}  
 \exp \bigl( \, i \, S(\lmp^+,\lmp^-,\a^\pm_L,\a^\pm_R) \, \bigr) \,, \end{split}\ee
with \be\begin{split} \qquad S(\lmp^+,\lmp^-,\a^\pm_L,\a^\pm_R) & \= \frac{\pi\,k}{\lmpt} 
\Bigl(  -2 \, \alphaR^- \, {\a}_L^+ +  \bigl( \alphaL^- \, {\a}_L^+ +  \alphaR^- \, {\a}_R^+ \bigr) 
\Bigr)\,.
\end{split}
\ee
The fermionic partition function is the same as for the~$SL(2)$ model and is given by~\eqref{eq:sl2ferdet}. 
The bosonic partition is given by the ratio,
\be 
Z^{\wt{SL}(2)}_\text{WZW}(\rho^+,\rho^-,\a^\pm_L, \a^\pm_R) \= 
\frac{Z^{\wt{SL}(2)}_\text{SWZW}(\lmp^+,\lmp^-,\a^\pm_L, \a^\pm_R) }{Z^{SL(2)}_\text{fer}(\lmp^+,\lmp^-,\a^\pm_L, \a^\pm_R) } \,.
\ee

\bigskip

\ndt {$\bullet$  $SL_\gamma(2)$}

\smallskip

This model is defined as the orbifold $SL_\gamma(2)=\wt{SL}(2)/\IZ$, where 
the generator of $\IZ$ acts by translation on the universal cover of
the circle~$x_1^2+x_2^2=1$. 
The localization analysis is similar to what it was for $SL(2,\R)$, but now the worldsheet is winding around
a circle of circumference $\gamma$ rather than a circle of circumference $2\pi$.
This effectively replaces~$(m-w\tau)$ by $\frac{\gamma}{2\pi}(m-w\tau)$ in all the expressions for~$SL(2,\R)$.  So we define
\be \label{eq:alphamw}
\alpha^{\gamma, -}_{R,m,w} \= \,\alphaR^- + \frac{\gamma}{4\pi}(m-w \, \lmp^-) \,, \qquad 
\alpha^{\gamma, +}_{L,m,w} \= \,\alphaL^+ - \frac{\gamma}{4\pi}(m-w \, \lmp^+) \,.
\ee

The signs in the above expressions follow from the form of the covariant derivative 
in~\eqref{eq:ggeWZWcurrents} and the form of the Lorentzian worldsheet with windings
as given in~\eqref{eq:gLormw}. 
The zero-mode~$h_0$ is now valued in~$U(1)$ of volume~$\gamma\sqrt{2|k|}$, so that we have 
\be  \label{eq:sl2gam}
\begin{split}
Z^{SL_\gamma(2)}_\text{SWZW}(\lmp^+,\lmp^-,\alpha^\pm_L,\alpha^\pm_R) & \= 
\frac{\gamma}{\pi} \sqrt{\frac{|k|}{\lmpt}}  
\sum_{m,w \in \IZ}  \exp \bigl(i \, S^\gamma_{m,w}(\lmp^+,\lmp^-,\a^\pm_L,\a^\pm_R)  \bigr) \,, \end{split}\ee
with
\be\begin{split} \quad S^\gamma_{m,w}(\lmp^+,\lmp^-,\a^\pm_L,\a^\pm_R)  & \= 
\frac{\pi\,k}{\lmpt} \Bigl( -4\alpha^{\gamma, -}_{R,m,w} \, \alpha^{\gamma, +}_{L,m,w}
+  2 \alphaR^- \, {\a}_L^+ +  \bigl( \alphaL^- \, {\a}_L^+ +  \alphaR^- \, {\a}_R^+ \bigr) 
\Bigr)\,.
\end{split}
\ee

This expression is invariant under modular transformations, since a modular transformation only relabels
the summation variables~$m$, $w$, and therefore leaves the lattice invariant. 
For~$\a^\pm_R = \a^\pm_L = \a^\pm$ and for~$\a^\pm_R = -\a^\pm_L = \a^\pm$, there is also an 
invariance under elliptic transformations~$\a^\pm \to \a^\pm + \frac{\gamma}{4\pi}(\IZ \rho^\pm + \IZ)$.

\smallskip

The fermionic determinant in the twisted sector is given by extending 
the formula~\eqref{eq:sl2ferdet} to the case of general~$\gamma$, 
\be   \label{eq:sl2gamfer}
\begin{split}
& Z^{SL_\gamma(2)}_{\text{fer},m,w}(\lmp^+,\lmp^-, \alpha^\pm_L,\alpha^\pm_R) \= \\
& \;  2 \,   e^{-\frac{2 \pi i}{\lmpt} \left( \left(\a_{R,m,w}^{\gamma,-}\right)^2 + \left(\a_{L,m,w}^{\gamma,+} \right)^2
- \a_{R,m,w}^{\gamma,-} \a_{R,m,w}^{\gamma,+} - \a_{L,m,w}^{\gamma,-} \a_{L,m,w}^{\gamma,+} \right) }  
\vth_1(\lmp^-,2\a_{R,m,w}^{\gamma,-}) \, \vth_1(-\lmp^+,-2\a_{L,m,w}^{\gamma,+}) \,.
\end{split}
\ee
We recall that for $G=SU(2)$, the theory of the adjoint-valued free fermions is equivalent to the WZW model 
at level 2, but nothing like that is true for a non-compact group such as $SL(2,\R)$.   
For example, the states obtained by quantizing fermions in the adjoint representation (or any finite-dimensional 
representation) of $SL(2,\R)$ are all in finite-dimensional representations of $SL(2,\R)$, but the states
of the $SL(2,\R)$ WZW model are in infinite-dimensional representations of $SL(2,\R)$, as analyzed 
in~\cite{Maldacena:2000hw}. Likewise, the fermion partition function \eqref{eq:sl2gamfer} does not coincide 
with the partition function of the $SL(2,\R)$ WZW model at any value of $k$.

Upon dividing by the fermionic determinant in each twisted sector, we obtain the bosonic partition function,
given by  
\be  \label{eq:SL2gambos}
\begin{split}
&Z^{SL_\gamma(2)}_\text{WZW}(\lmp^+,\lmp^-,\alpha^\pm_L,\alpha^\pm_R)\= \\
& \qquad \frac{\gamma}{2\pi} \sqrt{\frac{|k|}{\lmpt}} \; 
  e^{i\frac{\pi\,k}{\lmpt} \bigl(  2   \a_{R}^- \, {\a}_{L}^+  +  \a_{L}^- \, {\a}_{L}^+ 
  +  \a_{R}^- \, {\a}_{R}^+  \bigr)} \times \\
& \qquad \qquad  \sum_{m,w \in \IZ} 
\frac{ e^{-\frac{4\pi i k}{\lmpt}  \a_{m,w,R}^- \, {\a}_{m,w,L}^+ + 
\frac{2 \pi i}{\lmpt} \left( \left(\a_{R,m,w}^{\gamma,-}\right)^2 + \left(\a_{L,m,w}^{\gamma,+} \right)^2
- \a_{R,m,w}^{\gamma,-} \a_{R,m,w}^{\gamma,+} - \a_{L,m,w}^{\gamma,-} \a_{L,m,w}^{\gamma,+} \right)}
 }{ \vth_1(\lmp^-,2\,\a_{R,m,w}^{\gamma,-}) \, \vth_1(-\lmp^+,-2\, \a_{L,m,w}^{\gamma,+}) }\,.
\end{split}
\ee

For~$\gamma = 2\pi n, n \in \IZ$, it is possible to write a simpler formula because, as explained in section \ref{sltwo},
the fermions can be decoupled.    Concretely, for $\gamma=2\pi n$, the theta function arguments
$2\alpha^-_{R,m,w}$ and $-2\alpha^+_{R,m,w}$ differ from
$2\alpha^-_L$ and $-2 \alpha^+_R$ by lattice vectors $n(m-w\rho^\mp)$.   
Consequently, for $\gamma=2\pi n$,  it is
possible as follows to write a formula with $m,w$ omitted in the factors that came from the fermion partition function.
To this aim, we note that the arguments~$\a_{R,m,w}^{\gamma,-}, \a_{L,m,w}^{\gamma,+}$ given in~\eqref{eq:alphamw} are mapped 
to $\a_R^-$, $\a_L^+$ under a constant gauge transformation by~$n(m-w\rho^\mp)$  in opposite directions.  
The effect of such a constant gauge transformation on the fermionic partition function is given by 
the analytic continuation of~\eqref{eq:ferelliptic3}. 
Accordingly, the  partition function  of the bosonic $SL_\gamma(2,\R)$ model, for $\gamma=2\pi n$, takes the following form:
\be \label{eq:SL2gamdecoupling}
\begin{split}
&Z^{SL_{2\pi n}(2)}_\text{WZW}(\lmp^+,\lmp^-,\alpha^\pm_L,\alpha^\pm_R)\= \\
& \qquad \frac{\gamma}{2\pi} \sqrt{\frac{|k|}{\lmpt}} \; 
 \frac{  e^{i\frac{\pi\,(k-2)}{\lmpt} \bigl(  2   \a_{R}^- \, {\a}_{L}^+  +  \a_{L}^- \, {\a}_{L}^+ 
  +  \a_{R}^- \, {\a}_{R}^+  \bigr)} \; 
  e^{\frac{2 \pi i}{\lmpt} (\a_R^-+\a_L^+)^2}  }{ \vth_1(\lmp^-,2\,\a_{R}^{-}) \, \vth_1(-\lmp^+,-2\, \a_{L}^{+}) } \; \times \\
&\qquad \sum_{m,w \in \IZ} 
e^{\frac{4\pi i k}{\lmpt}  \left(\alpha_L^- + \frac{n}{2}(m-w \, \lmp^-) \right) \, \left(-\alpha_R^+ + \frac{n}{2} (m-w \, \lmp^+) \right)} \,
e^{i\frac{\pi n}{\rho_0} \left( (\alphaR^-+\alphaL^-)(m-w\lmp^+ ) - (\alphaR^+ +\alphaL^+)(m-w\lmp^- )\right)} \,.
\end{split}
\ee

If we set $n=1$, then $SL_\gamma(2,\R)$ reduces to $SL(2,\R)$, but the formula (\ref{eq:SL2gamdecoupling}) 
does not reduce to $Z^{SL(2)}_{\rm WZW}$ as computed in (\ref{eq:sl2bos}).
They differ by the presence on the right hand side of eqn.~(\ref{eq:SL2gamdecoupling}) of a 
factor $e^{i\frac{\pi n}{\rho_0} \left( (\alphaR^-+\alphaL^-)(m-w\lmp^+ ) - (\alphaR^+ +\alphaL^+)(m-w\lmp^- )\right)} $
with an exponent proportional to $n$ but independent of $k$.   The fact that two different derivations 
of the partition function of the $SL(2,\R)$ WZW model differ by this factor reflects
the fact that the coupling of abelian gauge fields to the $SL_\gamma(2,\R)$ model is not completely unique; 
it can be modified by the couplings of eqn.~(\ref{modac}).
Since those couplings do not have an analog for the coupling of the $SU(2)$ WZW model to gauge fields 
of $SU(2)_L\times SU(2)_R$, they will not appear in a derivation that starts there,
proceeds by specializing to gauge fields of $U(1)_L\times U(1)_R$, and then analytically continues to $SL(2,\R)$.  
That is the route that led to eqn.~(\ref{eq:sl2bos}).
By contrast, eqn.~(\ref{eq:SL2gamdecoupling}) was obtained in a derivation that started 
from $SL_\gamma(2,\R)$ with generic~$\gamma$.   In that context, there is no obstruction to
the presence of the couplings~(\ref{modac}), and, for~$\gamma=2\pi n$,  they are appearing with $u_L=u_R=n$. 
Concretely, in reducing the $SL_\gamma(2,\R)$ result (\ref{eq:SL2gambos}) to eqn.~(\ref{eq:sl2bos}) 
when $\gamma=2\pi$, we had to make a topologically non-trivial gauge transformation of the coupling 
to fermions of the background gauge field, and here one runs into the anomaly of eqn.~(\ref{wintrans}).

For~$-\a_L = \a_R = \a$, there is no anomaly and we obtain 
\be 
\begin{split}
&Z^{SL_{2\pi n}(2)}_\text{WZW}(\lmp^+,\lmp^-,-\alpha^\pm,\alpha^\pm)\= \\
& \qquad \frac{\gamma}{2\pi} \sqrt{\frac{|k|}{\lmpt}} \; \times 
 \frac{e^{\frac{2\pi i}{\rho_0} (\a^- - \a^+)^2}}{ \vth_1(\lmp^-,2\,\a^-) \, \vth_1(-\lmp^+,2\, \a^+) }
\sum_{m,w \in \IZ} 
e^{\frac{4\pi i k}{\lmpt}  \left(\alpha^- + \frac{n}{2}(m-w \, \lmp^-) \right) \, \left(\alpha^+ + \frac{n}{2} (m-w \, \lmp^+) \right)} \,.
\end{split}
\ee
For~$n=1$ we indeed recover the~$SL(2)$ partition function~\eqref{eq:sl2bos}.

\bigskip

\bigskip

\ndt {$\bullet$  $H_3^+$ and $H_3^+/\IZ$}

\smallskip

The $H_3^+$ and~$H_3^+/\IZ$ models are usually studied with $k<0$, in which case the target space has Euclidean signature (it has signature $---$ for $k>0$).
So the worldsheet
also naturally has  Euclidean signature and a complex modular
parameter $\tau$, related to the real parameters used in analyzing the $SL(2,\R)$ theory 
by $ (\lmp^-, \lmp^+)  \mapsto \bigl( \tau,\taubar \bigr) $.    Once one chooses the worldsheet to have Euclidean signature, the $H_3^+$ and $H_3^+/\Z$ actions defined by
analytic continuation from $SU(2)$ are positive for $k<0$ and negative for $k>0$.    So the path integral and the formulas that will come from the localization procedure
are really only well-defined for $k<0$.   Rather than write formulas in terms of a parameter $k$ that is supposed to be negative, we define ${\k}=-k$ and write formulas
in terms of $\k>0$.

Recall that the $H_3^+$ model has just a single $SL(2,\C)$ symmetry, in contrast to the 
independent $G_L$ and $G_R$ symmetries of the WZW model of a group $G$.  Accordingly, 
the gauged version of the model has a single $SL(2,\C)$ gauge field $A$, not the independent
left and right  gauge fields $\AL$ and $\AR$ of the gauged version of a WZW model.   For a flat gauge field 
on a torus, we can assume that $A$ is constant, valued in the Lie algebra of a maximal torus of $SL(2,\C)$.  
We can consider the maximal torus generated by Lie algebra elements that, in the notation of 
eqn.~(\ref{linzo}), act by
\be\label{tellmet} 
k_3\= x_1\frac{\partial}{\partial y_2}+y_2\frac{\partial}{\partial x_1} \,, \qquad 
j_3 \= y_3\frac{\partial}{\partial y_4}-y_4\frac{\partial}{\partial y_3} \,.
\ee
Thus $j_3$ generates a rotation of the $y_3-y_4$ plane, and $k_3$ generates a boost of 
the $x_1,y_2$ plane.  (We write $J_3$ and $K_3$ for the conserved charges corresponding to vector 
fields $j_3$ and $k_3$.)   A flat $\mathfrak{sl}(2,\C)$-valued gauge field will take the form
\be\label{firstone} 
A\=-\frac{\overline \alpha_j \, j_3  +\overline \alpha_k \, k_3}{2i\tau_2} \, \d z \,+ \, 
\frac{ \alpha_j \, j_3 + \alpha_k \, k_3}{2i\tau_2} \, \d \bar z \,.
\ee
The right-moving fermions $\psi$ transform in a holomorphic representation of $SL(2,\C)$, 
which is annihilated by $J_i-i K_i$, and the left-moving fermions $\widetilde\psi$
transform in an antiholomorphic representation, annihilated by $J_i+i K_i$.  Hence $\psi$ 
and $\widetilde\psi$ couple effectively to gauge fields $\AR$ and $\AL$ that
can be described as follows.  Let
\be\label{transfo} 
\alpha_j j_3+\alpha_k k_3\=\alpha'(j_3+i k_3)+\alpha''(j_3-i k_3)\,,
\ee
or equivalently
\be\label{ransfo} 
\alpha' \=\frac{\alpha_j-i\alpha_k}{2}\,, \qquad 
\alpha'' \=\frac{\alpha_j+i\alpha_k}{2} \,.
\ee
Then
\be\label{elfo}
A\=\AL+\AR
\ee
with
\be \label{tellme} 
\begin{split}
\AL &\= (j_3-ik_3) \, \frac{ -\bar\alpha'\, \d z+\alpha'' \d \bar z   } {2i\tau_2}  \\  
\AR &\= (j_3+ik_3) \, \frac{-\bar\alpha'' \d z +\alpha'  \d \bar z   }{2i\tau_2} \,.  
\end{split}
\ee

\smallskip

For~$H_3^+$, the supersymmetric partition function is the analytic continuation 
of the result~\eqref{eq:wtsl2}  for~$\wt{SL}(2)$ model,  and is given by
\be
Z^{H_3^+}_\text{SWZW}(\tau,\a',\a'') \= 
\frac{V}{(2 \pi^2\t_2)^{1/2}}  \exp \Bigl(-\frac{\pi\,{\rm k}}{2\t_2}  \bigl( - 2\a' \, \overline{\a'} + \a'' \, \overline{\a'} 
+   \a' \, \overline{\a''} \, \bigr) \Bigr) \,.
\ee
The factor~$V$ here refers to the (divergent) volume of the subgroup parameterized by 
the bosonic zero-mode~$h_0$ of the localizing solution.

\bigskip

The generator of the~$\IZ$-action on~$H_3^+$ that leads to the quotient~$H_3^+/\IZ$ 
is given by~\eqref{moraru}, which commutes with the twists just discussed. 
As in the~$SL_\gamma(2)$ model, the partition function of the quotient~$H_3^+/\IZ$ twisted by~$\a$ 
is given by  a sum over images of the above result for~$H_3^+$.
The parameters coupling to the generators of the twist action are related as\footnote{The factor of $i$ in this 
relation shows that the decoupling of the fermions that occurs in the $SL_\gamma(2,\R)$ model if $\gamma=2\pi n$ 
does not have an analog for real $\beta$.}~$ \gamma  \mapsto -i \beta$. 
The factor of~$i$ appears because the generator of the~$\IZ$-action is now hyperbolic instead of elliptic. 
The volume of the zero-mode is~$\beta \sqrt{2\,\text{k}}$. 
Assembling all these elements, we obtain the following expression for the 
supersymmetric partition function,
\be  \label{eq:ZsuperH3}
Z^{H_3^+/\mathbb{Z}}_\text{SWZW}(\tau,\beta, \a',\a'') 
\= \frac{\beta}{\pi} \sqrt{\frac{\k}{\tau_2}}  
\sum_{m,w \in \IZ}  \exp \bigl(- S^\beta_{m,w}(\t,\a',\a'')  \bigr) \,,
\ee
with 
\be 
S^\beta_{m,w}(\t,\a',\a'')   \= 
 \frac{\pi\,\k}{\t_2} 
 \bigl(-4\, \a'^{\beta}_{m,w}  \, \overline{\a'^{\beta}_{m,w}}  
 + 2\a' \, \overline{\a'} + \a'' \, \overline{\a'} +   \a' \, \overline{\a''}
\Bigr)\,,
\ee
and
\be
\a'^{\beta}_{m,w} \= \a' + \frac{\beta}{4\pi i}(m - w\,\t) \,.
\ee

We note that the action depends only on $\alpha'$ and $\bar\alpha'$, not $\alpha''$ or $\bar\alpha''$, 
up to terms independent of $m,w$. Something similar happens for the fermions, because positive chirality 
fermions couple only to  the $(0,1)$ part of $A^R$, which is proportional to $\alpha'$, and negative chirality 
fermions couple only to the $(1,0)$ part of $A^L$, which is proportional to $\bar\alpha'$. For the~$H_3^+/\IZ$ 
model, the background gauge field that couples to fermions in the twisted sector~$(m,w)$ is given by 
the ansatz (\ref{tellme}) with $\a'$ replaced by~$\a'^{\beta}_{m,w}$ as given above. 
The fermionic path integral can be read off from~\eqref{zeldico}, \eqref{fermcontx} to be 
\be
Z_{\text{fer},m,w}^{H_3^+/\IZ} \= 2 \, e^{- \frac{8 \pi}{\t_2} \text{Im}(\a'^{\beta}_{m,w})} \; 
\vth_1(\t,2\a'^{\beta}_{m,w} ) \; \overline{\vth_1(\t,2\a'^{\beta}_{m,w} )} \,.
\ee

Upon dividing by the fermionic determinant in each twisted sector, we obtain the bosonic partition function,  
\be \label{eq:PIsl2final}
\begin{split}
& Z^{H_3^+/\mathbb{Z}}_\text{WZW}(\tau,\beta,\a',\a'') \= \\
& \qquad \frac{\beta}{2\pi} \sqrt{\frac{\k}{\tau_2}} \; 
  e^{-\frac{\pi\,\k}{\t_2} \left( 2\a' \, \overline{\a'} + \a'' \, \overline{\a'} +   \a' \, \overline{\a''} \right)}
    \sum_{m,w \in \IZ} 
\frac{ \exp \Bigl(\frac{4\pi\,\k}{\t_2} \, \a'^{\beta}_{m,w}  \, \overline{\a'^{\beta}_{m,w}}  \, + \, 
 \frac{8 \pi}{\t_2} \text{Im}(\a'^{\beta}_{m,w})^2 \; \Bigr)
 }{ \vth_1(\t,2\a'^{\beta}_{m,w} ) \; \overline{\vth_1(\t,2\a'^{\beta}_{m,w} )} }\,.
\end{split}
\ee
These formulas make it clear that the $H_3^+/\Z$ partition function is a sort of analytic continuation of 
the $SL_\gamma(2,\R)$ partition function, with $\gamma\mapsto -i \beta$,
$\rho^-\to \tau$, $\rho^+\to \bar\tau$, and $\alpha_R^-\to \alpha'$, $\alpha_L^+\to \bar\alpha'$.  
For a summary of the correspondence between $SU(2)$, $SL(2,\R)$, and $H_3^+$, see Table~\ref{table:parameters}.

The $H_3^+/\Z$ partition function was previously computed by Maldacena-Ooguri-Son~\cite{Maldacena:2000kv} 
by generalizing a method that had been used earlier for $H^+_3$ in~\cite{Gawedzki:1988nj,Gawedzki:1991yu}.
The motivation in~\cite{Maldacena:2000kv} was precisely to test earlier results about the $SL(2,\R)$ WZW 
model~\cite{{Maldacena:2000hw}} by comparing to the partition function
of a Wick rotated version of the theory, involving $H_3^+$.  In our language, the analytic continuation 
between $SL_\gamma(2,\R)$ and $H_3^+/\Z$ was the motivation for studying~$H_3^+/\Z$.    
To compare our results for the $H_3^+/\Z$ partition function to eqn.~(27) of  \cite{Maldacena:2000kv}, 
we set $\alpha'=\alpha''=0$ (since those parameters are not included in \cite{Maldacena:2000kv}).
When this is done,  the~$m=w=0$ mode is  divergent. 
Regulating this by~$\alpha'=\varepsilon \to 0$, we obtain 
\be \label{eq:PIsl2finalzero}
\begin{split}
Z^{H_3^+/\mathbb{Z}}_\text{WZW}(\tau,\beta, \varepsilon) & \= 
\frac{\beta}{2\pi} \, \frac{1}{|2\pi\varepsilon|^2} \, \sqrt{\frac{\k}{\tau_2}} \, \frac{1}{|\eta(\t)|^2} \\
&  \qquad +  \frac{\beta}{2\pi} \sqrt{\frac{\k}{\tau_2}} \; 
   \sum_{m,w \in \IZ \atop (m,w) \neq (0,0)} 
\frac{ \exp \Bigl(\frac{k\,\beta^2}{4\pi \t_2} |m-w\t|^2 +\frac{2 \pi}{\t_2} \,
\bigl(\text{Im}(\frac{\beta}{2\pi i}(m-w\tau))\bigr)^2 \; \Bigr)
 }{ \vth_1(\t,\frac{\beta}{2\pi i}(m-w\tau)) \, \overline{\vth_1(\t,\frac{\beta}{2\pi i}(m-w\tau))} }\,.
\end{split}
\ee
This agrees with eqn.~(27) of ~\cite{Maldacena:2000kv}
up to an overall factor of~$\sqrt{\frac{|\k-2|}{\k}}$ (this is a subtlety that involves a possible quantum 
correction to the volume of the circle parametrized by $h_0$).

This (near) agreement is not a coincidence. The calculation in \cite{Maldacena:2000kv} for $H_3^+/\Z$ 
proceeds by expressing the path integral as a sum over critical
points with each critical point weighted by an iterated Gaussian integral. The localization fixed points in 
the language of the present article are the critical points that were relevant in \cite{Maldacena:2000kv}, 
and in our computation, each localization fixed point is weighted by a Gaussian integral.
It seems that the method of \cite{Gawedzki:1988nj,Gawedzki:1991yu, Maldacena:2000kv} could also be used to 
analyze the partition function of the $SL(2,\R)$ or $SL_\gamma(2,\R)$ WZW model, provided one is willing to 
work on a worldsheet of Lorentz signature and to compute a result that only makes sense as a distribution.  
It is really only when the Weyl group comes into play (which happens for a compact Lie group such as $SU(2)$, 
and also for noncompact semi-simple Lie groups of rank bigger than 1)  that the machinery of supersymmetric
 localization that we have applied in the present article seems to be needed.
 
In \cite{Maldacena:2000kv}, another parameter is included that we have omitted, corresponding to the 
angular momentum of the BTZ black hole.  To include this parameter, in the definition of $H_3^+/\Z$, 
one takes $\Z$ to act by a rotation of $y_3,y_4$ as well as a boost of $x_1,y_2$.   
 Similarly, as remarked in section \ref{sltwo}, another parameter can be included in the definition 
 of $SL_\gamma(2,\R)=\widetilde{SL}(2,\R)/\Z$, by taking $\Z$ to act
 by a rotation of $y_3,y_4$ as well as a shift of $\theta$.   Including these additional parameters 
 maintains the relation between the two models by analytic continuation.
\begin{table}[h!]
\centering
\begin{tabular}{|c || c | c | c|} 
 \hline
 Theory & Left gauge potentials &  Right gauge potentials & Level \\ [0.5ex] 
 \hline
 $SU(2)$ & $(\a_L, \overline{\a_L})$ & $(\a_R, \overline{\a_R})$ & $k$ \\ [0.5ex] 
 \hline
 $SL(2)$ & $(\a_L^-, \a_L^+)$ & $(\a_R^-, \a_R^+ ) $ & $k$ \\ [0.5ex] 
 \hline
 $H_3^+$ & $(\a'', \overline{\a'})$ & $(\a', \overline{\a''})$ & $-\k$ \\ [0.5ex] 
 \hline
\end{tabular}
\caption{\small{The analytic continuation between the parameters in the different theories. 
The entries are identified vertically, 
e.g.~$\a_L$ in the~$SU(2)$ theory is identified with~$\a_L^-$ in the~$SL(2)$ theory and with~$\a''$ in the~$H_3^+$ theory.}}
\label{table:parameters}
\end{table}

\section*{Acknowledgements}

We thank S.~Ashok, S.~Lester, Y.~L\"u, 
G.~W.~Moore, H. Ooguri, M.~Rangamani, N.~Seiberg, and  L.~Takhtajan for discussions.
S.M. acknowledges the support of the J.~Robert Oppenheimer 
Visiting Professorship 2023-24 at the Institute for Advanced Study, Princeton, USA and 
the STFC UK grants ST/T000759/1, ST/X000753/1.
This work was performed in part at Aspen Center for Physics, 
which is supported by National Science Foundation grant PHY-2210452,
and in part  during the KITP program, ``What is string theory? 
Weaving perspectives together,"which was supported by the grant NSF PHY-2309135 
to the Kavli Institute for Theoretical Physics (KITP).
S.M.~would like to thank the ACP, the KITP, and the Abdus Salam ICTP for hospitality
during the completion of this work.   
Research of E.W. is supported in part by NSF grant PHY-2207584.   E.W.~also acknowledges
the hospitality of the University of Washington, where this work was completed.

\section*{Statements and Declarations: Competing Interests and Data Availability}

On behalf of all authors, the corresponding author states that there is no
conflict of interest and data sharing is not applicable to this article as no 
datasets were generated or analysed during the current study.

\appendix 

\section{Solutions to the localization equations \label{sec:locsol}}

In this appendix, we will explain a few group theoretic details that were 
important in sections~\ref{sec:su2} and~\ref{sec:othergroups}.

In section~\ref{sec:locPIcalc}, we encountered eqn.~(\ref{eq:covconstcurrent}), which reads
\be\label{loca}0\= \partial_z \tilde \J_{\bar z}+[\AL_z,\tilde\J_{\bar z}]
\=\partial_z \tilde \J_{\bar z}+\frac{\alphaL}{2\tau_2}[\sigma^3,\wt\J_{\bar z}] \,. \ee
Diagonalizing $\sigma^3$ in the adjoint representation, and denoting the component of $\J_{\bar z}$ 
with eigenvalue $q$ as  $\J_{\bar z}^q$ ,   the equation becomes
\be\label{zoca}0\=\left(\partial_z +q\frac{\alphaL}{2\tau_2}
\right)\J_{\bar z}^q
\ee
for each $q$.
The values of $q$ are 0 for the Cartan subalgebra and $\pm 2$ for the part orthogonal to it.
For $q=0$, the equation (\ref{zoca}) implies that $\J_{\bar z}^0$ is an antiholomorphic function 
on $T^2$ and hence must be constant.  
We want to show that  for generic $\alphaL$, there are
no nonzero solutions with $q\not=0$.

This is actually a basic fact about  line bundles on a torus.   In explaining this,  because  it is more natural to discuss 
holomorphic line bundles rather than antiholomorphic ones, we replace $z$ with $u=\overline z$.   
Thus for $q\not=0$, setting $\alpha=q\alphaL/2\tau_2$ and $f=\J_{\bar z}^q$,
 the equation we have to 
look at is
\be\label{moca} \left(\frac{\partial}{\partial \bar u} + \alpha\right) f \=0. \ee
This equation tells us that $f$ is a holomorphic section of a holomorphic line bundle 
of degree zero over $T^2$, which is defined by the $\bar\partial$ operator
$\frac{\partial}{\partial \bar u} + \alpha$.  
(The line bundle has degree zero and is topologically trivial because the connection form $\alpha$ is globally defined.)
Such a line bundle is associated to a point in the Jacobian of $T^2$. 
For generic $\alpha$, this line bundle is holomorphically nontrivial.
  A classic result says that a nontrivial holomorphic line
bundle over $T^2$ has no nonzero holomorphic section.    So for generic $\alpha$, eqn.~(\ref{moca}) implies that $f=0$.
Thus, in a solution of eqn.~(\ref{loca}), $h=\J_{\bar z}$ is a constant valued in the Cartan subalgebra $\mathfrak h$.

We also encountered eqn.~\eqref{eq:constcurrent} in section 4.3:
\be\label{poca} D_{\bar z} g \, g^{-1} \= h \ee
or in more detail
\be\label{details}\partial_{\bar z}g\cdot g^{-1}-g A^R_{\bar z}g^{-1} \=-\wt A^L_{\bar z},\ee
where $\wt A^L_{\bar z}= A_{\bar z}^L-h$.
We can write this
\be\label{coca} g\left(\partial_{\bar z} +A^R_{\bar z}\right)g^{-1} \= \partial_{\bar z} +\wt A^L_{\bar z}.  \ee
Here $\partial_{\bar z}+A^R_{\bar z}$ is a $\bar\partial$ operator that defines a holomorphic 
structure on a rank two trivial\footnote{The bundle is trivial because
the connection form $ A^R$ is  globally defined as a one-form.}    vector bundle over $T^2$.
Since $ A^R_{\bar z}$ is valued in the Cartan subalgebra, this holomorphic vector bundle 
is isomorphic to $\L\oplus \L^{-1}$, for some holomorphic
line bundle $\L\to T^2$. Generically $\L$ is nontrivial; we assume that in what follows.  
Likewise, $\partial_{\bar z} +\wt A^L_{\bar z}$   defines a holomorphic
structure on the same trivial rank two vector bundle over $T^2$.   As $\wt A^L$ is valued 
in the Cartan subalgebra, this bundle is isomorphic to $\M\oplus \M^{-1}$ for some holomorphic line bundle $\M$.
Eqn.~(\ref{coca}) asserts that $\L\oplus \L^{-1}$ is isomorphic to $\M\oplus \M^{-1}$ via the gauge transformation~$g$.  
This is only possible if either $\L\cong \M$
or $\L\cong \M^{-1}$.   If $\L\cong \M$, then $g$ must be a diagonal gauge transformation.  
If $\L$ is nontrivial and 
$\L\cong \M^{-1}$, then $g=p \, \wt g$ where $p=\biggl(\,\begin{matrix}0 & -1\cr 1&0\end{matrix} \,\biggr)$ 
(so that $p$ exchanges $\M$ and $\M^{-1}$)  and $\wt g$ is diagonal.  
Here $p$ represents the non-trivial element of the Weyl group of $SU(2)$.   
This analysis this leads to the structure claimed in eqn.~(\ref{eq:Weyltwist}).

With only minor changes, the same arguments apply for any compact, connected and simply-connected simple Lie group $G$.  
In analyzing eqn.~(\ref{zoca}), we decompose $\wt\J_{\bar z}$
in one-dimensional representations of the chosen Cartan subalgebra $\mathfrak h$.   
Then we reason as before for each such representation, showing as before that  the part of $\J_{\bar z}$ 
valued in $\mathfrak h$
is constant, and the part orthogonal to $\mathfrak h$ vanishes.   In analyzing eqn.~(\ref{coca}), 
we use the fact that if $\bar\partial_1$ and $\bar\partial_2$ are $\bar\partial$ operators over $T^2$
that define holomorphic $G$-bundles $E_1,E_2\to T^2$ with the same structure group $\mathfrak h$, 
then a gauge transformation that maps one to the other is the product of a Weyl group element and
a diagonal gauge transformation.   Hence $g=\wt g \, \omega$, where $\omega$ is a Weyl group element and $\wt g$ 
is valued in the maximal torus $H$.  Once one knows this, we can set $\wt g = e^y$ with
$y\in \mathfrak h$, and eqn.~(\ref{coca}) becomes a linear equation for $y$ that  leads immediately 
to the form claimed in eqn.~(\ref{eq:WeyltwistsuN}).  If $G$ is  simple, compact and connected but 
not simply-connected, we still reduce in the same way to the case that $g=\omega \, \wt g $ 
with $\wt g$ valued in $H$, but the description of the resulting solutions is slightly different 
because the condition for $g$ to be single-valued on $T^2$
is modified. Accordingly the ``winding numbers'' generalizing $(m,w)$ for $G=SU(2)$ are not integer-valued 
but can have certain fractional values.

For a noncompact group such as $SL(2,\R)$, the localization equations lead to a similar result, with 
some slight differences in the derivation. We recall from section \ref{formulas} that a Lorentz signature torus 
is described by real light cone coordinates $v^\pm$ and  real moduli $\rho^\pm$.
The localization equation $D_+ (g^{-1} D_- g)=0$ says that the current $g^{-1} D_- g$ is covariantly constant 
along the orbits of the vector field $\frac{\partial}{\partial v^+}$.
That vector field has closed orbits if $\rho^+/\rho^-$ is a rational number, and otherwise its orbits are dense.
Either way, for a generic $\frak h$-valued  background gauge field $A^R$, covariant constancy along the orbits implies
that $g^{-1} D_-g$ is a constant valued in  $\mathfrak h$.
Thus we are left with the equation $g^{-1} D_- g = h$, where $h$ is a constant element of $\mathfrak h$.
As in  the previous derivation, this equation implies that $g$ can be viewed as a gauge transformation that 
conjugates one $\mathfrak h$-valued gauge field to another.
For $SL(2,\R)$ and for  generic $A_L$ and $A_R$, this implies that $g$ must be valued in $H$, 
leading to the picture claimed in section~\ref{formulas}.
For a more general non-compact group $G$, this is modified slightly by the appearance of a  Weyl group.  
For example, if  $H$ is a compact maximal torus of $G$, then $g=g_0 \, \omega$ where $g_0$ is valued in $H$  
and $\omega$ is an element of the Weyl group of a maximal compact subgroup of $G$.  
That Weyl group is trivial for $G=SL(2,\R)$ but is non-trivial for non-compact semisimple Lie groups of higher rank.

\section{Equivalence of twisted~$su(2)_1$ and the twisted self-dual boson \label{app:u1su2eq}}

Here we tie up a loose end mentioned at the end of section~\ref{sec:twisting}. 
We want to show that the partition function of the~$su(2)_1$ model equals the partition function of the 
free boson at~$R=\sqrt{2}$ even in the presence of the external gauge field.  
In fact we will show that the characters of the two theories are equal, from which the 
equality of partition function follows.

More precisely, we show below that the~$su(2)_1$ characters~$\chi^1_\ell (\t,\alpha)$, as given 
by eqn.~\eqref{eq:su2char},\footnote{The $SU(2)$ characters have theta functions of index~3 because 
the $SU(2)$ model at bosonic level~$k=1$ has~$k=3$ as the corresponding supersymmetric level.} 
are related to the~${\mathfrak u}(1)$ characters for~$R^2=2$, as given by eqn.~\eqref{eq:fthetarel}, as 
\be \label{eq:theta31id}
i\chi^1_\ell (\t,\alpha) \= \frac{\vth_{3,\ell} (\t,\a) -\vth_{3,\ell} (\t,-\a)}{-i\vth_1(\t,2\a)}
\=  \frac{\vth_{1,\ell-1} (\t,\a)}{\eta(\t)} \= \chi_{{\mathfrak u}(1),\ell} (\sqrt{2};\tau, \a) \,, \quad \ell \= 1,2 \,.
\ee
The theta functions in this equation are defined in eqn.~\eqref{eq:defthml}.

To prove the identity~\eqref{eq:theta31id}, one writes it as, for~$\ell \= 1,2$,  
\be \label{eq:theta3id}
\begin{split}
 \eta(\tau) \, \bigl(  \vth_{3,\ell} (\t,\a) -\vth_{3,\ell} (\t,-\a)  \bigr)
 \= -i  \vth_1(\t,2\a) \, \vth_{1,\ell-1} (\t,\a) \,.
\end{split}
\ee
Now, using Euler's formula which expresses the~$\eta$-function as a theta-series,
\be \label{eq:Eulereta}
\eta(\tau)\=\frac12 \sum_{\mu \in \mathbb{Z}} \chi_{12}(\mu) \, q^{\,\mu^2/24} \,, \qquad 
\chi_{12}(\mu)\= 
\begin{cases}
+1 \,, & \mu = \pm 1 \; \text{mod} \; 12 \,, \\
-1 \,, & \mu = \pm 5 \; \text{mod} \; 12 \,, \\
0 \,, & \text{otherwise} \,,
\end{cases}
\ee
we write the left-hand side of~\eqref{eq:theta3id} as a sum over~$\lambda,\mu \in \mathbb{Z}$
\be \label{eq:defL}
L \= \biggl( \sum_{\mu, \lambda \in \IZ \atop  \lambda \, \equiv \, \ell \, \text{mod} \; 6} \, - \,
\sum_{\mu, \lambda \in \IZ \atop \lambda \, \equiv \, -\ell \, \text{mod} \; 6} \biggr) \, 
\chi_{12}(\mu) \, q^{\frac{\mu^2}{24}+\frac{\lambda^2}{12}} \, e^{2 \pi i \a \lambda} \,. 
\ee
Writing the odd Jacobi theta function as an odd combination of~$\vartheta_{m,\ell}$
\be
-i \vth_1(\t,2\a) \=  \vth_{2,1} (\t,\a) -\vth_{2,-1} (\t,-\a)  \,,
\ee
we write the right-hand side of~\eqref{eq:theta3id} as a sum over~$r,s \in \mathbb{Z}$, 
\be \label{eq:defR}
R \= \biggl(\sum_{s \, \equiv \, 1 \; \text{mod} \; 4 \atop r \, \equiv \, (\ell-1) \; \text{mod} \; 2} \, - \,
\sum_{s \, \equiv \, -1 \; \text{mod} \; 4 \atop r \, \equiv \, (\ell-1) \; \text{mod} \; 2}\biggr) 
q^{\frac{r^2}{4}+\frac{s^2}{8}} \, e^{2 \pi i \a(r+s)}  \,.
\ee
Now, upon identifying~$\lambda=s+r$, $\mu=s-2r$, we see that the exponents of~$q$ and~$e^{2 \pi i \alpha}$
in~$L$ and~$R$ agree. To show the equalities of the ranges of summation, we solve the congruences,
using the Chinese remainder theorem, as follows. 

From the definition of~$\lambda$ and~$\mu$ we see that~$\lambda \equiv \mu \; \text{mod} \; 3$. 
When~$\ell=1$ we have~$\lambda \equiv \pm 1 \, \text{mod} \; 6$ from~\eqref{eq:defL}, and~$r$ is 
even from~\eqref{eq:defR}, and hence~$\mu \equiv s \; \text{mod} \; 4$.
Then, for~$s \equiv 1 \; \text{mod} \; 4$, we have  
\be
\begin{split}
 \lambda \, \equiv \, \mu \,  \equiv \, 1 \; \text{mod} \; 3 \quad & \Rightarrow \quad 
	\lambda \, \equiv \, 1 \; \text{mod} \; 6  \,, \quad \mu \, \equiv \, 1 \; \text{mod} \; 12  \,, \\
\lambda \, \equiv \, \mu \,  \equiv \, -1 \; \text{mod} \; 3 \quad & \Rightarrow \quad 
	\lambda \, \equiv \, -1 \; \text{mod} \; 6  \,, \quad \mu \, \equiv \, 5 \; \text{mod} \; 12  \,, 
\end{split}
\ee
and for~$s \equiv -1 \; \text{mod} \; 4$, we have 
\be 
\begin{split}
 \lambda \, \equiv \, \mu \,  \equiv \, 1 \; \text{mod} \; 3 \quad & \Rightarrow \quad 
	\lambda \, \equiv \, 1 \; \text{mod} \; 6  \,, \quad \mu \, \equiv \, -5 \; \text{mod} \; 12  \,, \\
 \lambda \, \equiv \, \mu \,  \equiv \, -1 \; \text{mod} \; 3 \quad & \Rightarrow \quad 
	\lambda \, \equiv \, -1 \; \text{mod} \; 6  \,, \quad \mu \, \equiv \, -1 \; \text{mod} \; 12  \,.
\end{split}
\ee
When~$ \lambda \equiv \mu \equiv \pm 1 \; \text{mod} \; 3$,  
the solutions of~$s=+1  \; \text{mod} \; 4$ ($s=-1  \; \text{mod} \; 4$), 
correspond precisely to the conditions given 
in~\eqref{eq:Eulereta} and~\eqref{eq:defL} that give the sign~$+1$ ($-1$), in~$L$. 
When~$\ell=2$, 
we can solve the congruences similarly and we find, once again, that the expressions for~$L$ and~$R$ agree.

\section{Twisted Fermions, Current Algebras, and Anomalies}\label{orbifoldtheory}

\def\s{s}
In section \ref{sec:HamTracesu2}, in the comparison between path integrals and a Hamiltonian
formalism, we needed certain details about ground state quantum numbers in a twisted Hilbert space.  
Those points will be explained here.    

In that derivation, we considered positive chirality fermions $\psi, \overline{\psi}$ of charge~$\pm 2$ coupled to an 
effective $u(1)$ gauge field $a_R$ and negative chirality fermions $\tilde\psi, \overline{\tilde\psi}$ of charge~$\pm 2$ coupled 
to another effective $u(1)$ gauge field $a_L$.  Here we slightly generalize and consider $\psi$ and $\tilde\psi$
to have an arbitrary charge $\s$.    In section \ref{sec:HamTracesu2}, the  gauge
field holonomies  around a spatial circle $v_1\cong v_1+2\pi$ were $e^{i\varphi_{L/R}} $ with
\be\label{varph} \varphi_{L/R}\=2\pi \frac{{\rm Im}\,\alpha^{L/R}}{\tau_2}.\ee
The analysis of ground state quantum numbers  is slightly more simple in an anomaly-free case, 
because in that case the coupling to background gauge fields
is uniquely determined by gauge invariance.   So, to begin with,
we assume that\footnote{The other anomaly-free case $\alphaL=-\alphaR$ can be reduced 
to $\alphaL=\alphaR$  by exchanging $\tilde\psi$ with its hermitian conjugate.}
$\alphaL=\alphaR=\alpha$ and hence $\varphi_L=\varphi_R=\varphi$.    
Instead of coupling the fermions to a flat background gauge field
with  holonomy~$\varphi$, it is equivalent to eliminate the gauge field and modify the periodicity condition 
satisfied by the fermions. For fermions of charge $s$, the modified periodicity condition is
\be\label{twistferm}\psi(v_1+2\pi)\=e^{i \s\varphi}\psi(v_1)\,, \qquad \tilde\psi(v_1+2\pi)\=e^{i\s\varphi}\tilde \psi(v_1)\,.\ee

The Hilbert space of this system of twisted fermions is straightforwardly constructed as a 
Fock space.\footnote{If $e^{i\s\varphi}=1$, the Hilbert space is not quite a standard Fock space; 
in constructing it, one has to take into account the zero-modes of $\psi$, $\tilde\psi$, and their adjoints.}  
The subtlety of interest here is to understand the quantum numbers of the ground state.   
The ground state energy can be found by computing the expectation value in the ground state of the 
normal-ordered energy-momentum tensor, and similarly the ground state charge can be determined
from the ground state expectation value of the normal ordered current operator.  
Alternatively, one can bosonize the fermions and then the values of the charges can be read off classically. 
This is actually the route that we will follow in the anomalous case.

Restricting to $\alphaL=\alphaR=\alpha$ means that to evaluate the partition function (\ref{hilbtr2}), we only 
need to know the diagonal charge of the ground state, that is the sum of the left and right charges $Q_{3,L}$ 
and $Q_{3,R}$.   This sum vanishes regardless of $\varphi$ precisely because we have restricted to an anomaly-free
case.  Varying $\varphi$ can be accomplished by varying the background gauge fields $a_L$ and $a_R$.   
As these are neutral, any change in the ground state charge that results from varying them represents an anomaly.  
In the anomaly-free case  $\alphaL=\alphaR$,  there is no such effect.   

On the other hand, the $\varphi$-dependent shifts in $L_0$ and $\barL_0$ are equal, so these shifts 
cancel in the momentum $P=L_0-\barL_0$, but add in the Hamiltonian
$H=L_0+\bar L_0$.   The ground state value of $L_0$ or $\barL_0$ for a positive or negative chirality 
fermion $\psi$ or $\tilde\psi$ of charge~$\s$ twisted as in eqn.~(\ref{twistferm}) is,
by a standard formula (see for example~ \cite{Polchinski:1998rr})
\be\label{standform}\frac{1}{12}-\frac{|\s\varphi|}{4\pi}+\frac{\s^2\varphi^2}{8\pi^2}. \ee

Here, the constant $1/12$ is the ground state energy of the untwisted fermion $\psi$ and 
its hermitian conjugate $ \bar\psi$ in the Ramond sector.   Together with a similar contribution~$1/24$
from the neutral fermion of the $SU(2)$ WZW model, this constant accounts for the factor~$q^{1/8}$ in eqn.~(\ref{trace1}).   

The  term $-\frac{|\s\varphi|}{4\pi}$ may be puzzling at first sight as it is not analytic at $\varphi=0$.  
The nonanalyticity arises as follows.  At $\varphi=0$, $\psi$ and $\bar\psi$ have one zero-mode each.  
The quantization of these two modes gives two quantum states, of charges $\pm \s/2$. When $\varphi\not=0$, 
the two states have unequal energies.  Which of these states is the ground state depends on the sign
of $\varphi$, and hence the ground state energy is nonanalytic as a function of $\varphi$.   
This nonanalyticity is actually visible in eqn.~(\ref{trace1}), which contains a factor  $\zeta-\zeta^{-1}$ 
(with $\zeta=e^{2\pi i\alpha}$).   For general $\s$, that factor becomes $\zeta^{\s/2}-\zeta^{-\s/2}$, or 
in more detail, \be\label{notr}e^{\pi i\s\,{\rm Re}\,\alpha-\pi \s\,{\rm Im}\,\alpha}-e^{-\pi i\s{{\rm Re\,\alpha}+\pi\s {\rm Im}\,\alpha}}
\=e^{\pi i\s{\rm Re}\,\alpha}e^{-2\pi \tau_2\s(\varphi/4\pi)}-e^{-\pi i\s{\rm Re}\,\alpha}e^{2\pi \tau_2\s(\varphi/4\pi)}\,.\ee   
This factor represents the contribution to the trace in (\ref{hilbtr2}) of a pair of states, one with charge $\s/2$  
and $L_0=\s\varphi/4\pi$  and one with charge $-\s/2$ and $L_0=-\s\varphi/4\pi$.  
Which state has lower energy (and therefore dominates the trace for large $\tau_2$) 
depends on the sign of $\varphi$, so the ground state energy is nonanalytic in~$\varphi$.

Accordingly, the only contribution to $L_0$ in eqn.~(\ref{standform}) that is not already contained in the 
oscillator sum (\ref{trace1}) and must instead be interpreted as an additional contribution to the value 
of $L_0$ for the ground state is the last term, $\frac{\s^2\varphi^2}{8\pi^2}= 2({\rm Im}\,\alpha)^2/\tau_2^2$.  
With $\alphaL=\alphaR=\alpha$, this is also the contribution to $\barL_0$ that is not already included in the 
oscillator sum (\ref{trace1}).  Altogether then, the additional factor in the trace (\ref{hilbtr2}) that is not
already contained in the oscillator sum (\ref{trace1}) is $\exp(-8\pi ({\rm Im}\,\alpha)^2/\tau_2)$, 
in agreement with eqn.~(\ref{zeldico}).

It is less straightforward to extend this calculation to the anomalous case $\alphaL\not=\pm \alphaR$, 
because if complete gauge invariance cannot be achieved, it is tricky to specify what couplings to background 
gauge fields $a^L$, $a^R$ are desired.  Quadratic terms in $a^L,a^R$ can be added and would contribute to the shifts
we are interested in. We want to couple to background gauge fields in such a way that the failure of gauge 
invariance takes precisely the form described in eqn.~(\ref{anomaly}).  
Though this can certainly be done in the fermionic language, a particularly simple procedure is to 
bosonize the fermions, after which the ground state quantum numbers
can be read off classically.   In the bosonized version, $\psi$ and $\tilde\psi$ are replaced by a 
circle-valued field $X$ (obeying $X\cong X+2\pi$), whose action can be described precisely in the form
of eqn.~(\ref{abreduction}), but with $k$ replaced by a value appropriate for free fermions, 
namely\footnote{The free fermion radius is $R=1$ in eqn.~(\ref{eq:wsaction}), which corresponds 
to $k=1/2$ in eqn.~(\ref{abreduction}). This non-integral value of $k$ means that the action we will study does
not arise as an abelian reduction of the $SU(2)$ WZW model, though it is perfectly consistent on its own.}  
$k=1/2$, and with $a^{L/R}$ multiplied by the charge~$\s$ of  $\psi$ and $\tilde\psi$.
The following computation is more transparent if we work on a worldsheet of Lorentz signature.    
So as in section \ref{formulas}, we Wick rotate to Lorentz signature by setting $v_0=-i v_2$,
so that $z=v_1+i v_2$ becomes $v_1-v_0$, $\partial_z = \frac{1}{2}(\partial_1-i \partial_2) $ 
becomes $\frac{1}{2}(\partial_1-\partial_0)$, etc.   
The action becomes
\begin{align}\label{acbec} \notag 
S\=\frac{1}{2\pi}\int \d^2v &\left( \frac{1}{2}(\partial_0X)^2-\frac{1}{2}(\partial_1 X)^2  
 -\frac{\s}{2}(a_{1}^L-a_{0}^L)(\partial_1 X +\partial_0 X) \right.\\
&\left. +\frac{\s}{2}(a_{1}^R+a_{0}^R)(\partial_1 X-\partial_0 X)-\frac{\s^2}{4}
\bigl( (a_{1}^L)^2 -(a_{0}^L)^2+(a_{1}^R)^2- (a_{0}^R)^2 \bigr)\right. \notag\\
&\left.+\frac{\s^2}{2}(a_{1}^L-a_{0}^L)(a_{1}^R+a_{0}^R)\right).
\end{align}
The canonical momentum is then
\be\label{canonmom}
\Pi \=\frac{\delta S}{\delta \partial_0 X}\=
\frac{1}{4\pi}  \Bigl(\partial_0X -\s (a_{1}^L-a_{0}^L) -\s(a_{1}^R+a_{0}^R)\Bigr) \,.
\ee
To determine how the Hamiltonian  and the conserved charges that couple to $a^L$ and $a^R$ 
depend on the background gauge fields, we will write all of these quantities in canonical 
variables.\footnote{The momentum~$P=L_0-\overline{L_0}$ has no dependence on the background 
field, since its eigenvalues are integers.}
We will denote the Hamiltonian and the left and right conserved charges in the presence of the background 
fields as $\h H$, $\h J_L$, and $\h J_R$, and
write simply $H$, $J_L$, and $J_R$ for the corresponding objects in the absence of the background fields.
We have
\be\label{leftcharge}
\h J_L\=\int_0^{2\pi}\d v_1\,\frac{\delta S}{\delta a_{0}^L}\=
\frac{\s}{4\pi}\int_0^{2\pi} \d v_1 \left(4\pi\Pi +\partial_1X +\s \, a_{1}^L\right)
\=J_L+\frac{\s^2}{2}a_{1}^L\,.
\ee
Similarly
\be\label{rightcharge}
\h J_R\=\int_0^{2\pi} \d v_1\,\frac{\delta S}{\delta a_{0}^R}\=
\frac{\s}{4\pi}\int_0^{2\pi} \d v_1 \left(-4\pi\Pi +\partial_1X -\s \, a_{1}^R\right)
\=J_R-\frac{\s^2}{2}a_{1}^R \,.
\ee
Finally,
\begin{align}\label{hamshift}\notag 
\h H &\= \frac{1}{8\pi}\int_0^{2\pi } \d v_1
\left(\bigl(4\pi \Pi +\s(a_{1}^L+a_{1}^R)\bigr) ^2+\bigl(\partial_1 X+\s(a_{1}^L-a_{1}^R)\bigr)^2\right)    \\
    &\=H+a_{1}^L J_L-a_{1}^R J_R+\frac{\s^2}{2}\left( (a_{1}^L)^2+ (a_{1}^R)^2\right).          
\end{align}
A check on these calculations is that $\h J_L$, $\h J_R$, and $\h H$, when written in canonical variables, 
depend only on $a_{1}^{L/R}$ and not on $a_{0}^{L/R}$.
Since $a_{0}^{L/R}$ can be gauged away in a neighborhood of any initial value surface, quantities that 
can be computed from data on an initial value surface, such as the energies
and charges of states, cannot depend on $a_{0}^{L/R}$.  

In eqns.~(\ref{leftcharge}), (\ref{rightcharge}) we have  shifts of $J_L$ and $J_R$ by $\frac{\s^2}{2}a_{1}^L$
and $-\frac{\s^2}{2} a_{1}^R$, respectively.   
For $\s=-2$,\footnote{The fermions have charges~$\pm 2$; $s=-2$ agrees with our conventions in 
section~\ref{sec:HamTracesu2}.} 
with $\varphi_{L/R}/2\pi = a_{1}^{L/R}$ as in 
eqn.~(\ref{varph}),  $J_L$ and $J_R$ are shifted respectively by $-\varphi_L/\pi$ and $\varphi_R/\pi$,
as claimed in section~\ref{sec:HamTracesu2}  in the derivation of eqn.~(\ref{zeldico}).   
Likewise, we  see in eqn.~(\ref{hamshift}) a shift\footnote{
The linear terms $a_{1}^L J_L-a_{1}^R J_R$
are already contained in the naive formula~\eqref{hilbtr} given at the beginning of 
section~\ref{sec:HamTracesu2}, so only the quadratic shift is relevant here. 
} 
in the ground state value of $H=L_0+\barL_0$  
that for $\s=\pm 2$ is $\frac{1}{2\pi^2}(\varphi_L^2+\varphi_R^2)$,
as also claimed in the derivation of eqn.~(\ref{zeldico}).

Apart from the fact that this did not create any extra complication, the reason that we did this calculation 
with a general value of $\s$, rather than setting $\s=\pm 2$ from the beginning,
is that this makes it straightforward to determine the analogous formula for a representation of the affine 
algebra $\frak{su}(2)_k$ at any level.    Chiral fermions in any representation $R$
provide a representation of $\frak{su}(2)_\kappa$, where $\kappa$  is determined by the value of the 
quadratic Casimir operator of $SU(2)$ in the representation $R$.
In general, after picking a maximal torus  $U(1)\subset SU(2)$, $R$ decomposes as the sum of pairs of 
fermions of $U(1)$ charges $\pm \s_i$, for some positive integers $\s_i$,
along with some possibly unpaired neutral fermions.    Relative to this decomposition,
\be\label{thelevel}\kappa\=\frac{1}{2}\sum_i\s_i^2.\ee  Any positive integer $\kappa$ is possible, 
with a suitable choice of $R$.
All formulas found earlier  for shifts of ground state quantum numbers are simply proportional to $\s^2$.   
So for $\frak{su}(2)_\kappa$ for any $\kappa$, the shifts in ground state quantum numbers associated 
with twisting in a spatial direction can be found by simply replacing $\s^2/2$
in the preceding formulas  with the level~$\kappa$.   This was claimed in writing eqn.~(\ref{bosan}).
 
Finally, we should explain the factor of $i$ in the formula (\ref{trace1}) for a certain Hilbert space trace of free fermions.  
The presence or absence of this factor depends on precisely how one defines the operator $(-1)^F$.  
In the Ramond sector, the fermion~$\psi$ and its hermitian adjoint each have one zero-mode, making two 
Majorana fermion zero-modes $\psi_1$ and $\psi_2$ in all.   These modes obey a rank~2 Clifford 
algebra $\{\psi_i,\psi_j\}=2\delta_{ij}$,  which can be represented in a two-dimensional Hilbert space $\H_0$.
They also are expected to commute with the symmetry $\CPT$.   Up to a unitary transformation, we can 
assume that $\CPT$ acts  on $\H_0$ simply by complex conjugation, in which case
for $\psi_1,\psi_2$ to commute with $\CPT $ means simply that they are real $2\times 2$ matrices, for 
instance $\psi_1=\sigma_1$, $\psi_2=\sigma_3$.
The operator $(-1)^F$ should anticommute with $\psi_1,\psi_2$, so it must act in $\H_0$ as a multiple of $\psi_1\psi_2$.  
Classically, we expect two properties of $(-1)^F$: (a) its square should equal 1, and (b) it should commute with $\CPT$.  
However, it is impossible to satisfy both of these conditions.   We can satisfy (b) with $(-1)^F=\pm \psi_1\psi_2$,
or we can satisfy (a) with $(-1)^F=\pm i \psi_1\psi_2$.   The formula (\ref{trace1}) corresponds 
to a choice that satisfies~(a).



\begin{thebibliography}{10}

\bibitem{Choi:2021yuz}
C.~Choi and L.~A. Takhtajan, \emph{{Supersymmetry and trace formulas. Part I.
  Compact Lie groups}},
  \href{http://dx.doi.org/10.1007/JHEP06(2024)026}{\emph{JHEP} {\bf 06} (2024)
  026}, [\href{http://arxiv.org/abs/2112.07942}{{\tt 2112.07942}}].

\bibitem{DiVecchia:1984nyg}
P.~Di~Vecchia, V.~G. Knizhnik, J.~L. Petersen and P.~Rossi, \emph{{A
  Supersymmetric Wess-Zumino Lagrangian in Two-Dimensions}},
  \href{http://dx.doi.org/10.1016/0550-3213(85)90554-1}{\emph{Nucl. Phys. B}
  {\bf 253} (1985) 701--726}.

\bibitem{Pestun_2017}
V.~Pestun, M.~Zabzine, F.~Benini, T.~Dimofte, T.~T. Dumitrescu, K.~Hosomichi
  et~al., \emph{Localization techniques in quantum field theories},
  \href{http://dx.doi.org/10.1088/1751-8121/aa63c1}{\emph{Journal of Physics A:
  Mathematical and Theoretical} {\bf 50} (Oct., 2017) 440301}.


\bibitem{Eskin}
L.~D. \`Eskin, \emph{Heat equation on {L}ie groups},  in \emph{{In Memoriam: N.
  G. Cebotarev (Russian)}}, pp.~113--132.
\newblock Izdat. Kazan. Univ., Kazan$^\prime$, 1964.

\bibitem{Choi:2025iqp}
C.~Choi and L.~A. Takhtajan, \emph{{Supersymmetry and trace formulas III.
  Frenkel trace formula}},  \href{http://arxiv.org/abs/2502.10210}{{\tt
  2502.10210}}.

\bibitem{Cecotti:1992qh}
S.~Cecotti, P.~Fendley, K.~A. Intriligator and C.~Vafa, \emph{{A New
  supersymmetric index}},
  \href{http://dx.doi.org/10.1016/0550-3213(92)90572-S}{\emph{Nucl. Phys. B}
  {\bf 386} (1992) 405--452}, [\href{http://arxiv.org/abs/hep-th/9204102}{{\tt
  hep-th/9204102}}].

\bibitem{Maldacena:1999bp}
J.~M. Maldacena, G.~W. Moore and A.~Strominger, \emph{{Counting BPS black holes
  in toroidal Type II string theory}},
  \href{http://arxiv.org/abs/hep-th/9903163}{{\tt hep-th/9903163}}.

\bibitem{Kiritsis:1997gu}
E.~Kiritsis, \emph{{Introduction to nonperturbative string theory}},
  \href{http://dx.doi.org/10.1063/1.54695}{\emph{AIP Conf. Proc.} {\bf 419}
  (1998) 265--308}, [\href{http://arxiv.org/abs/hep-th/9708130}{{\tt
  hep-th/9708130}}].

\bibitem{Witten:1992xu}
E.~Witten, \emph{{Two-dimensional gauge theories revisited}},
  \href{http://dx.doi.org/10.1016/0393-0440(92)90034-X}{\emph{J. Geom. Phys.}
  {\bf 9} (1992) 303--368}, [\href{http://arxiv.org/abs/hep-th/9204083}{{\tt
  hep-th/9204083}}].

\bibitem{Polchinski:1998rr}
J.~Polchinski, \emph{{String theory. Vol. 2: Superstring theory and beyond}}.
\newblock Cambridge Monographs on Mathematical Physics. Cambridge University
  Press, 12, 2007.
\newblock 10.1017/CBO9780511618123.

\bibitem{Polchinski:1998rq}
J.~Polchinski, \emph{{String theory. Vol. 1: An introduction to the bosonic
  string}}.
\newblock Cambridge Monographs on Mathematical Physics. Cambridge University
  Press, 12, 2007.
\newblock 10.1017/CBO9780511816079.

\bibitem{DiFrancesco:1997nk}
P.~Di~Francesco, P.~Mathieu and D.~Senechal, \emph{{Conformal Field Theory}}.
\newblock Graduate Texts in Contemporary Physics. Springer-Verlag, New York,
  1997.
\newblock 10.1007/978-1-4612-2256-9.

\bibitem{Witten:1983ar}
E.~Witten, \emph{{Nonabelian Bosonization in Two-Dimensions}},
  \href{http://dx.doi.org/10.1007/BF01215276}{\emph{Commun. Math. Phys.} {\bf
  92} (1984) 455--472}.

\bibitem{Gawedzki:1988hq}
K.~Gawedzki and A.~Kupiainen, \emph{{G/h Conformal Field Theory from Gauged WZW
  Model}}, \href{http://dx.doi.org/10.1016/0370-2693(88)91081-7}{\emph{Phys.
  Lett. B} {\bf 215} (1988) 119--123}.

\bibitem{Gawedzki:1988nj}
K.~Gawedzki and A.~Kupiainen, \emph{{Coset Construction from Functional
  Integrals}},
  \href{http://dx.doi.org/10.1016/0550-3213(89)90015-1}{\emph{Nucl. Phys. B}
  {\bf 320} (1989) 625--668}.

\bibitem{Witten:1991mm}
E.~Witten, \emph{{On Holomorphic factorization of WZW and coset models}},
  \href{http://dx.doi.org/10.1007/BF02099196}{\emph{Commun. Math. Phys.} {\bf
  144} (1992) 189--212}.

\bibitem{Dei:2024uyx}
A.~Dei and E.~J. Martinec, \emph{{NS5-brane backgrounds and coset CFT partition
  functions}}, \href{http://dx.doi.org/10.1007/JHEP06(2024)147}{\emph{JHEP}
  {\bf 06} (2024) 147}, [\href{http://arxiv.org/abs/2403.17258}{{\tt
  2403.17258}}].

\bibitem{Witten:1991mk}
E.~Witten, \emph{{The N matrix model and gauged WZW models}},
  \href{http://dx.doi.org/10.1016/0550-3213(92)90235-4}{\emph{Nucl. Phys. B}
  {\bf 371} (1992) 191--245}.

\bibitem{Henningson:1993nr}
M.~Henningson, \emph{{N=2 gauged WZW models and the elliptic genus}},
  \href{http://dx.doi.org/10.1016/0550-3213(94)90614-9}{\emph{Nucl. Phys. B}
  {\bf 413} (1994) 73--83}, [\href{http://arxiv.org/abs/hep-th/9307040}{{\tt
  hep-th/9307040}}].


\bibitem{MurZag}
S.~Murthy and D.~Zagier, ``To appear.''

\bibitem{Zhao:2025hen}
B.~Zhao, \emph{{WZW Partition Functions from Supersymmetric Localization}},
  \href{http://arxiv.org/abs/2507.11673}{{\tt 2507.11673}}.

\bibitem{Lu:2025}
Y.~L{\"u}, ``Private communication, and to appear.''

\bibitem{Maldacena:2000hw}
J.~M. Maldacena and H.~Ooguri, \emph{{Strings in AdS(3) and SL(2,R) WZW model
  1.: The Spectrum}}, \href{http://dx.doi.org/10.1063/1.1377273}{\emph{J. Math.
  Phys.} {\bf 42} (2001) 2929--2960},
  [\href{http://arxiv.org/abs/hep-th/0001053}{{\tt hep-th/0001053}}].

\bibitem{Choi:2023pjn}
C.~Choi and L.~A. Takhtajan, \emph{{Supersymmetry and trace formulas II.
  Selberg trace formula}},  \href{http://arxiv.org/abs/2306.13636}{{\tt
  2306.13636}}.

\bibitem{Ashok:2022thd}
S.~K. Ashok and J.~Troost, \emph{{Path integrals on sl(2, R) orbits}},
  \href{http://dx.doi.org/10.1088/1751-8121/ac802c}{\emph{J. Phys. A} {\bf 55}
  (2022) 335202}, [\href{http://arxiv.org/abs/2204.00232}{{\tt 2204.00232}}].

\bibitem{Bostelmann_2009}
H.~Bostelmann and C.~J. Fewster, \emph{Quantum inequalities from operator
  product expansions},
  \href{http://dx.doi.org/10.1007/s00220-009-0853-x}{\emph{Communications in
  Mathematical Physics} {\bf 292} (July, 2009) 761--795},
  [\href{http://arxiv.org/abs/0812.4760}{{\tt 0812.4760}}].

\bibitem{Maldacena:2000kv}
J.~M. Maldacena, H.~Ooguri and J.~Son, \emph{{Strings in AdS(3) and the SL(2,R)
  WZW model. Part 2. Euclidean black hole}},
  \href{http://dx.doi.org/10.1063/1.1377039}{\emph{J. Math. Phys.} {\bf 42}
  (2001) 2961--2977}, [\href{http://arxiv.org/abs/hep-th/0005183}{{\tt
  hep-th/0005183}}].

\bibitem{Gawedzki:1991yu}
K.~Gawedzki, \emph{{Noncompact WZW conformal field theories}},  in \emph{{NATO
  Advanced Study Institute: New Symmetry Principles in Quantum Field Theory}},
  pp.~0247--274, 10, 1991.
\newblock \href{http://arxiv.org/abs/hep-th/9110076}{{\tt hep-th/9110076}}.

\end{thebibliography}

\bigskip 

\bigskip

\providecommand{\href}[2]{#2}\begingroup\raggedright\endgroup

\end{document}